\documentclass[twocolumn,showpacs,preprintnumbers,amsmath,amssymb]{revtex4}
\usepackage{graphicx}
\usepackage{dcolumn}
\usepackage{bm}

\DeclareMathOperator{\Tr}{Tr}

\usepackage[usenames]{color}

\begin{document}


\title{Criteria of off-diagonal long-range order in Bose and Fermi systems \\
based on the Lee-Yang cluster expansion method}
\author{Naoyuki~Sakumichi$^1$, Norio~Kawakami$^1$ and Masahito~Ueda$^{2,3}$}
\affiliation{$^1$Department of Physics, Kyoto University, Kyoto 606-8502, Japan \\
$^2$Department of Physics, The University of Tokyo, 7-3-1 Hongo, Tokyo 113-0033, Japan \\
$^3$ERATO Macroscopic Quantum Project, JST, Tokyo 113-0033, Japan
}%
\date{\today}
\begin{abstract}
The quantum-statistical cluster expansion method of Lee and Yang is extended to investigate off-diagonal long-range order (ODLRO) in one- and multi-component mixtures of bosons or fermions.
Our formulation is applicable to both a uniform system and a trapped system without local-density approximation and allows systematic expansions of one- and multi-particle reduced density matrices in terms of cluster functions which are defined for the same system with Boltzmann statistics.
Each term in this expansion can be associated with a Lee-Yang graph.
We elucidate a physical meaning of each Lee-Yang graph;
in particular, for a mixture of ultracold atoms and bound dimers, an infinite sum of the ladder-type Lee-Yang $0$-graphs is shown to lead to Bose-Einstein condensation of dimers below the critical temperature.
In the case of Bose statistics, an infinite series of Lee-Yang $1$-graphs is shown to converge and gives the criteria of ODLRO at the one-particle level.
Applications to a dilute Bose system of hard spheres are also made.
In the case of Fermi statistics, an infinite series of Lee-Yang $2$-graphs is shown to converge and gives the criteria of ODLRO at the two-particle level.
Applications to a two-component Fermi gas in the tightly bound limit are also made.
\end{abstract}


\maketitle

\section{INTRODUCTION}

In 1957, Lee and Yang showed that the thermodynamic functions for quantum-mechanical systems of particles obeying Bose-Einstein or Fermi-Dirac statistics can be expressed in terms of cluster functions for the same systems obeying Boltzmann statistics \cite{LY57-0, LY58-I, LY60-IV, LY60-V}.
This method is based on a cumulant expansion of the grand partition function, and hence referred to as the Lee-Yang cluster expansion method.
They subsequently applied this method to a dilute system of hard-sphere bosons and discussed the $\lambda$-transition of liquid helium-4 \cite{LY60-IV, LY60-V}.
However, since a cluster expansion is equivalent to a virial expansion, it is more suitable for dilute quantum many-body systems than dense systems such as liquid helium-4.
Unfortunately, back in those days there was no such dilute quantum many-body system.
Since then, the Lee-Yang cluster expansion method has been almost completely forgotten, with only a few exceptions \cite{M61, dD62, PK63, I64, OU06, SK10}.
Now, however, it seems that the method is best suited to describe ultracold dilute atomic gases.
In the present paper, we extend the Lee-Yang cluster expansion method, which was originally developed to describe a system of hard-core bosons and fermions with small scattering lengths, to describe ultracold atomic gases of bosons and fermions with large scattering lengths.

The Lee-Yang cluster expansion method is a cumulant expansion of the grand partition function in terms of fugacity $z:=\exp(\beta \mu)$,
and enables us to treat a quantum many-body problem systematically,
where $\mu$ is the chemical potential and $\beta := 1/(k_{\rm B} T)$ is the inverse temperature with the Boltzmann constant $k_{\rm B}$ .
Although we usually apply the Matsubara Green's function method or the Feynman graphs to tackle quantum many-body problems, the ranges of validity of these approximations are often not well controlled.
On the other hand, the Lee-Yang method allows us to make a perturbation expansion in terms of the number of the interacting atoms, and utilize the fugacity expansion of the grand potential density, which is equivalent to a virial expansion.
Moreover, the cluster expansion (or virial expansion) for the unitary Fermi gas is currently of great interest 
\cite{R07, LHD, KS11, DB12}, because the cluster integrals (or virial coefficients) $b^{(3)}$ and $b^{(4)}$ have recently been measured using a mixture of $^6$Li in the two lowest hyperfine states
\cite{NNJCS10, KSCZ11}.
Here, the virial expansion is
$\beta p \lambda^3 = 2z + \frac{3\sqrt{2}}{4} z^2 + b^{(3)} z^3 + b^{(4)} z^4 +\cdots$, with the pressure $p$ and the thermal de Broglie length $\lambda := (2\pi \hbar^2 \beta/m)^{1/2}$.
The primary purpose of this paper is to identify the criteria of the off-diagonal long-range order (ODLRO) of the reduced density matrices in Bose and Fermi systems \cite{PO56, Y62} based on the Lee-Yang cluster expansion method.

This paper is organized as follows.
In Sec. II, the quantum cluster expansion method is extended for the case of a two-component Bose or Fermi system of interacting particles.
We represent the grand partition function and the one- and two-particle reduced density matrices in terms of cluster functions obeying Bose-Einstein or Fermi-Dirac statistics.
Our extension is formally applicable to a uniform system and a trapped system without local-density approximation.
The derivation and mathematical justification of our method is given in Appendix \ref{app:CE}. 
In Sec. III, we review the rules by which quantum-statistical cluster functions can be computed from the knowledge of quantum cluster functions obeying Boltzmann

classical- (i.e., Boltzmann-) statistical cluster functions.
This formulation was first developed by Lee and Yang for the case of a one-component system \cite{LY58-I} and it was extended by Pathria and Kawatra for a multi-component system \cite{PK63}.
In Sec. IV, we show that the logarithm of the grand partition function as well as the one- and two-particle reduced density matrices can be expressed as a sum over a particular set of Lee-Yang graphs.
For the case of a one-component uniform system, the expressions of the grand partition function and the one-particle reduced density matrix were obtained by Lee and Yang \cite{LY60-IV}, and the two-particle density matrix was obtained by de Dominicis \cite{dD62}.
We generalize these results to the case of multi-component uniform and trapped systems.
In Sec. V, we elucidate a physical meaning of Lee-Yang primary graphs,
in particular, that of an infinite series of the ladder-type Lee-Yang primary graphs, the sum of which leads to Bose-Einstein condensation (BEC) of tightly bound diatomic molecules below a critical temperature \cite{OU06}.
In Sec. VI, we give the criteria of ODLRO based on the Lee-Yang cluster expansion method for the case of Bose-Einstein statistics.
Applications to a Bose gas of hard spheres are also made.
In Sec. VII,
the corresponding criteria for Fermi-Dirac statistics are given.
Applications to a two-component Fermi gas in the tightly bound limit are also made.
In Sec. VIII, we summarize the main results of this paper.
The detailed proofs of several formula are relegated to the appendices to avoid digressing from the main subject.
The relations between the Lee-Yang and our notations are listed in Appendix \ref{app:notation}. 


\section{formulation for a two-component quantum system}
\label{sec:QCE}

\subsection{Grand partition function}

We consider a two-component gas of bosons or fermions
whose components are either internal states or different atomic species,
and distinguished by indices $\sigma = \uparrow, \downarrow$.
A generalization to a general multi-component system is straightforward [see, for example, Ref.~\cite{PK63}].
To make the wave function square integrable,
a system is considered to be in a uniform finite volume $V=L^3$
or in a trap with an effective volume $v_{\text{trap}}$.
In the case of a uniform system,
the Hamiltonian is given by
\begin{equation}
\begin{split}
 H^{(N_\uparrow, N_\downarrow)} = 
& - \sum_{\sigma=\uparrow, \downarrow} \frac{1}{2m_\sigma}
 \sum_{i_\sigma=1}^{N_\sigma} \nabla_{i_\sigma}^2 \\
& + \sum_{\sigma=\uparrow, \downarrow} \sum_{\sigma'=\uparrow, \downarrow} \sum_{i_\sigma,j_{\sigma'}} v_{\sigma\sigma'} \left(\left|\bold{r}_{i_\sigma}-\bold{r}_{j_{\sigma'}} \right| \right), 
\end{split}
\end{equation}
where $m_\sigma$ and $N_\sigma$ are the atomic mass and the particle number in state $\sigma$.
Here we set $\hbar=1$.
In the case of a trapped system, the Hamiltonian should include additional terms
$\sum_{i_\uparrow}v_{\text{trap}}(\bold{r}_{i_\uparrow})+\sum_{i_\downarrow}v_{\text{trap}}(\bold{r}_{i_\downarrow})$.
We assume that the system is kept at constant inverse temperature $\beta=(k_{\rm B} T)^{-1}$,
and that the system is allowed to exchange two kinds of atoms with a reservoir at given fugacities $z_\uparrow$ and $z_\downarrow$,
where the fugacity $z_\sigma$ ($\sigma=\uparrow,\downarrow$) is related to the chemical potential per atom $\mu_\sigma$ by
$z_\sigma = \exp(\beta\mu_\sigma)$.

We introduce the function
\begin{align}
& \langle 1_\uparrow',\dots , N_\uparrow'; 1_\downarrow',\dots , N_\downarrow'
| W_{\alpha}^{(N_\uparrow, N_\downarrow)}  | 
  1_\uparrow,\dots , N_\uparrow; 1_\downarrow,\dots , N_\downarrow \rangle \notag\\
& := (N_\uparrow !) (N_\downarrow !) \!\! \!\!\!\!\!\!\!
\sum_{\quad\psi_i \in \mathcal{H}_{\alpha_\uparrow}^{(N_\uparrow)} \otimes \mathcal{H}_{\alpha_\downarrow}^{(N_\downarrow)}} \!\!\!\!\!
 \psi_i \left( 1_\uparrow',\dots , N_\uparrow'; 1_\downarrow',\dots , N_\downarrow' \right) \notag\\
& \qquad\qquad\qquad
 \times \psi_i \left( 1_\uparrow,\dots , N_\uparrow; 1_\downarrow,\dots , N_\downarrow \right)  e^{-\beta E_i},
\label{eq:def:WA}
\end{align}
where
$\alpha := (\alpha_\uparrow, \alpha_\downarrow)$,
$1_\uparrow := \bold{r}_{1\uparrow}
 := (r_{1\uparrow,x},r_{1\uparrow,y},r_{1\uparrow,z})$,
and
$1'_\uparrow := \bold{r}_{1'\uparrow}
:= (r_{1'\uparrow,x},r_{1'\uparrow,y},r_{1'\uparrow,z})$,
etc., 
and $\psi_i$ and $E_i$ are the normalized eigenfunction and the corresponding eigenvalue of 
 $H^{(N_\uparrow, N_\downarrow)}$, respectively.
The summation in Eq.~(\ref{eq:def:WA})
extends over all eigenfunctions $\psi_i \in \mathcal{H}_{\alpha_\uparrow}^{(N_\uparrow)} \otimes \mathcal{H}_{\alpha_\downarrow}^{(N_\downarrow)}$,
where the subscript $\alpha_\sigma=$ S or A 
indicates that the $N_\sigma$-particle Hilbert space $\mathcal{H}_{\alpha_\sigma}^{(N_\sigma)}$ is symmetrized or antisymmetrized.

To obtain the logarithm of the grand partition function, 
we define the matrix elements of cluster functions $U_{\alpha}^{(N_\uparrow, N_\downarrow)}$ in coordinate representation as follows \cite{U27}:
\begin{equation}
\begin{split}
    \langle 1_\uparrow' | W_{\alpha}^{(1,0)} | 1_\uparrow \rangle
    & \equiv \langle 1_\uparrow' | U_{\alpha}^{(1,0)} | 1_\uparrow \rangle, \\
    \langle 1_\downarrow' | W_{\alpha}^{(0,1)} | 1_\downarrow \rangle
    & \equiv \langle 1_\downarrow' | U_{\alpha}^{(0,1)} | 1_\downarrow \rangle, \\
    \langle 1_\uparrow', 2_\uparrow' | W_{\alpha}^{(2,0)} | 1_\uparrow, 2_\uparrow \rangle
    & \equiv  \langle 1_\uparrow', 2_\uparrow' | U_{\alpha}^{(2,0)} | 1_\uparrow, 2_\uparrow \rangle \\
    & \quad +  \langle 1_\uparrow' | U_{\alpha}^{(1,0)} | 1_\uparrow \rangle \, \langle 2_\uparrow' | U_{\alpha}^{(1,0)} | 2_\uparrow \rangle,\\
    \langle 1_\uparrow'; 1_\downarrow' | W_{\alpha}^{(1,1)} | 1_\uparrow; 1_\downarrow \rangle
    & \equiv  \langle 1_\uparrow'; 1_\downarrow' | U_{\alpha}^{(1,1)} | 1_\uparrow; 1_\downarrow \rangle \\
    & \quad +  \langle 1_\uparrow' | U_{\alpha}^{(1,0)} | 1_\uparrow \rangle \, \langle 1_\downarrow' | U_{\alpha}^{(0,1)} | 1_\downarrow \rangle,\\
    \langle 1_\downarrow', 2_\downarrow' | W_{\alpha}^{(0,2)} | 1_\downarrow, 2_\downarrow \rangle
    & \equiv  \langle 1_\downarrow', 2_\downarrow' | U_{\alpha}^{(0,2)} | 1_\downarrow, 2_\downarrow \rangle \\
    & \quad +  \langle 1_\downarrow' | U_{\alpha}^{(0,1)} | 1_\downarrow \rangle \, \langle 2_\downarrow' | U_{\alpha}^{(0,1)} | 2_\downarrow \rangle,\\
      \langle 1_\uparrow', 2_\uparrow' , 3_\uparrow' | W_{\alpha}^{(3,0)} |  1_\uparrow, &2_\uparrow, 3_\uparrow \rangle \\
      \equiv \langle 1_\uparrow',  2_\uparrow' , 3_\uparrow' |  U_{\alpha}^{(3,0)} & | 1_\uparrow, 2_\uparrow , 3_\uparrow \rangle  \\
      \quad +  \langle 1_\uparrow' |  U_{\alpha}^{(1,0)}   | 1_\uparrow \rangle & \langle 2_\uparrow', 3_\uparrow' | U_{\alpha}^{(2,0)} | 2_\uparrow, 3_\uparrow \rangle \\
      \quad + \langle 2_\uparrow' |  U_{\alpha}^{(1,0)}  | 2_\uparrow \rangle & \langle 3_\uparrow', 1_\uparrow' | U_{\alpha}^{(2,0)} | 3_\uparrow, 1_\uparrow \rangle \\
      \quad + \langle 3_\uparrow' |  U_{\alpha}^{(1,0)}  | 3_\uparrow \rangle & \langle 1_\uparrow', 2_\uparrow' | U_{\alpha}^{(2,0)} | 1_\uparrow, 2_\uparrow \rangle \\
      \quad +\langle 1_\uparrow' |  U_{\alpha}^{(1,0)}  | 1_\uparrow \rangle & \langle 2_\uparrow' | U_{\alpha}^{(1,0)} | 2_\uparrow \rangle \langle 3_\uparrow' | U_{\alpha}^{(1,0)} | 3_\uparrow \rangle , 
 {\rm etc.}
\label{eq:def:UA}
\end{split}
\end{equation}
%
The grand partition function is given by
\begin{equation}
\begin{split}
\Xi
= & \sum_{N_\uparrow = 0}^\infty \sum_{N_\downarrow = 0}^\infty
   \frac{z_\uparrow^{N_\uparrow}}{N_\uparrow!}
   \frac{z_\downarrow^{N_\downarrow}}{N_\downarrow!} 
\int \! d^{3N_\uparrow}\bold{r}_{\uparrow} \int d^{3N_\downarrow}\bold{r}_{\downarrow} \\
& \times
 \langle 1_\uparrow,\dots , N_\uparrow; 1_\downarrow,\dots , N_\downarrow| \\
& \qquad W_{\alpha}^{(N_\uparrow, N_\downarrow)}  | 
  1_\uparrow,\dots , N_\uparrow; 1_\downarrow,\dots , N_\downarrow \rangle ,
\label{eq:def:Xi}
\end{split}
\end{equation}
where
$\int \! d^{3N_\uparrow}\bold{r}_{\uparrow} :=
\int \! d^3\bold{r}_{1\uparrow} \dotsi \int d^3\bold{r}_{N\uparrow}$
and
$\int \! d^{3N}\bold{r}_{\downarrow} :=
\int \! d^3\bold{r}_{1\downarrow} \dotsi \int d^3\bold{r}_{N\downarrow}$.

We can show that
\begin{equation}
\begin{split}
\log \Xi
 &= \sum_{l_\uparrow = 0}^\infty \sum_{l_\downarrow = 0}^\infty
   z_\uparrow^{l_\uparrow}
   z_\downarrow^{l_\downarrow}
 B^{(l_\uparrow, l_\downarrow)},
 \label{eq:Xi-b}
\end{split}
\end{equation}
where cluster integrals are
\begin{align}
& B^{(0,0)} := 0, \notag\\
& B^{(l_\uparrow, l_\downarrow)}
 := \frac{1}{(l_\uparrow !) (l_\downarrow !)} 
\int \! d^{3l_\uparrow}\bold{r}_{\uparrow} \int d^{3l_\downarrow}\bold{r}_{\downarrow} 
\label{eq:b-UA} \\
& \quad \times 
 \langle 1_\uparrow,\dots , l_\uparrow; 1_\downarrow,\dots , l_\downarrow| 
   U_{\alpha}^{(l_\uparrow, l_\downarrow)}  | 
  1_\uparrow,\dots , l_\uparrow; 1_\downarrow,\dots , l_\downarrow \rangle . \notag
\end{align}
The proof of Eq.~(\ref{eq:Xi-b}) is given in in Appendix \ref{app:CE}-1.


Once the cluster integrals are obtained,
we can evaluate thermodynamic quantities.
For the case of a uniform system in a finite volume $V=L^3$,
the equilibrium pressure $p$ and the particle-number density $\rho_\sigma := \langle \hat{n}_\sigma \rangle $ of the system are given by
\begin{equation}
p 
  = \frac{1}{\beta} \lim_{V\to\infty} \frac{1}{V} 
  \sum_{l_\uparrow = 0}^\infty \sum_{l_\downarrow = 0}^\infty
   z_\uparrow^{l_\uparrow}
   z_\downarrow^{l_\downarrow}
 B^{(l_\uparrow, l_\downarrow)},
\end{equation}
and
\begin{equation}
 \rho_\sigma 
  =   \lim_{V\to\infty} \frac{1}{V} 
  \sum_{l_\uparrow = 0}^\infty \sum_{l_\downarrow = 0}^\infty
  l_\sigma
   z_\uparrow^{l_\uparrow}
   z_\downarrow^{l_\downarrow}
 B^{(l_\uparrow, l_\downarrow)}.
\end{equation}
Similarly, for the case of a trapped system,
the particle number $N_\sigma$ of the system is given by
\begin{equation}
 N_\sigma 
  =  
  \sum_{l_\uparrow = 0}^\infty \sum_{l_\downarrow = 0}^\infty
  l_\sigma
   z_\uparrow^{l_\uparrow}
   z_\downarrow^{l_\downarrow}
 B^{(l_\uparrow, l_\downarrow)}.
\end{equation}

\subsection{One- and two-particle reduced density matrices}

In terms of second-quantized field operators,
the one- and two-particle reduced density matrices can be expressed as 
\begin{equation}
\langle \bold{r} \sigma | \rho^{(1)} | \bold{r}' \sigma' \rangle
 := \langle \hat{\Psi}_ {\sigma}^{\dagger} (\bold{r}) \, \hat{\Psi}_ {\sigma'} (\bold{r}') \rangle,
\end{equation}
\begin{equation}
\begin{split}
& \langle \bold{r}_1 \sigma_1, \bold{r}_2 \sigma_2 | \rho^{(2)} | \bold{r}_1' \sigma_1', \bold{r}_2' \sigma_2' \rangle  \\
& := \langle \hat{\Psi}_ {\sigma_1}^{\dagger} (\bold{r}_1) \hat{\Psi}_{\sigma_2}^{\dagger} (\bold{r}_2   )
  \hat{\Psi}_ {\sigma_2'} (\bold{r}'_2) \hat{\Psi}_{\sigma_1'} (\bold{r}_1')  \rangle,
\end{split}
\end{equation}
where 
$\langle \hat{\mathcal{O}} \rangle$ 
is the expectation value of $\hat{\mathcal{O}}$ over the grand canonical ensemble defined by
\begin{align}
 \langle \hat{\mathcal{O}} \rangle
 : = \Xi^{-1}
   \Tr_{\mathcal{H}_{\alpha}}\left[ z_\uparrow^{\hat{N}_\uparrow} z_\downarrow^{\hat{N}_\downarrow}
 e^{-\beta \hat{H}}  \hat{\mathcal{O}} \right],
\end{align}
and
$\hat{\Psi}_ {\sigma}^{\dagger} (\bold{r})$ and $\hat{\Psi}_ {\sigma} (\bold{r})$
are the creation and annihilation operators.
Here, $\hat{H}$ and $\hat{N}_\sigma :=\int d^3 \bold{r} \, \hat{\Psi}_ {\sigma}^{\dagger} (\bold{r}) \, \hat{\Psi}_ {\sigma} (\bold{r}) $
are, respectively, the second-quantized operators for the Hamiltonian and the total number of particles with spin $\sigma$.
[See, for example, Ref.~\cite{Leggett_QL}, in particular, Eqs.~(2.1.6) and (2.4.2)]

The one-particle reduced density matrix can be shown to be related to 
$U_{\alpha}^{(l_\uparrow, l_\downarrow)}$
by 
\begin{equation}
\begin{split}
& \langle \bold{x} \uparrow | \rho^{(1)} | \bold{y} \uparrow \rangle  \\
& = 
 \sum_{l_\uparrow = 1}^\infty \sum_{l_\downarrow = 0}^\infty
   \frac{z_\uparrow^{l_\uparrow}}{(l_\uparrow -1)!}
   \frac{z_\downarrow^{l_\downarrow}}{l_\downarrow!}
 \int \! d^{3(l_\uparrow-1)}\bold{r}_{\uparrow} \int d^{3l_\downarrow}\bold{r}_{\downarrow} \\
& \quad \times 
 \langle \bold{y}, 1_\uparrow,\dots , l_\uparrow -1; 1_\downarrow,\dots , l_\downarrow | \\
& \qquad \quad U_{\alpha}^{(l_\uparrow, l_\downarrow)}  | 
 \bold{x}, 1_\uparrow,\dots , l_\uparrow -1; 1_\downarrow,\dots , l_\downarrow \rangle ,
\label{eq:DM1-UA-u}
\end{split}
\end{equation}
and
\begin{equation}
\begin{split}
& \langle \bold{x} \downarrow | \rho^{(1)} | \bold{y} \downarrow \rangle  \\
& = 
 \sum_{l_\uparrow = 0}^\infty \sum_{l_\downarrow = 1}^\infty
   \frac{z_\uparrow^{l_\uparrow}}{l_\uparrow!}
   \frac{z_\downarrow^{l_\downarrow}}{(l_\downarrow -1)!}
 \int \! d^{3l_\uparrow}\bold{r}_{\uparrow} \int d^{3(l_\downarrow-1)}\bold{r}_{\downarrow} \\
& \quad \times 
 \langle 1_\uparrow,\dots , l_\uparrow; \bold{y}, 1_\downarrow,\dots , l_\downarrow -1 | \\
& \qquad \quad U_{\alpha}^{(l_\uparrow, l_\downarrow)}  | 
 1_\uparrow,\dots , l_\uparrow ;\bold{x}, 1_\downarrow,\dots , l_\downarrow -1\rangle .
\label{eq:DM1-UA-d}
\end{split}
\end{equation}
The proofs of Eqs.~(\ref{eq:DM1-UA-u}) and (\ref{eq:DM1-UA-d}) are given in Appendix \ref{app:CE}-2.
Clearly,
$\langle \bold{x} \uparrow | \rho^{(1)} | \bold{y} \downarrow \rangle
= \langle \bold{x} \downarrow | \rho^{(1)} | \bold{y} \uparrow \rangle =0$.

The two-particle reduced density matrix can also be shown to be related to 
$U_{\alpha}^{(l_\uparrow, l_\downarrow)}$
by 
\begin{equation}
\begin{split}
& \langle \bold{x}_1 \uparrow, \bold{x}_2 \uparrow | \rho^{(2)} | \bold{y}_1 \uparrow, \bold{y}_2 \uparrow \rangle   \\
& = \langle \bold{x}_1 \uparrow | \rho^{(1)} | \bold{y}_1 \uparrow \rangle 
 \langle \bold{x}_2 \uparrow | \rho^{(1)} | \bold{y}_2 \uparrow \rangle  \\
& + 
 \sum_{l_\uparrow = 2}^\infty \sum_{l_\downarrow = 0}^\infty
   \frac{z_\uparrow^{l_\uparrow}}{(l_\uparrow -2)!}
   \frac{z_\downarrow^{l_\downarrow}}{l_\downarrow !} 
    \int \! d^{3(l_\uparrow-2)}\bold{r}_{\uparrow} \int d^{3l_\downarrow}\bold{r}_{\downarrow} \\
& \quad \times 
 \langle \bold{y}_1, \bold{y}_2, 1_\uparrow,\dots , l_\uparrow -2;1_\downarrow,\dots , l_\downarrow | \\
& \qquad \quad U_{\alpha}^{(l_\uparrow, l_\downarrow)}  | 
 \bold{x}_1, \bold{x}_2,1_\uparrow,\dots , l_\uparrow -2;  1_\downarrow,\dots , l_\downarrow  \rangle ,
\label{eq:DM2-UA-uu}
\end{split}
\end{equation}
\begin{equation}
\begin{split}
& \langle \bold{x}_1 \downarrow, \bold{x}_2 \downarrow | \rho^{(2)} | \bold{y}_1 \downarrow, \bold{y}_2 \downarrow \rangle   \\
& = \langle \bold{x}_1 \downarrow | \rho^{(1)} | \bold{y}_1 \downarrow \rangle 
 \langle \bold{x}_2 \downarrow | \rho^{(1)} | \bold{y}_2 \downarrow \rangle  \\
& + 
 \sum_{l_\uparrow = 0}^\infty \sum_{l_\downarrow = 2}^\infty
   \frac{z_\uparrow^{l_\uparrow}}{l_\uparrow !}
   \frac{z_\downarrow^{l_\downarrow}}{(l_\downarrow -2)!} 
    \int \! d^{3l_\uparrow}\bold{r}_{\uparrow} \int d^{3(l_\downarrow-2)}\bold{r}_{\downarrow} \\
& \quad \times 
 \langle 1_\uparrow,\dots , l_\uparrow ; \bold{y}_1, \bold{y}_2, 1_\downarrow,\dots , l_\downarrow-2 | \\
& \qquad \quad U_{\alpha}^{(l_\uparrow, l_\downarrow)}  | 
 1_\uparrow,\dots , l_\uparrow ; \bold{x}_1, \bold{x}_2, 1_\downarrow,\dots , l_\downarrow -2 \rangle ,
\label{eq:DM2-UA-dd}
\end{split}
\end{equation}
and
\begin{align}
& \langle \bold{x}_1 \uparrow, \bold{x}_2 \downarrow | \rho^{(2)} | \bold{y}_1 \uparrow, \bold{y}_2 \downarrow \rangle  \notag \\
& =\epsilon \langle \bold{x}_1 \uparrow, \bold{x}_2 \downarrow | \rho^{(2)} | \bold{y}_2 \downarrow, \bold{y}_1 \uparrow \rangle  \notag \\
& = \epsilon \langle \bold{x}_2 \downarrow, \bold{x}_1 \uparrow | \rho^{(2)} | \bold{y}_1 \uparrow, \bold{y}_2 \downarrow \rangle  \notag \\
& = \langle \bold{x}_2 \downarrow, \bold{x}_1 \uparrow | \rho^{(2)} | \bold{y}_2 \downarrow, \bold{y}_1 \uparrow \rangle  \notag \\
& =   \langle \bold{x}_1 \uparrow | \rho^{(1)} | \bold{y}_1 \uparrow \rangle 
 \langle \bold{x}_2 \downarrow | \rho^{(1)} | \bold{y}_2 \downarrow \rangle  \notag\\
& \,\,+ \sum_{l_\uparrow = 1}^\infty \sum_{l_\downarrow = 1}^\infty
   \frac{z_\uparrow^{l_\uparrow}}{(l_\uparrow -1)!}
   \frac{z_\downarrow^{l_\downarrow}}{(l_\downarrow -1)!}
     \int \! d^{3(l_\uparrow-1)}\bold{r}_{\uparrow} \int \! d^{3(l_\downarrow-1)}\bold{r}_{\downarrow} \notag \\
& \quad \times 
 \langle \bold{y}_1, 1_\uparrow,\dots , l_\uparrow -1; \bold{y}_2,1_\downarrow,\dots , l_\downarrow -1 | \label{eq:DM2-UA-ud}\\
& \qquad \quad U_{\alpha}^{(l_\uparrow, l_\downarrow)}  | 
 \bold{x}_1, 1_\uparrow,\dots , l_\uparrow -1; \bold{x}_2, 1_\downarrow,\dots , l_\downarrow -1 \rangle ,\notag
\end{align}
where $\epsilon =+1$ for $\alpha =$ S (Bose-Einstein statistics) and
$\epsilon =-1$ for $\alpha =$ A (Fermi-Dirac statistics).
The proofs of Eqs.~(\ref{eq:DM2-UA-uu})-(\ref{eq:DM2-UA-ud}) are given in Appendix \ref{app:CE}-3.
Clearly, the other components such as
$\langle \bold{x}_1 \uparrow, \bold{x}_2 \downarrow | \rho^{(2)} | \bold{y}_1 \downarrow, \bold{y}_2 \downarrow \rangle$ vanish.
 
A generalization to an $N$-particle reduced density matrix is straightforward.
For example, the three-particle reduced density matrix is
\begin{align}
& \langle \bold{x}_1 \uparrow, \bold{x}_2 \uparrow, \bold{x}_3 \downarrow | \rho^{(3)} | \bold{y}_1 \uparrow, \bold{y}_2 \uparrow, \bold{y}_3 \downarrow \rangle   \notag\\
& = 
 \langle \bold{x}_1 \uparrow | \rho^{(1)} | \bold{y}_1 \uparrow \rangle 
 \langle \bold{x}_2 \uparrow | \rho^{(1)} | \bold{y}_2 \uparrow \rangle 
 \langle \bold{x}_3 \downarrow | \rho^{(1)} | \bold{y}_3 \downarrow \rangle  \notag\\
&\, + 
 \langle \bold{x}_1 \uparrow | \rho^{(1)} | \bold{y}_1 \uparrow \rangle 
 \langle \bold{x}_2 \uparrow, \bold{x}_3 \downarrow | \rho^{(2)} | \bold{y}_2 \uparrow , \bold{y}_3 \downarrow \rangle  \notag\\
&\, + 
 \langle \bold{x}_2 \uparrow | \rho^{(1)} | \bold{y}_2 \uparrow \rangle 
 \langle \bold{x}_1 \uparrow, \bold{x}_3 \downarrow | \rho^{(2)} | \bold{y}_1 \uparrow , \bold{y}_3 \downarrow \rangle  \notag\\
&\, + 
 \langle \bold{x}_3 \downarrow | \rho^{(1)} | \bold{y}_3 \downarrow \rangle  
  \langle \bold{x}_1 \uparrow, \bold{x}_2 \uparrow | \rho^{(2)} | \bold{y}_1 \uparrow , \bold{y}_2 \uparrow \rangle  \notag\\
&\, +
 \sum_{l_\uparrow = 2}^\infty \sum_{l_\downarrow = 1}^\infty
   \frac{z_\uparrow^{l_\uparrow}}{(l_\uparrow -2)!}
   \frac{z_\downarrow^{l_\downarrow}}{(l_\downarrow -1)!}
     \int \! d^{3(l_\uparrow-2)}\bold{r}_{\uparrow} \int \! d^{3(l_\downarrow-1)}\bold{r}_{\downarrow} \notag \\
& \,\, \times 
 \langle \bold{y}_1, \bold{y}_2, 1_\uparrow,\dots , l_\uparrow -2; \bold{y}_3,1_\downarrow,\dots , l_\downarrow -1 | \label{eq:DM3-UA-uud}\\
& \quad\,\,\, U_{\alpha}^{(l_\uparrow, l_\downarrow)}  | 
 \bold{x}_1, \bold{x}_2, 1_\uparrow,\dots , l_\uparrow -2; \bold{x}_3, 1_\downarrow,\dots , l_\downarrow -1 \rangle .\notag
\end{align}

The formulation given in this section is expressed in the form of an infinite series.
Thus, one might question the mathematical rigor of the formulation.
Actually, the derivations of the formulation can be mathematically justified, as shown in Appendix \ref{app:CE}-4.

\section{$U_{\alpha}^{(l_\uparrow, l_\downarrow)}$
    in terms of $U^{(l_\uparrow, l_\downarrow)}$}
\label{sec:U}

In the preceding section, we derived the formulas of the grand partition function and the one- and two-particle reduced density matrices.
To evaluate these quantities,
we must calculate $U_{\alpha}^{(l_\uparrow, l_\downarrow)}$.
In this section we review the rules by which
$U_{\alpha}^{(l_\uparrow, l_\downarrow)}$ can be computed
from the knowledge of certain $U^{(l_\uparrow, l_\downarrow)}$ functions
for the same system obeying Boltzmann statistics.
These rules were first established by Lee and Yang for the case of a one-component system \cite{LY58-I} 
and was extended by Pathria and Kawatra for a multi-component system \cite{PK63}.


\subsection{Boltzmann statistics}

For Boltzmann statistics,
we introduce the function
\begin{align}
& \langle 1_\uparrow',\dots , N_\uparrow'; 1_\downarrow',\dots , N_\downarrow'
| W^{(N_\uparrow, N_\downarrow)}  | 
  1_\uparrow,\dots , N_\uparrow; 1_\downarrow,\dots , N_\downarrow \rangle \notag\\
& 
:= \!\! \!\!\!\!\!\!\!
\sum_{\quad\psi_i \in \mathcal{H}^{(N_\uparrow)} \otimes \mathcal{H}^{(N_\downarrow)}} \!\!\!\!\!
 \psi_i \left( 1_\uparrow',\dots , N_\uparrow'; 1_\downarrow',\dots , N_\downarrow' \right) \notag \\
& \qquad\qquad\,
 \times \psi_i \left( 1_\uparrow,\dots , N_\uparrow; 1_\downarrow,\dots , N_\downarrow \right)  e^{-\beta E_i}.
\label{eq:def:W}
\end{align}
The summation 
on the right-hand side of Eq.~(\ref{eq:def:W}) runs over all eigenfunctions $\psi_i \in \mathcal{H}^{(N_\uparrow)} \otimes \mathcal{H}^{(N_\downarrow)}$,
where $\mathcal{H}^{(N_\sigma)}$ is the $N_\sigma$-particle Hilbert space.

Similarly to Eq.~(\ref{eq:def:UA}),
we define the matrix elements of cluster functions $U^{(N_\uparrow, N_\downarrow)}$ in the coordinate representation as
\begin{equation}
\begin{split}
    \langle 1_\uparrow' | W^{(1,0)} | 1_\uparrow \rangle
    & = \langle 1_\uparrow' | U^{(1,0)} | 1_\uparrow \rangle, \\
    \langle 1_\downarrow' | W^{(0,1)} | 1_\downarrow \rangle
    & = \langle 1_\downarrow' | U^{(0,1)} | 1_\downarrow \rangle, \\
    \langle 1_\uparrow', 2_\uparrow' | W^{(2,0)} | 1_\uparrow, 2_\uparrow \rangle
    & =  \langle 1_\uparrow', 2_\uparrow' | U^{(2,0)} | 1_\uparrow, 2_\uparrow \rangle \\
    & \quad +  \langle 1_\uparrow' | U^{(1,0)} | 1_\uparrow \rangle \, \langle 2_\uparrow' | U^{(1,0)} | 2_\uparrow \rangle,\\
    \langle 1_\uparrow'; 1_\downarrow' | W^{(1,1)} | 1_\uparrow; 1_\downarrow \rangle
    & =  \langle 1_\uparrow'; 1_\downarrow' | U^{(1,1)} | 1_\uparrow; 1_\downarrow \rangle \\
    & \quad +  \langle 1_\uparrow' | U^{(1,0)} | 1_\uparrow \rangle \, \langle 1_\downarrow' | U^{(0,1)} | 1_\downarrow \rangle,\\
    \langle 1_\downarrow', 2_\downarrow' | W^{(0,2)} | 1_\downarrow, 2_\downarrow \rangle
    & =  \langle 1_\downarrow', 2_\downarrow' | U^{(0,2)} | 1_\downarrow, 2_\downarrow \rangle \\
    & \quad +  \langle 1_\downarrow' | U^{(0,1)} | 1_\downarrow \rangle \, \langle 2_\downarrow' | U^{(0,1)} | 2_\downarrow \rangle,\\
      \langle 1_\uparrow', 2_\uparrow' , 3_\uparrow' | W^{(3,0)} |  1_\uparrow, &2_\uparrow, 3_\uparrow \rangle \\
      = \langle 1_\uparrow',  2_\uparrow' , 3_\uparrow' |  U^{(3,0)} & | 1_\uparrow, 2_\uparrow , 3_\uparrow \rangle  \\
      +  \langle 1_\uparrow' |  U^{(1,0)}   | 1_\uparrow \rangle \langle & 2_\uparrow', 3_\uparrow' | U^{(2,0)} | 2_\uparrow, 3_\uparrow \rangle \\
      + \langle 2_\uparrow' |  U^{(1,0)}  | 2_\uparrow \rangle \langle &  3_\uparrow', 1_\uparrow' | U^{(2,0)} | 3_\uparrow, 1_\uparrow \rangle \\
      + \langle 3_\uparrow' |  U^{(1,0)}  | 3_\uparrow \rangle \langle & 1_\uparrow', 2_\uparrow' | U^{(2,0)} | 1_\uparrow, 2_\uparrow \rangle \\
      +\langle 1_\uparrow' |  U^{(1,0)}  | 1_\uparrow \rangle \langle & 2_\uparrow' | U^{(1,0)} | 2_\uparrow \rangle \langle 3_\uparrow' | U^{(1,0)} | 3_\uparrow \rangle , 
 {\rm etc.}
\label{eq:def:U}
\end{split}
\end{equation}

\subsection{$U_{\alpha}^{(l_\uparrow, l_\downarrow)}$
    in terms of $U^{(l_\uparrow, l_\downarrow)}$}

We now formulate the rules by which $U_{\alpha}^{(l_\uparrow, l_\downarrow)}$
can be computed from the knowledge of $U^{(l_\uparrow, l_\downarrow)}$.
Such rules result from the fact that
$U_{\alpha}^{(l_\uparrow, l_\downarrow)}$ and $U^{(l_\uparrow, l_\downarrow)}$
are defined in terms of
$W_{\alpha}^{(l_\uparrow, l_\downarrow)}$ and $W^{(l_\uparrow, l_\downarrow)}$,
respectively, which, in turn, are related to each other through 
\begin{align}
& \langle 1_\uparrow',\dots , N_\uparrow'; 1_\downarrow',\dots , N_\downarrow'
| W_{\alpha}^{(N_\uparrow, N_\downarrow)}  | 
  1_\uparrow,\dots , N_\uparrow; 1_\downarrow,\dots , N_\downarrow \rangle \notag\\
& =  \!\!\! \sum_{P \in S_{N_\uparrow}} \! \sum_{Q \in S_{N_\downarrow}} \!\!\! \epsilon^P \epsilon^Q  
 \bigl\langle P(1_\uparrow'),\dots , P(N_\uparrow'); Q(1_\downarrow'),\dots , Q(N_\downarrow') \big| \notag\\
& \qquad\qquad\qquad    W^{(N_\uparrow, N_\downarrow)} 
  \big|\, 1_\uparrow,\dots , N_\uparrow; 1_\downarrow,\dots , N_\downarrow  \, \bigr\rangle,
\label{eq:W-WA}
\end{align}
where $\epsilon =+1$ for $\alpha =$ S (Bose-Einstein statistics) and
$\epsilon =-1$ for $\alpha =$ A (Fermi-Dirac statistics).
Here, $P$ and $Q$ denote permutations,
and $(-1)^P$ and $(-1)^Q$ take on $1$ or $-1$ for even or odd permutations.
Equation (\ref{eq:W-WA}) is proved in Appendix \ref{app:W}.
It can be seen from the structure of 
Eqs.~(\ref{eq:def:U}) and (\ref{eq:W-WA})
that $U_\alpha^{(l_\uparrow, l_\downarrow)}$
may, in general, be expressed 
in terms of the sum over connected products of $\Upsilon_{\alpha}^{(l_\uparrow, l_\downarrow)}$ functions,
where 
\begin{align}
& \langle 1_\uparrow',\dots , l_\uparrow'; 1_\downarrow',\dots , l_\downarrow'
| \Upsilon_{\alpha}^{(l_\uparrow, l_\downarrow)}  | 
  1_\uparrow,\dots , l_\uparrow; 1_\downarrow,\dots , l_\downarrow \rangle \notag\\
& :=  \sum_{P \in S_{l_\uparrow}} \sum_{Q \in S_{l_\downarrow}} \epsilon^P \epsilon^Q  
 \bigl\langle P(1_\uparrow'),\dots , P(l_\uparrow'); Q(1_\downarrow'),\dots , Q(l_\downarrow') \big| \notag\\
& \qquad\qquad\qquad\quad     U^{(l_\uparrow, l_\downarrow)} 
  \big|\, 1_\uparrow,\dots , l_\uparrow; 1_\downarrow,\dots , l_\downarrow  \, \bigr\rangle .
\label{eq:def:UpsA}
\end{align}

Here, we only show the results:

{\it Rule}---
To calculate
$\langle \bold{r}_{1\uparrow}',\dots , \bold{r}_{l\uparrow}'; \bold{r}_{1\downarrow}',\dots , \bold{r}_{l\downarrow}' | U_{\alpha}^{(l_\uparrow, l_\downarrow)}$
$  |  \bold{r}_{1\uparrow},\dots , \bold{r}_{l\uparrow}; \bold{r}_{1\downarrow},\dots , \bold{r}_{l\downarrow} \rangle$,
we first consider a grouping of the $l_\uparrow + l_\downarrow$ integers
$1_\uparrow,\dots , l_\uparrow, 1_\downarrow,\dots , l_\downarrow$:
\begin{align}
& \bigl\{ (a_\uparrow) (b_\uparrow) \dots \bigr\}
   \bigl\{ (c_\uparrow d_\uparrow) (e_\uparrow f_\uparrow) \dots \bigr\}
   \bigl\{ (g_\uparrow h_\uparrow i_\uparrow) \dots \bigr\}\dots \notag\\
&\bigl\{ (j_\uparrow k_\downarrow) (l_\uparrow m_\downarrow) \dots \bigr\}
   \bigl\{ (n_\uparrow o_\uparrow p_\downarrow) \dots \bigr\}
   \bigl\{ (q_\uparrow r_\downarrow s_\downarrow) \dots \bigr\}\dots \notag\\
& \bigl\{ (t_\downarrow) (u_\downarrow) \dots \bigr\}
   \bigl\{ (v_\downarrow w_\downarrow) (x_\downarrow y_\downarrow) \dots \bigr\}
   \dots ,
\label{eq:Rule}
\end{align}
where $a_\uparrow , b_\uparrow , \dots , q_\uparrow , \dots$
and
$k_\downarrow , m_\downarrow , \dots , y_\downarrow , \dots$
are permutations of these $l_\uparrow$ and $l_\downarrow$ integers,
respectively.
In each pair of the curly brackets, there are $m_{n_\uparrow, n_\downarrow}( = 0,1,2,\dots)$ round brackets with $n_\uparrow$ up-spin integers and $n_\downarrow$ down-spin integers which are subject to
\begin{equation}
\begin{split}
& \sum_{n_\uparrow = 0}^{l_\uparrow}
   \sum_{n_\downarrow = 0}^{l_\downarrow}
   n_\uparrow m_{n_\uparrow, n_\downarrow}
  = l_\uparrow, \\
& \sum_{n_\uparrow = 0}^{l_\uparrow}
   \sum_{n_\downarrow = 0}^{l_\downarrow}
   n_\downarrow m_{n_\uparrow, n_\downarrow}
  = l_\downarrow .
\end{split}
\end{equation}
Within each pair of the round brackets in (\ref{eq:Rule}), the integers for the same spin
are arranged in ascending order, i.e., $c<d$, $e<f$, $g<h<i$, $o<p$, etc.,
and within each pair of the curly brackets, 
the round brackets are arranged in such a manner that their first
entries are placed in ascending order. 
We then form the sum
\begin{equation}
\begin{split}
&\sum
 \bigg(
 \left\{
 \langle \bold{r}_{A\uparrow}' |  \Upsilon_{\alpha}^{(1,0)} | \bold{r}_{a\uparrow} \rangle
 \langle \bold{r}_{B\uparrow}' |  \Upsilon_{\alpha}^{(1,0)} | \bold{r}_{b\uparrow} \rangle
 \cdots \right\} \\
& \qquad\quad\times \left\{
 \langle \bold{r}_{C\uparrow}', \bold{r}_{D\uparrow}' |  \Upsilon_{\alpha}^{(2,0)} | \bold{r}_{c\uparrow}, \bold{r}_{d\uparrow}\rangle
 \cdots \right\} \cdots \\
& \quad \times \left\{
 \langle \bold{r}_{J\uparrow}'; \bold{r}_{K\downarrow}' |  \Upsilon_{\alpha}^{(1,1)} | \bold{r}_{j\uparrow}; \bold{r}_{k\downarrow} \rangle
 \cdots
 \right\} \cdots \\
& \quad \times 
\left\{
 \langle \bold{r}_{T\downarrow}' |  \Upsilon_{\alpha}^{(0,1)} | \bold{r}_{t\downarrow} \rangle
 \langle \bold{r}_{U\downarrow}' |  \Upsilon_{\alpha}^{(0,1)} | \bold{r}_{u\downarrow} \rangle
 \cdots \right\} \cdots
 \bigg),
\end{split}
\label{eq:UA-UpsA}
\end{equation}
where $A, B, \dots , J, \dots$ and $K, \dots , T, U, \dots$ are permutations of $1, 2, \dots , l_\uparrow$ and
$1, 2, \dots , l_\downarrow$, respectively.
Because we use $\Upsilon_{\alpha}^{(l_\uparrow, l_\downarrow)}$ in Eq.~(\ref{eq:def:UpsA}),
only one permutation will be included in the sum in Eq.~(\ref{eq:UA-UpsA})
 among all the permutations $A, B, \dots , J, \dots$ which differ from each other only in the relative positions of numbers within the same bra
(e.g., $\langle \bold{r}_{C\uparrow}', \bold{r}_{D\uparrow}' |$ and $\langle \bold{r}_{D\uparrow}', \bold{r}_{C\uparrow}' |$ ).
The sum in Eq.~(\ref{eq:UA-UpsA}) then runs over all permutations,
provided that upon setting $\bold{r}_{i \uparrow}'=\bold{r}_{i\uparrow}$ and $\bold{r}_{i \downarrow}'=\bold{r}_{i\downarrow}$ for all $i$, the summand in Eq.~(\ref{eq:UA-UpsA}) cannot be written as a product of two factors,
one of which depends only on some, but not all, of the coordinates 
$\bold{r}_{1\uparrow},\dots , \bold{r}_{l\uparrow}$
and
$\bold{r}_{1\downarrow},\dots , \bold{r}_{l\downarrow}$,
while the other depends only on the rest of these coordinates. 
The sum of all expressions in Eq.~(\ref{eq:UA-UpsA}) over the different groupings in Eq.~(\ref{eq:def:UpsA}) then gives $U_{\alpha}^{(l_\uparrow, l_\downarrow)}$.

We give some examples.\\

\noindent
\textit{Example 1}:
\begin{equation}
    \langle 1_\uparrow' | U_\alpha^{(1,0)} | 1_\uparrow \rangle
    = \langle 1_\uparrow' | \Upsilon_\alpha^{(1,0)} | 1_\uparrow \rangle.
\end{equation}
\textit{Example 2}:
\begin{equation}
\begin{split}
    \langle 1_\uparrow', 2_\uparrow' | U_\alpha^{(2,0)} | 1_\uparrow, 2_\uparrow &\rangle
     =  \langle 1_\uparrow', 2_\uparrow' | \Upsilon_\alpha^{(2,0)} | 1_\uparrow, 2_\uparrow \rangle \\
    & + \epsilon \langle 2_\uparrow' | \Upsilon_\alpha^{(1,0)} | 1_\uparrow \rangle \, \langle 1_\uparrow' | \Upsilon_\alpha^{(1,0)} | 2_\uparrow \rangle.
\end{split}
\end{equation}
\textit{Example 3}:
\begin{equation}
    \langle 1_\uparrow'; 1_\downarrow' | U_\alpha^{(1,1)} | 1_\uparrow; 1_\downarrow \rangle
     =  \langle 1_\uparrow'; 1_\downarrow' | \Upsilon_\alpha^{(1,1)} | 1_\uparrow; 1_\downarrow \rangle.
\end{equation}
\textit{Example 4}:
\begin{equation}
\begin{split}
      \langle 1_\uparrow', 2_\uparrow' ; 3_\downarrow' | U_\alpha^{(2,1)} | 1_\uparrow, & 2_\uparrow ; 3_\downarrow \rangle 
      = \langle 1_\uparrow',  2_\uparrow' ; 3_\downarrow' |  \Upsilon_\alpha^{(2,1)} | 1_\uparrow, 2_\uparrow ; 3_\downarrow \rangle  \\
      &+  \langle 2_\uparrow' |  \Upsilon_\alpha^{(1,0)}  | 1_\uparrow \rangle  \langle 1_\uparrow', 3_\downarrow' | \Upsilon_\alpha^{(1,1)} | 2_\uparrow, 3_\downarrow \rangle \\
     &  + \langle 1_\uparrow' |  \Upsilon_\alpha^{(1,0)}  | 2_\uparrow \rangle \langle 2_\downarrow', 3_\uparrow' | \Upsilon_\alpha^{(1,1)} | 1_\downarrow, 3_\uparrow \rangle.
\end{split}
\end{equation}


For reasons which will become clear later,
it is useful to define a function $\Upsilon_{\alpha}^{(l)}$
related to
$\Upsilon_{\alpha}^{(l_\uparrow, l_\downarrow)}$
as follows.
First, we define
$( \tilde{\bold{r}}_{1},\dots , \tilde{\bold{r}}_{l} ) :=
( \bold{r}_{1\uparrow},\dots , \bold{r}_{l\uparrow}, \bold{r}_{1\downarrow},\dots , \bold{r}_{l\downarrow} )$,
$( \tilde{\bold{r}}_{1}',\dots , \tilde{\bold{r}}_{l}' ) :=
( \bold{r}_{1\uparrow}',\dots , \bold{r}_{l\uparrow}', \bold{r}_{1\downarrow}',\dots , \bold{r}_{l\downarrow}' )$,
$\sigma_1= \dots=\sigma_{l\uparrow}=\uparrow$
and
$\sigma_{l\uparrow +1}= \dots=\sigma_{l}=\downarrow$,
where $l:=l_\uparrow + l_\downarrow$.
Then, we define a function $\Upsilon_{\alpha}^{(l)}$ as
\begin{equation}
\begin{split}
& 
\langle \tilde{\bold{r}}_{Q(1)}', \sigma_{Q(1)};
\dots ;
\tilde{\bold{r}}_{Q(l)}', \sigma_{Q(l)} |\\
&\qquad \,\, \Upsilon_{\alpha}^{(l)}
  |  \tilde{\bold{r}}_{P(1)}, \sigma_{P(1)};
 \dots ; \tilde{\bold{r}}_{P(l)}, \sigma_{P(l)} \rangle  \\
& := \epsilon^P \epsilon^Q 
\langle \tilde{\bold{r}}_{1}',\dots , \tilde{\bold{r}}_{l}' | \Upsilon_{\alpha}^{(l_\uparrow, l_\downarrow)}  |  \tilde{\bold{r}}_{1},\dots , \tilde{\bold{r}}_{l} \rangle  \\
& = \epsilon^P \epsilon^Q 
\langle \bold{r}_{1\uparrow}',\dots , \bold{r}_{l\uparrow}'; \bold{r}_{1\downarrow}',\dots , \bold{r}_{l\downarrow}' | \\
&\qquad \quad \Upsilon_{\alpha}^{(l_\uparrow, l_\downarrow)} 
  |  \bold{r}_{1\uparrow},\dots , \bold{r}_{l\uparrow}; \bold{r}_{1\downarrow},\dots , \bold{r}_{l\downarrow} \rangle,
\end{split}
\end{equation}
where 
$P$ and $Q$ are any one of $l!$ permutations.

We give a few examples.
\\

\noindent
\textit{Example 1}:
\begin{equation}
    \langle \bold{r}' , \uparrow | \Upsilon_\alpha^{(1)} | \bold{r} , \uparrow \rangle
    = \langle \bold{r}' | \Upsilon_\alpha^{(1,0)} | \bold{r} \rangle.
\end{equation}
\begin{equation}
    \langle \bold{r}' , \downarrow | \Upsilon_\alpha^{(1)} | \bold{r} , \downarrow \rangle
    = \langle \bold{r}' | \Upsilon_\alpha^{(0,1)} | \bold{r} \rangle.
\end{equation}
\textit{Example 2}:
\begin{equation}
\begin{split}
  &    \langle \bold{r}'_1, \uparrow; \bold{r}'_2, \uparrow | \Upsilon_\alpha^{(2)} | \bold{r}_1, \uparrow; \bold{r}_2, \uparrow \rangle \\
  &   =  \epsilon 
  \langle \bold{r}'_2, \uparrow; \bold{r}'_1, \uparrow | \Upsilon_\alpha^{(2)} | \bold{r}_1, \uparrow; \bold{r}_2, \uparrow \rangle \\
  &   =  \epsilon 
  \langle \bold{r}'_1, \uparrow; \bold{r}'_2, \uparrow | \Upsilon_\alpha^{(2)} | \bold{r}_2, \uparrow; \bold{r}_1, \uparrow \rangle \\
  &  =  \langle \bold{r}'_2, \uparrow; \bold{r}'_1, \uparrow | \Upsilon_\alpha^{(2)} | \bold{r}_2, \uparrow; \bold{r}_1, \uparrow \rangle \\
  &   =  \langle \bold{r}'_1, \bold{r}'_2 | \Upsilon_\alpha^{(2,0)} | \bold{r}_1, \bold{r}_2 \rangle .
\end{split}
\end{equation}
\textit{Example 3}:
\begin{equation}
\begin{split}
  &    \langle \bold{r}'_1, \uparrow; \bold{r}'_2, \downarrow | \Upsilon_\alpha^{(2)} | \bold{r}_1, \uparrow; \bold{r}_2, \downarrow \rangle \\
  &   =  \epsilon 
  \langle \bold{r}'_2, \downarrow; \bold{r}'_1, \uparrow | \Upsilon_\alpha^{(2)} | \bold{r}_1, \uparrow; \bold{r}_2, \downarrow \rangle \\
  &   =  \epsilon 
  \langle \bold{r}'_1, \uparrow; \bold{r}'_2, \downarrow | \Upsilon_\alpha^{(2)} | \bold{r}_2, \downarrow; \bold{r}_1, \uparrow \rangle \\
  &  =  \langle \bold{r}'_2, \downarrow; \bold{r}'_1, \uparrow | \Upsilon_\alpha^{(2)} | \bold{r}_2, \downarrow; \bold{r}_1, \uparrow \rangle \\
  &   =  \langle \bold{r}'_1; \bold{r}'_2 | \Upsilon_\alpha^{(1,1)} | \bold{r}_1; \bold{r}_2 \rangle .
\end{split}
\end{equation}

\section{graphical representation}
\label{sec:Graph}

\subsection{Primary $\zeta$-graphs}
\label{subsec:PriG}

Using Eqs.~(\ref{eq:Xi-b}), (\ref{eq:b-UA}), (\ref{eq:DM1-UA-u})-(\ref{eq:DM2-UA-ud}) and
the rules introduced in the preceding section, 
the grand partition function and the one- and two-particle reduced density matrices can be expressed in terms of the sum over connected products of  $\Upsilon_\alpha$ functions.
The exact character of this sum is most simply described in terms of primary $\zeta$-graphs introduced by Lee and Yang \cite{LY60-IV} for the grand partition function and the one-particle reduced density matrix,
and by de Dominisis \cite{dD62} for the two-particle reduced density matrix.
A primary $\zeta$-graph is defined as follows:

\textit{Definition.}---
A primary $\zeta$-graph ($\zeta = 0,1,2, \dots$) is a graphical structure which consists of a collection of vertices connected by directed lines, with $\zeta$ external incoming lines and $\zeta$ external outgoing lines.
Here, a line that has vertices at both ends is called an internal line; otherwise, it is called an external line.
All external lines are considered distinguishable.
Each vertex, called the $l$-vertex ($l=2,3,\dots$), 
connects $l$ incoming lines and $l$ outgoing lines.
A primary $\zeta$-graph must include at least one vertex and one line, and all parts must be connected (i.e., there must be a path from any one vertex to any other vertex).
Two primary graphs are different if their topological structures are different.

To each of these graphs we assign a term which is determined by the following procedures:
\begin{enumerate}
\item[(i)] Associate with each internal line a different integer $i \; (i=1,\dots,N)$ 
and the corresponding coordinate and spin $(\bold{r}_i,\sigma_i)$.
Associate with each external line some prescribed coordinate and spin.
\item[(ii)] Assign to each $l$-vertex, a factor 
\begin{equation}
\begin{picture}(38,20)
\put(-2,17){\small $B_1$}
\put(5,-12){\line(5,4){30}}
\put(-2,-21){\small $A_1$}
\put(12,-6){\vector(4,3){1.5}}
\put(29,7){\vector(4,3){1.5}}
\put(34,17){\small $B_l$}
\put(35,-12){\line(-5,4){30}}
\put(34,-21){\small $A_l$}
\put(28,-6){\vector(-4,3){1.5}}
\put(11,7){\vector(-4,3){1.5}}
\put(20,0){\line(-2,-5){4.8}}
\put(17,-8){\vector(1,2){1.5}}
\put(10,-21){\small $A_2$}
\put(22,-21){\small $\dots$}
\put(20,0){\line(-2,5){4.8}}
\put(17,7){\vector(-1,2){1.5}}
\put(10,17){\small $B_2$}
\put(22,17){\small $\dots$}
\put(20,0){\circle*{5}}
\end{picture}
 = z_\uparrow^{l_\uparrow} z_\downarrow^{l_\downarrow} 
\langle B_1, \dots, B_l |  \Upsilon_{\alpha}^{(l)} | A_1, \dots, A_l \rangle,
\end{equation}\\ 
where $A_i$ and $B_i$ represent the coordinates and spins associated with its incoming and outgoing $i$th lines $(i=1,\dots,l)$, respectively.
The number of up-spins associated with its incoming lines is the same as that of outgoing lines,
and it is denoted as  $l_\uparrow$.
The same is true for down-spins $l_\downarrow$, with $l:=l_\uparrow + l_\downarrow$
\footnote{We use the rules different from those of Lee and Yang \cite{LY60-IV}.
The relationships between Lee-Yang and our notation are listed in Appendix \ref{app:notation}.
}.

\item[(iii)] Assign a factor $1/S$ to the entire graph,
where $S$ is the symmetry number and is defined as follows:

Consider all $N!$ permutations of the positions of $N$ integers associated with the lines.
The total number of permutations that leave the graph topologically
unchanged gives the symmetry number of the graph.
The symmetry numbers are listed under each graph 
in Figs. \ref{fig:Xi}, \ref{fig:Eta1Con}, etc.
\item[(v)] Assign a factor $-1$ to the entire graph,
if the permutation 
\begin{equation}
A_1 \rightarrow B_1,\; A_2\rightarrow B_2,\; \dots ,\; A_\alpha \rightarrow B_\alpha,\;\dots
\end{equation}
from all the initial coordinates into all the final coordinates of all the vertex function 
$\Upsilon^{(l)}_{\rm{A}}$ taken together is odd. 
\end{enumerate}

The term that corresponds to each graph is given by
\begin{equation}
\begin{split}
 \sum_{\sigma_1,\dots, \sigma_l} 
& \int d^{3l_\uparrow}\bold{r}_{\uparrow} 
\int d^{3l_\downarrow}\bold{r}_{\downarrow} \\
& \times [ {\rm \, product\;of\;all\;factors\;in\; (ii)-(v) \,} ].
 \label{eq:Pri-Term}
\end{split}
\end{equation}


\subsection{Primary $0$-graphs}

\begin{figure}
\begin{picture}(220,358)
\put(0,348){(a)}
\put(0,263){\includegraphics[width=205pt]{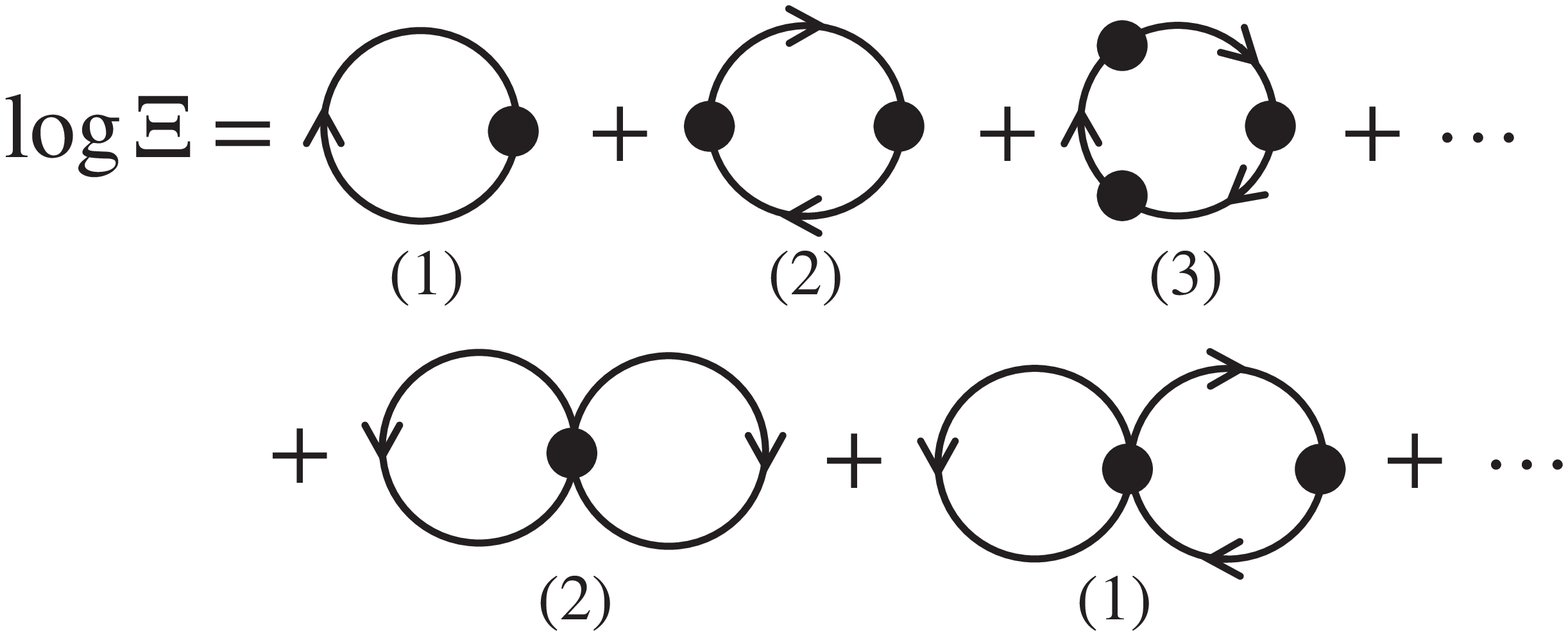}}
\put(36,163){\includegraphics[width=160pt]{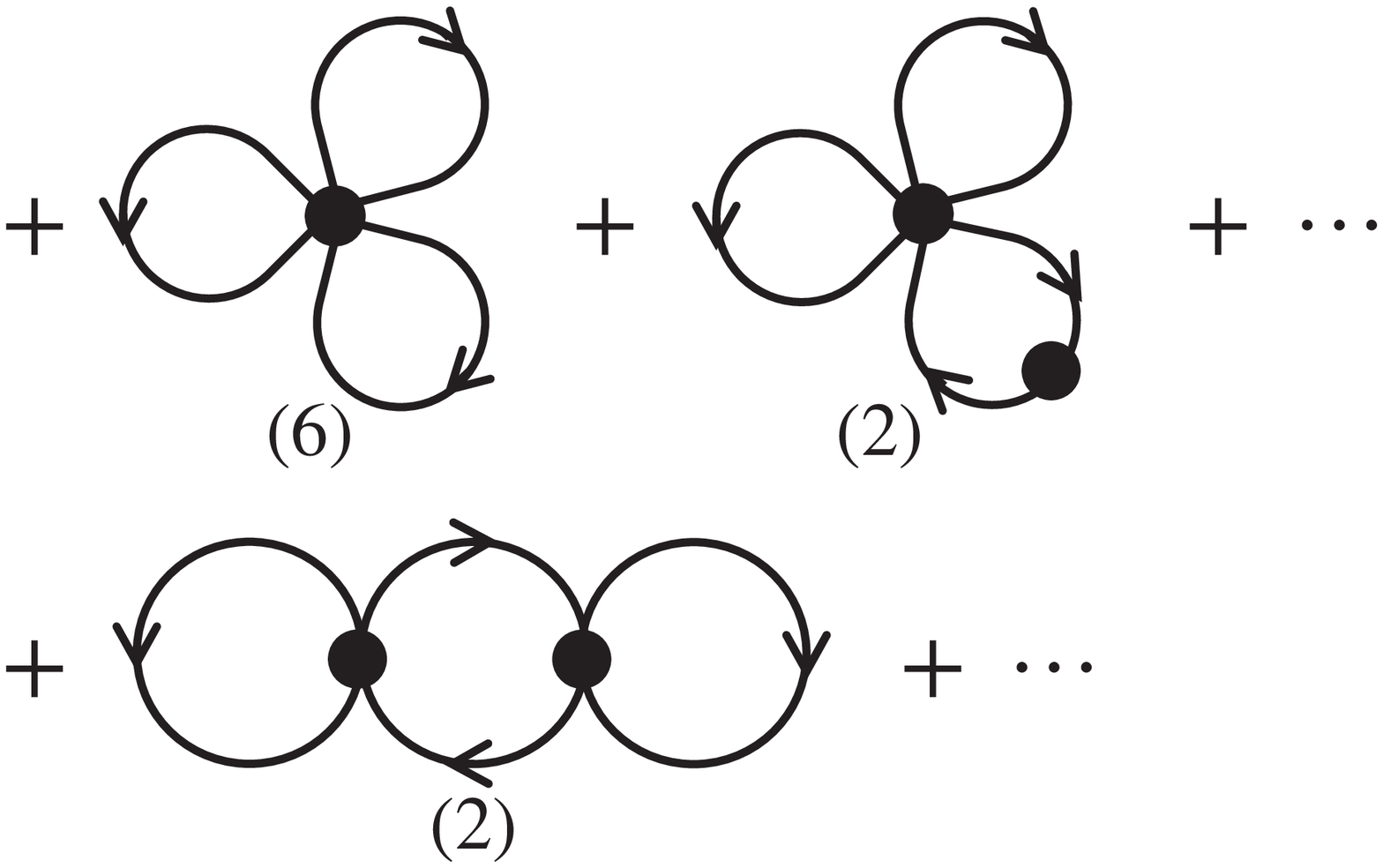}}
\put(31,76){\includegraphics[width=180pt]{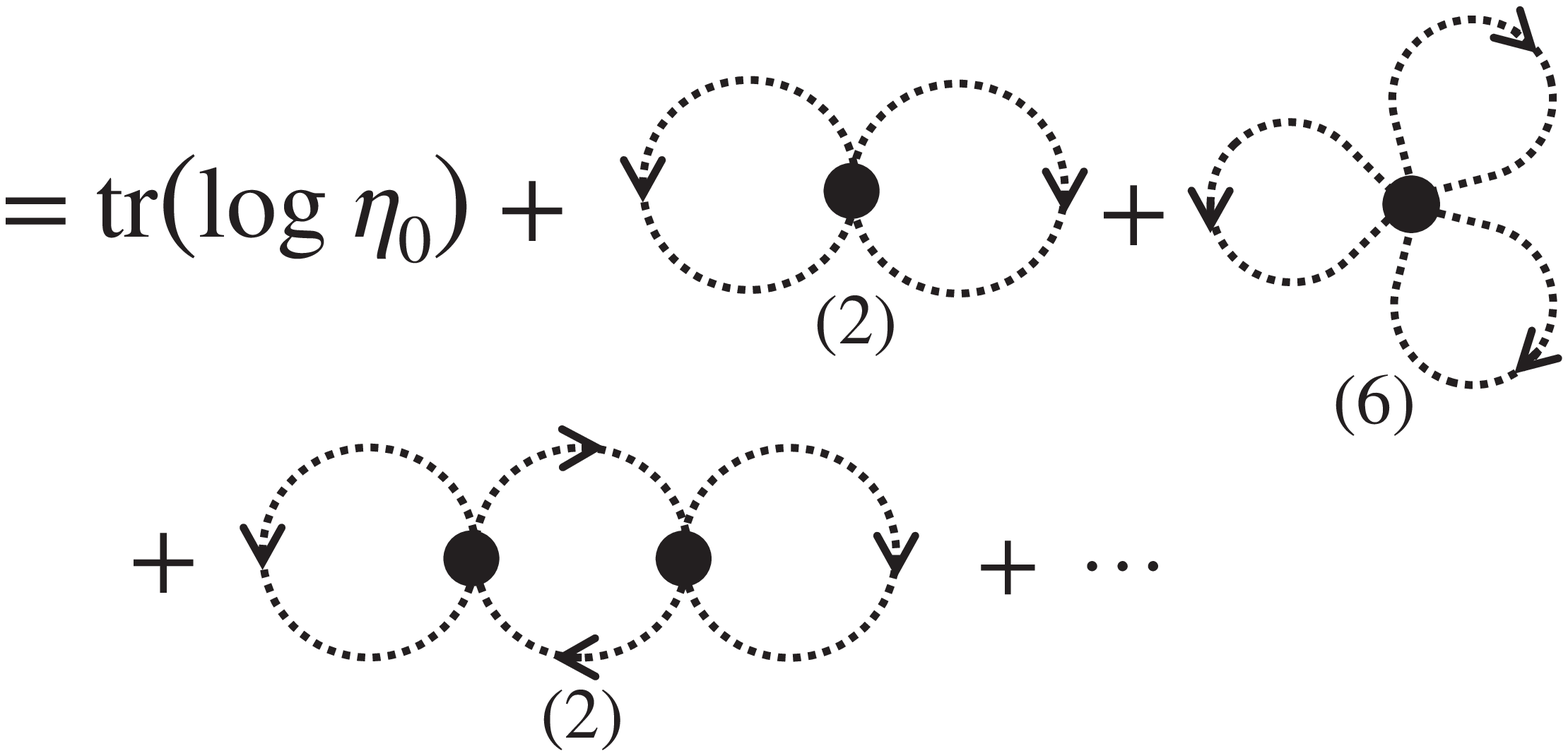}}
\put(0,69){(b)}
\put(7,-5){\includegraphics[width=215pt]{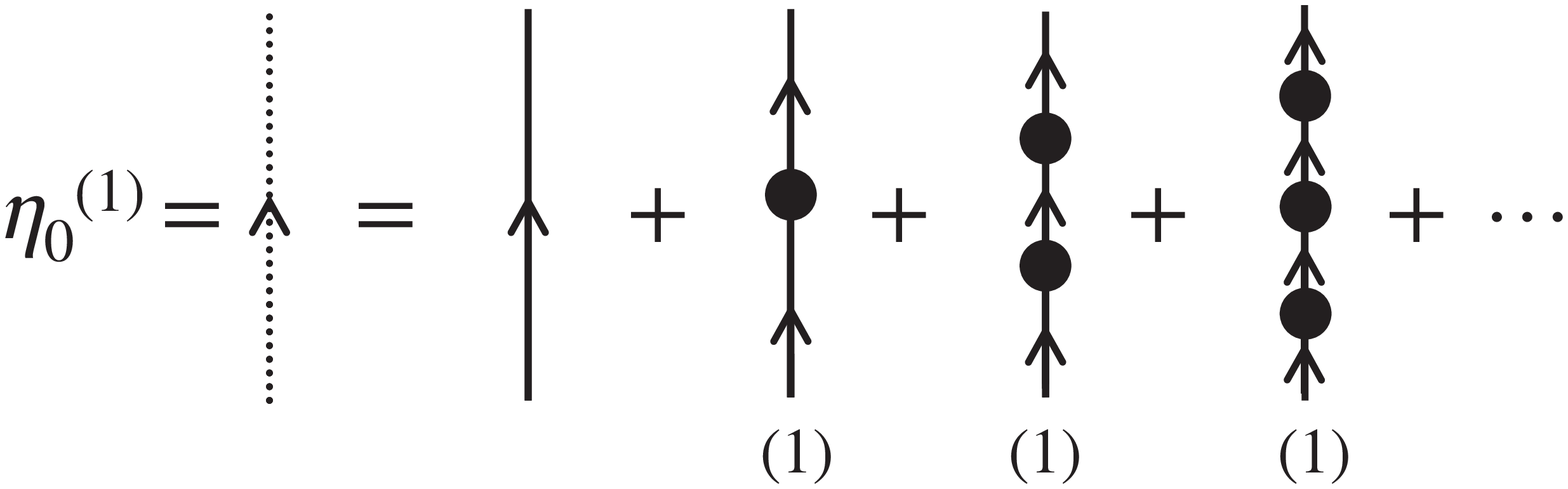}}
\end{picture}
\caption{
(a)
Expression of $\log \Xi$ as the sum over all different primary $0$-graphs (solid curves) and
as the sum over all different contracted $0$-graphs (dotted curves). 
The symmetry number is shown under each graph.
(b)
Expression of
$\langle \bold{r}'\sigma' | \eta_0^{(1)} | \bold{r}\sigma \rangle$
as the sum over different primary $1$-graphs. }
\label{fig:Xi}
\end{figure}

In terms of these primary graphs, 
we can write the grand partition function \cite{LY60-IV} as
\begin{equation}
\begin{split}
& \log \Xi \left( \beta,z \right) 
 = \sum \left[ \text{all different primary $0$-graphs} \right], \\
\end{split}
\label{eq:XiPri}
\end{equation}
to which each graph contributes a term given by Eq.~(\ref{eq:Pri-Term}).
Equation (\ref{eq:XiPri}) is illustrated in Fig.~\ref{fig:Xi}.
We can write Eq.~(\ref{eq:XiPri}) explicitly as 
\begin{align}
& \log \Xi \left( \beta,z \right) \notag\\
& =
                  z_\uparrow \! \Tr \, ( U^{(1,0)} )
     + \frac{z_\uparrow^2}{2} \Tr \, ( U^{(1,0)} )^2
     + \frac{z_\uparrow^3}{3} \Tr \, ( U^{(1,0)} )^3
     + \cdots    \notag\\
&     +             z_\downarrow \!  \Tr \, ( U^{(0,1)} )
     + \frac{z_\downarrow^2}{2} \Tr \, ( U^{(0,1)} )^2
     + \frac{z_\downarrow^3}{3} \Tr \, ( U^{(0,1)} )^3
     + \cdots    \notag\\
& + \int \! d^3\bold{r}_1 \int \! d^3\bold{r}_2 \,
     \langle \bold{r}_1,\bold{r}_2 | \Upsilon^{(2,0)}_{\alpha} | \bold{r}_1,\bold{r}_2 \rangle \notag\\
& \qquad \times \left(
       \frac{z_\uparrow^2}{2}
     + z_\uparrow^3 \langle \bold{r}_1 | U^{(1,0)} | \bold{r}_1 \rangle
     + \cdots    \right) \label{eq:XiPriEx}\\
& + \cdots , \notag
\end{align}
where
$\Tr X := \int \! d^3\bold{r} \, \langle \bold{r} | X | \bold{r} \rangle $,
$\Tr X^2 := \int \! d^3\bold{r}_1 \int \! d^3\bold{r}_2 \, $
$\langle \bold{r}_1 | X | \bold{r}_2 \rangle \langle \bold{r}_2 | X | \bold{r}_1 \rangle $,
etc.

\subsection{Primary $1$- and $2$-graphs}

%
It is useful to define functions $\langle \bold{r}'\sigma' | \eta^{(1)} | \bold{r}\sigma \rangle$ and $\langle \bold{r}_1'\sigma_1',\bold{r}_2'\sigma_2' | \eta^{(2)} | \bold{r}_1\sigma_1,\bold{r}_2\sigma_2 \rangle$
related to the one- and two-particle reduced density matrices by
\begin{align}
\langle \bold{r}'\sigma' | \eta^{(1)} | \bold{r}\sigma \rangle 
 :=  \delta (\bold{r}' - \bold{r}) \delta_{\sigma\sigma'}
 + \epsilon \langle \bold{r}\sigma | \rho^{(1)} | \bold{r}'\sigma' \rangle,
\label{eq:def:Eta1}
\end{align}
and
\begin{align}
&  \langle 1'2' | \eta^{(2)} | 12 \rangle   \notag\\
&    \quad  - 
    \langle 1' | \eta^{(1)} | 1 \rangle
      \langle 2' | \eta^{(1)} | 2 \rangle
 - \epsilon
    \langle 2' | \eta^{(1)} | 1 \rangle
      \langle 1' | \eta^{(1)} | 2 \rangle \notag\\
& \equiv  
  \langle 1'2' | \rho^{(2)} | 12 \rangle   \\
&    \quad  - 
    \langle 1' | \rho^{(1)} | 1 \rangle
      \langle 2' | \rho^{(1)} | 2 \rangle
 - \epsilon
    \langle 2' | \rho^{(1)} | 1 \rangle
      \langle 1' | \rho^{(1)} | 2 \rangle,
       \notag
\end{align}
where $1,2$ and $1',2'$ represent both the coordinate and spin of the particle
such as $1 := (\bold{r}_1,\sigma_1)$ and $1' := (\bold{r}_1',\sigma_1')$.

In terms of second-quantized field operators,
$ \eta^{(1)}$ and $\eta^{(2)}$ can be written as
\begin{align}
\langle \bold{r}'\sigma | \eta^{(1)} | \bold{r}\sigma \rangle 
 = \langle \hat{\Psi}_ {\sigma} (\bold{r}') \, \hat{\Psi}_ {\sigma}^{\dagger} (\bold{r})\rangle,
\end{align}
\begin{equation}
\begin{split}
& \langle \bold{r}_1 \sigma, \bold{r}_2 \sigma' | \eta^{(2)} | \bold{r}_1' \sigma, \bold{r}_2' \sigma' \rangle  \\
&  = \langle
 \hat{\Psi}_ {\sigma'} (\bold{r}_2') \hat{\Psi}_ {\sigma} (\bold{r}_1') \, \hat{\Psi}_ {\sigma}^{\dagger} (\bold{r}_1) \hat{\Psi}_ {\sigma'}^{\dagger} (\bold{r}_2)
   \rangle.
\end{split}
\end{equation}
These functions can be shown to be related to  $1$- and $2$-graphs \cite{LY60-IV, dD62} by
\begin{equation}
\begin{split}
 \langle \bold{r}'\sigma' | \eta^{(1)} | \bold{r}\sigma \rangle 
 = \delta (\bold{r}' - \bold{r}) \delta_{\sigma\sigma'}  + & \epsilon \sum \bigl[ \text{all different}\\
&  \text{primary $1$-graphs} \bigr] , 
\label{eq:Eta1Pri}
\end{split}
\end{equation}
and
\begin{equation}
\begin{split}
&  \langle 1'2' | \eta^{(2)} | 12 \rangle   \\
&    \quad  - 
    \langle 1' | \eta^{(1)} | 1 \rangle
      \langle 2' | \eta^{(1)} | 2 \rangle
 - \epsilon
    \langle 2' | \eta^{(1)} | 1 \rangle
      \langle 1' | \eta^{(1)} | 2 \rangle \\
& =  
 \sum \bigl[ \text{all different primary $2$-graphs} \bigr].
\label{eq:Eta2Pri}
\end{split}
\end{equation}

\subsection{Contracted $\zeta$-graphs}
\label{subsec:ConG}

In Fig.~\ref{fig:Xi} (a), we have arranged the sum so that the different
primary graphs in the same row have the same
number of $l$-vertices for $l \geq 2$ and they differ only in the
total number and relative positions of $1$-vertices.
We see that all the terms in the same row in Eq.~(\ref{eq:XiPriEx}) 
can easily be summed up. 
Thus, Eq.~(\ref{eq:XiPriEx}) becomes
\begin{equation}
\begin{split}
 \log \Xi 
 & =   \Tr \left[  \log   \left( \eta_{0}^{(1)}  \right)  \right]  \\
& + \frac{z_\uparrow^2}{2} 
     \int \! d^3\bold{r}_1 \! \int \! d^3\bold{r}_2 \,
     \langle \bold{r}_1,\bold{r}_2 | \Upsilon^{(2,0)}_{\alpha} | \bold{r}_1,\bold{r}_2 \rangle \\
&  \qquad\times
           \langle \bold{r}_1 \uparrow | \eta_{0}^{(1)} | \bold{r}_1 \uparrow \rangle \!
           \langle \bold{r}_2 \uparrow | \eta_{0}^{(1)} | \bold{r}_2 \uparrow \rangle \!   \\
& +  z_\uparrow z_\downarrow 
     \int \! d^3\bold{r}_1 \! \int \! d^3\bold{r}_2 \,
     \langle \bold{r}_1,\bold{r}_2 | \Upsilon^{(1,1)}_{\alpha} | \bold{r}_1,\bold{r}_2 \rangle \\
&  \qquad\times
          \langle \bold{r}_1 \uparrow | \eta_{0}^{(1)} | \bold{r}_1 \uparrow \rangle \!
           \langle \bold{r}_2 \downarrow | \eta_{0}^{(1)} | \bold{r}_2 \downarrow \rangle \!   \\
& + \frac{z_\downarrow^2}{2} 
     \int \! d^3\bold{r}_1 \! \int \! d^3\bold{r}_2 \,
     \langle \bold{r}_1,\bold{r}_2 | \Upsilon^{(0,2)}_{\alpha} | \bold{r}_1,\bold{r}_2 \rangle \\
&  \qquad\times
           \langle \bold{r}_1 \downarrow | \eta_{0}^{(1)} | \bold{r}_1 \downarrow \rangle \!
           \langle \bold{r}_2 \downarrow | \eta_{0}^{(1)} | \bold{r}_2 \downarrow \rangle \!   \\
& + \cdots,
\label{eq:XiConEx}
\end{split}
\end{equation}
where 
\begin{equation}
\begin{split}
 \langle \bold{r}' & \uparrow | \eta_{0}^{(1)} | \bold{r} \uparrow \rangle
 :=  \langle \bold{r}' | ( I -  \epsilon z U^{(1,0)} )^{-1} | \bold{r} \rangle \\
& = \delta (\bold{r}' - \bold{r})
+ \epsilon z_\uparrow \langle \bold{r}' | U^{(1,0)} | \bold{r} \rangle \\
& \quad + z_\uparrow^2 
     \int \! d^3\bold{r}_1
\langle \bold{r}' | U^{(1,0)} | \bold{r}_1 \rangle
                            \langle \bold{r}_1 | U^{(1,0)} | \bold{r} \rangle  
+ \cdots,
\end{split}
\end{equation}
\begin{equation}
\begin{split}
 \langle \bold{r}' & \downarrow  | \eta_{0}^{(1)} | \bold{r} \downarrow \rangle
 :=  \langle \bold{r}' | ( I -  \epsilon z U^{(0,1)} )^{-1} | \bold{r} \rangle \\
& = \delta (\bold{r}' - \bold{r})
+ \epsilon z_\downarrow \langle \bold{r}' | U^{(0,1)} | \bold{r} \rangle \\
& \quad + z_\downarrow^2 
     \int \! d^3\bold{r}_1
\langle \bold{r}' | U^{(0,1)} | \bold{r}_1 \rangle
                            \langle \bold{r}_1 | U^{(0,1)} | \bold{r} \rangle  
+ \cdots.
\end{split}
\end{equation}
It is therefore convenient to represent the primary
graphs in the same row as in Fig.~\ref{fig:Xi} (a) by a single structure
which is obtained from any one of these primary graphs
by simply eliminating all $1$-vertices (see Fig. 1 (b)).

We now introduce a contracted
graph. A contracted graph (or, contracted $\zeta$-graph) has
the same topological structure as a primary graph (or,
primary $\zeta$-graph) except that a contracted graph does
not have any $1$-vertex. 
To each contracted graph, 
we assign a term which is determined by the same procedures
(i) -(v) and the following additional rule:
\begin{enumerate}
\item[(vi)] Assign a factor $\eta_{0}^{(1)}$ to the $i$th internal line.
\end{enumerate}
In terms of these graphs, Eq.~(\ref{eq:XiConEx}) reduces to
\begin{equation}
\begin{split}
 \log \Xi & \left( \beta,z \right) 
  =  \Tr \left[  \log   \left( \eta_{0}^{(1)}  \right)  \right]  \\
& \quad + \sum \left[ \text{all different  contracted $0$-graphs} \right] .\\
\end{split}
\end{equation}
This sum is also illustrated in Fig.~\ref{fig:Xi} (b) in which solid lines represent the primary
graphs and dotted lines represent the contracted graphs. 

\begin{figure}
\begin{picture}(220,427)
\put(0,420){(a)}
\put(3,295){\includegraphics[width=230pt]{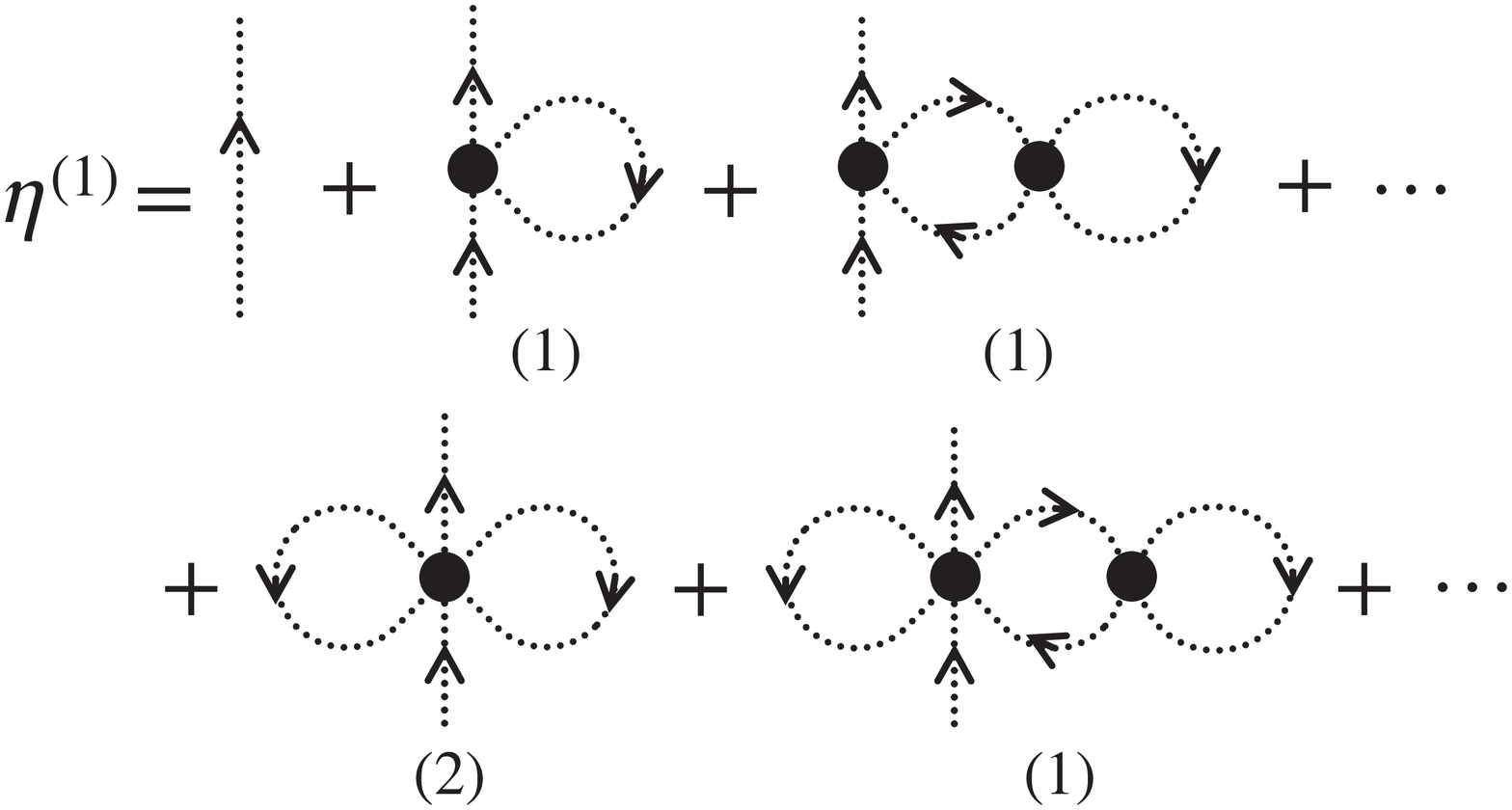}}
\put(26,208){\includegraphics[width=160pt]{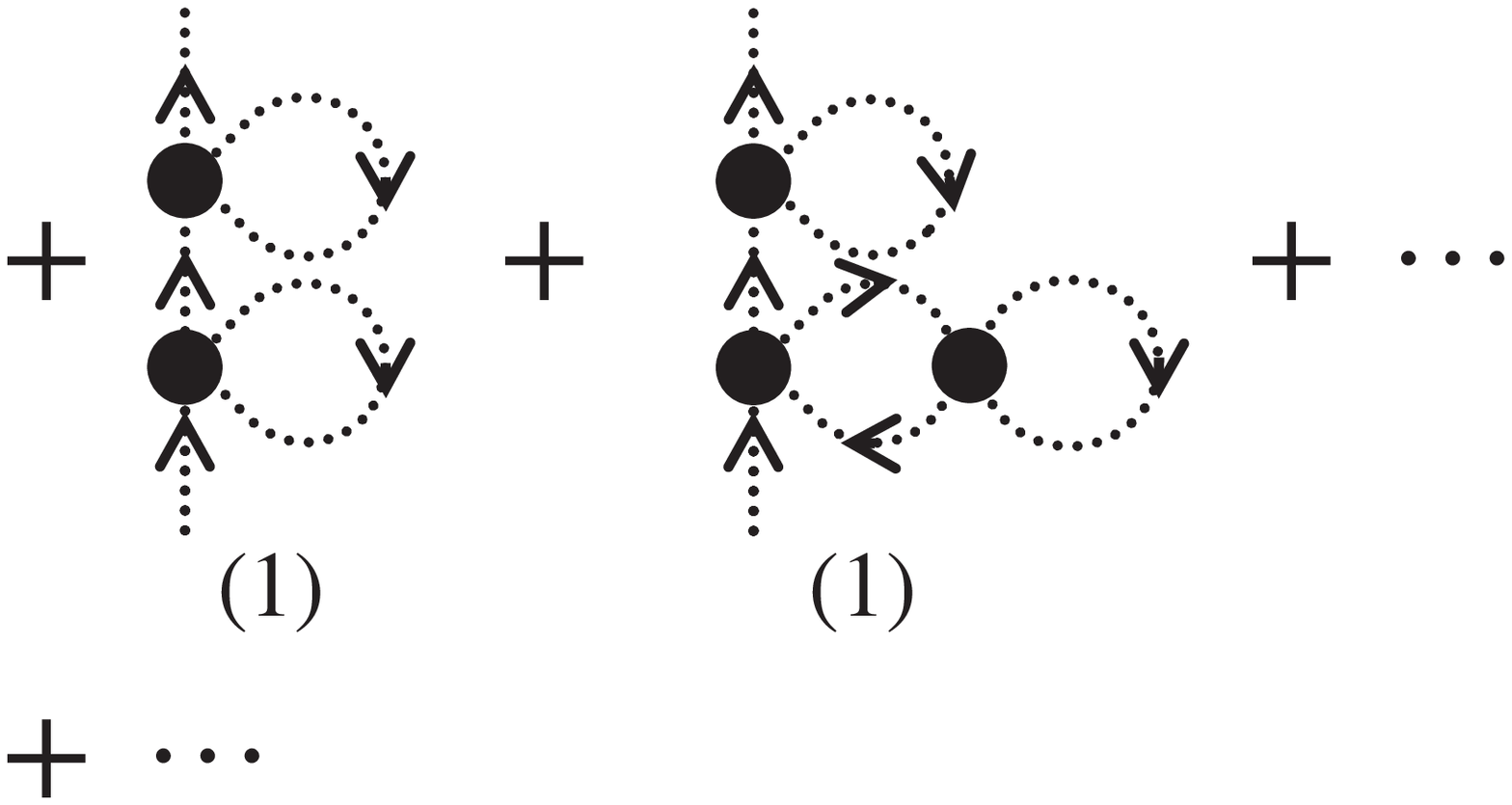}}
\put(23,135){\includegraphics[width=190pt]{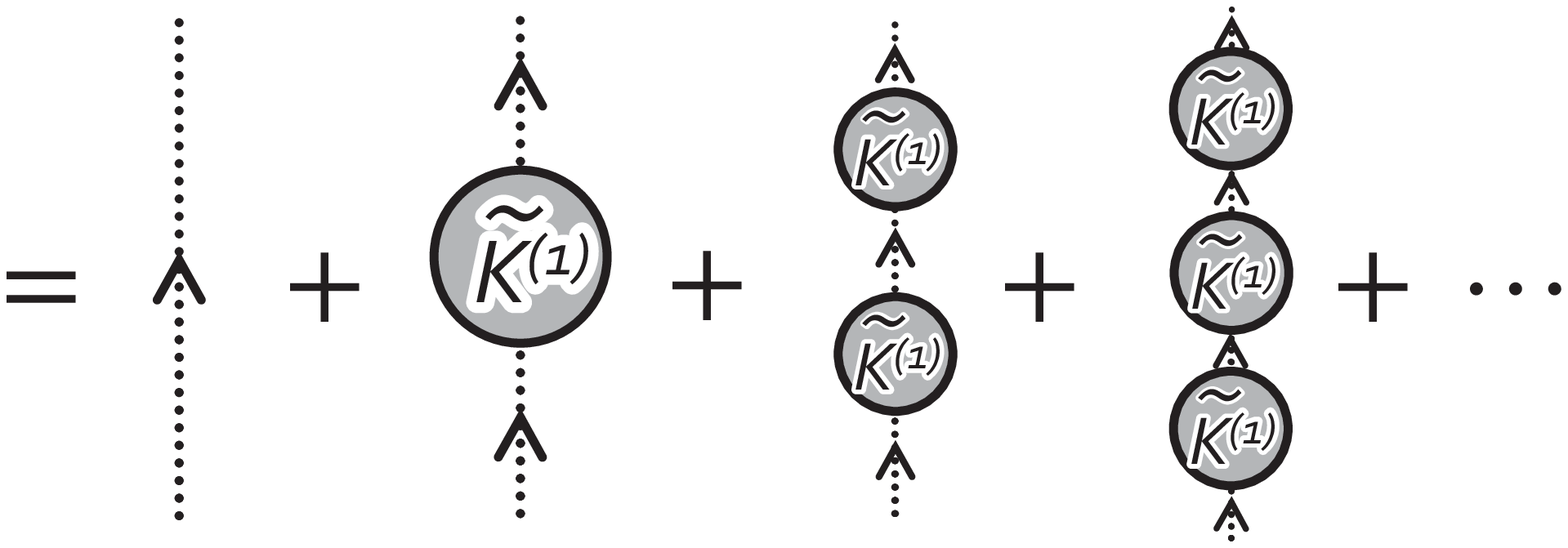}}
\put(0,130){(b)}
\put(15,-5){\includegraphics[width=190pt]{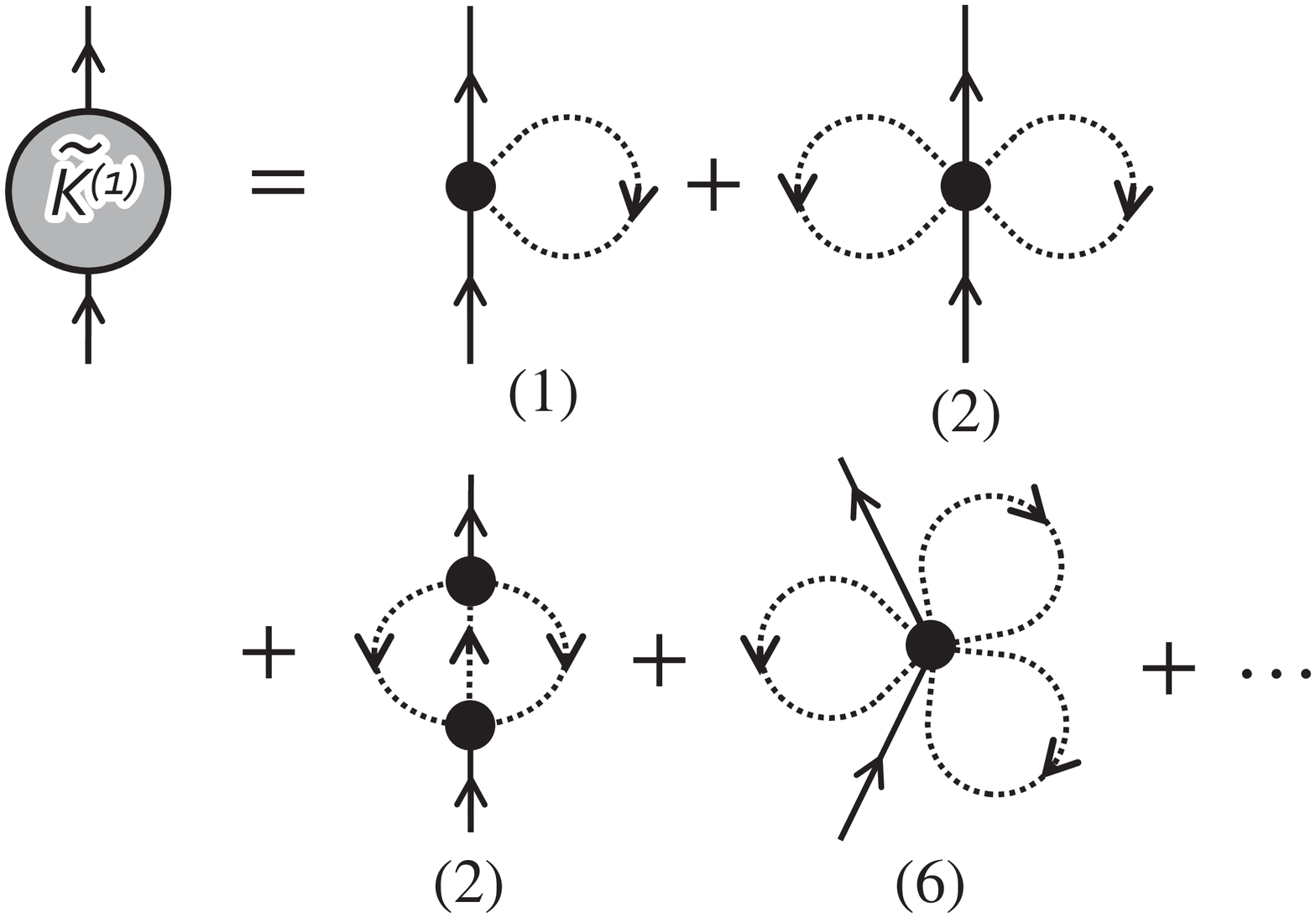}}
\end{picture}
\caption{
(a)
Expression of
$\langle \bold{r}'\sigma' | \eta^{(1)} | \bold{r}\sigma \rangle$
as the sum over all different contracted $1$-graphs, where the sum is rewritten in the form of a geometric series (last line).
%
(b)
Expression of
$\langle \bold{r}'\sigma' | \tilde{K}^{(1)} | \bold{r}\sigma \rangle$
as the sum over all different irreducible contracted $1$-graphs. 
The corresponding symmetry numbers are shown under the graphs.}
\label{fig:Eta1Con}
\end{figure}

Similarly, Eqs.~(\ref{eq:Eta1Pri}) and (\ref{eq:Eta2Pri}) become
\begin{align}
 \langle \bold{r}'\sigma' | & \eta^{(1)} | \bold{r}\sigma \rangle
    = \delta (\bold{r}' - \bold{r}) \delta_{\sigma\sigma'} \notag \\
& + \epsilon \sum \left[ \text{all different contracted $1$-graphs} \right] ,
\label{eq:Eta1Con}
\end{align}
and
\begin{equation}
\begin{split}
&  \langle 1'2' | \eta^{(2)} | 12 \rangle   \\
&    \quad  - 
    \langle 1' | \eta^{(1)} | 1 \rangle
      \langle 2' | \eta^{(1)} | 2 \rangle
 - \epsilon
    \langle 2' | \eta^{(1)} | 1 \rangle
      \langle 1' | \eta^{(1)} | 2 \rangle \\
& =  
 \sum \bigl[ \text{all different contracted $2$-graphs} \bigr].
\end{split}
\end{equation}
Equation (\ref{eq:Eta1Con}) is illustrated in Fig.~\ref{fig:Eta1Con}.

For further discussions as well as examples of primary and contracted $\zeta$-graphs, 
the reader is referred to Sec. 3 of Ref. \cite{LY60-IV} and Sec. II - (ii) of Ref. \cite{dD62}.

\subsection{One-component system}

In this subsection, we consider a one-component Bose or Fermi gas.
We can establish the formulae for the one-component gas
by substituting $z_{\downarrow}=0$
in the corresponding formulae of the two-component gas.
From Eqs.~(\ref{eq:XiPriEx}) and (\ref{eq:XiConEx}), the grand partition function is
\begin{align}
& \log \Xi \left( \beta,z \right) \notag\\
& =
                  z \Tr \, ( U^{(1)} )
     + \frac{1}{2}z^2 \Tr \, ( U^{(1)} )^2
     + \frac{1}{3}z^3 \Tr \, ( U^{(1)} )^3
     + \cdots    \notag\\
& \quad + \int \! d^3\bold{r}_1 d^3\bold{r}_2 \,
     \langle \bold{r}_1,\bold{r}_2 | \Upsilon^{(2)}_{\alpha} | \bold{r}_1,\bold{r}_2 \rangle \notag\\
& \qquad \times \left(
       \frac{1}{2}z^2
     + z^3 \langle \bold{r}_1 | U^{(1)} | \bold{r}_1 \rangle
     + \cdots    \right) \notag\\
& \quad + \cdots \notag\\
& = \Tr \left[  \log   \left( \eta_{0}^{(1)}  \right)  \right]  \\
& \quad + \frac{1}{2} \int \! d^3\bold{r}_1 d^3\bold{r}_2 \,
     \langle \bold{r}_1,\bold{r}_2 | \Upsilon^{(2)}_{\alpha} | \bold{r}_1,\bold{r}_2 \rangle \notag\\
& \qquad \times 
 \langle \bold{r}_1 | \eta_{0}^{(1)} | \bold{r}_1 \rangle
 \langle \bold{r}_2 | \eta_{0}^{(1)} | \bold{r}_2 \rangle \notag\\
& \quad + \cdots, \notag
\end{align}
where
$z := z_\uparrow$
$U^{(l)} := U^{(l,0)}$
$\Upsilon^{(l)}_{\alpha} := \Upsilon^{(l,0)}_{\alpha}$ and
\begin{equation}
 \langle \bold{r}'  | \eta_{0}^{(1)} | \bold{r} \rangle
 :=  \langle \bold{r}' | ( I -  \epsilon z U^{(1,0)} )^{-1} | \bold{r} \rangle .
\end{equation}

\section{Physical meaning of primary graphs}
\label{sec:Phys}


For a uniform system in a finite volume $V$,
the cluster functions in momentum representation are
\begin{equation}
\begin{split}
& \langle \bold{k}'_1, \dots, \bold{k}'_l | U^{(l_\uparrow, l_\downarrow)}  | \bold{k}_1, \dots, \bold{k}_l \rangle \\
& :=
\frac{1}{V^l}
\prod_{i=1}^l
\biggl(
\int \! d^3 \bold{r}_i \! \int \! d^3 \bold{r}'_i
e^{i (\bold{k}'_i \cdot \bold{r}_i' - \bold{k}_i \cdot \bold{r}_i )}
\biggr) \\
& \qquad \times \langle \bold{r}'_1, \dots, \bold{r}'_l | U^{(l_\uparrow, l_\downarrow)}  | \bold{r}_1, \dots, \bold{r}_l \rangle,
\label{eq:def:U_Mom}
\end{split}
\end{equation}
where $(\bold{k}_1, \dots, \bold{k}_l) = (\bold{k}_1, \dots, \bold{k}_{l_\uparrow}; \bold{k}_{l_\uparrow + 1}, \dots, \bold{k}_l)$, etc.

For later use, we evaluate the cluster functions $U^{(1)}$ for a one-component system
and $U^{(1,1)}$ for a two-component system.
We can evaluate $U^{(1)}$ for the one-component system as
\begin{equation}
   \langle \bold{k}' | U^{(1)}  | \bold{k} \rangle  
   = \delta_{\bold{k},\bold{k}'} e^{-\beta \bold{k}^2/(2m)}.
\label{eq:U1_Mom}
\end{equation}
Next, we evaluate $U^{(1,1)}$ for the two-component system.
The two-particle Hamiltonian is
\begin{equation}
H^{(1,1)}= \frac{1}{2m_\uparrow} \bold{k}_1^2 + \frac{1}{2m_\downarrow} \bold{k}_2^2 + v =\frac{1}{2(m_\uparrow + m_\downarrow)}\bold{K}^2+H^{({\rm rel})},
\end{equation}
\begin{equation}
H^{({\rm rel})} = \frac{1}{2m^*}\bold{k}^2 + v,
\end{equation}
where we introduce the center-of-mass momentum
$\bold{K}:=\bold{k}_1+\bold{k}_2$,
 the relative momentum
$\bold{k}:=(\bold{k}_1-\bold{k}_2)/2$,
and reduced mass $m^*:= m_\uparrow m_\downarrow/(m_\uparrow + m_\downarrow)$.
The cluster function is
\begin{equation}
\begin{split}
& \langle\bold{k}_1'\bold{k}_2'|U^{(1,1)}|\bold{k}_1\bold{k}_2\rangle  \\
& = \frac{8\pi^3}{V} \delta_{\bold{K},\bold{K}'} \,
     e^{- \beta\bold{K}^2/\{2(m_\uparrow + m_\downarrow ) \}} 
      \,  \langle\bold{k}'| U^{({\rm rel})}  |\bold{k}\rangle,
\end{split}
\label{eq:U11_Mom}
\end{equation}
where
\begin{equation}
\langle\bold{k}'| U^{({\rm rel})}  |\bold{k}\rangle
:= \langle\bold{k}'| e^{-\beta H^{({\rm rel})}}  |\bold{k}\rangle -  \delta_{\bold{k},\bold{k}'} \, e^{-\beta \bold{k}^2/(2m^*)}.
\label{eq:Urel_mom} 
\end{equation}
The function $\langle\bold{k}'| U^{({\rm rel})}  |\bold{k}\rangle$,
which is independent of the volume,
includes the effect of the interaction 
and can be calculated by the eigenfunction and the eigenvalue of $H^{({\rm rel})}$.


In the following,
we list four simple examples for a uniform and equal-mass ($m_\uparrow = m_\downarrow =m$) system in a finite volume $V$.
\\

\noindent
\textit{Example 1} (Quantum exchange and interaction):

\begin{figure}
\begin{picture}(220,125)
\put(10,-5){\includegraphics[width=190pt]{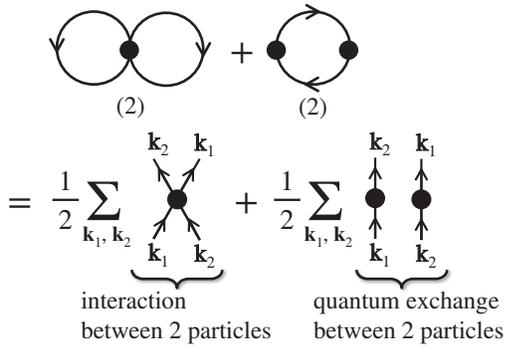}}
\end{picture}
\caption{
The second-order terms in fugacity $z$ for the primary $0$-graphs [see Eq.~(\ref{eq:XiPri-2nd})].}
\label{fig:XiPri-2nd}
\end{figure}
The grand partition function of a one-component Bose or Fermi gas is expressed as a sum of primary $0$-graphs.
Let us consider any one of the primary $0$-graphs.
It has several $l$-vertices and their orders are denoted as $l_1, \dots , l_n$.
The total number of the orders, $\sum_{i=1}^{n} l_i$,
corresponds to the order of fugacity $z$ of the corresponding primary $0$-graph.
In general a contribution from $N$ particles is expressed as the $N$th-order term in $z$.
Hence, the primary $0$-graph represents the many-body effect among $\sum_{i=1}^{n} l_i$ particles.
The one-particle cluster function is given by
\begin{equation}
\langle \bold{r}' | U^{(1)}  | \bold{r} \rangle
\xrightarrow[V \to \infty]{}
\lambda^{-3} \exp [-(\bold{r}' - \bold{r})^2/(4\beta) ],
\label{eq:U1_Uni}
\end{equation}
where $\lambda := (2\pi \beta/m)^{1/2}$
is the thermal de Broglie length.
This expression can be better understood if we separate the effect of the interaction from the quantum-exchange effect.
To provide a concrete example,
we consider the second-order in fugacity $z$ for the primary $0$-graph [see Fig.~\ref{fig:XiPri-2nd}]:
\begin{align}
   \frac{z^2}{2}  \int \! d^3\bold{r}_1 d^3\bold{r}_2 \,
&      \Bigl(
      \langle \bold{r}_1,\bold{r}_2 | \Upsilon^{(2)}_{\alpha} | \bold{r}_1,\bold{r}_2 \rangle  \notag\\
&     +  \epsilon\langle \bold{r}_2 | U^{(1)}  | \bold{r}_1 \rangle
      \langle \bold{r}_1 | U^{(1)}  | \bold{r}_2 \rangle 
    \Bigr) .
\label{eq:XiPri-2nd}
\end{align}
Here, the first term in the integrand is
\begin{equation}
\begin{split}
&     \langle \bold{r}_1,\bold{r}_2 | \Upsilon^{(2)}_\alpha | \bold{r}_1,\bold{r}_2 \rangle \\
&   =
     \langle \bold{r}_1,\bold{r}_2 | U^{(2)} | \bold{r}_1,\bold{r}_2 \rangle
     + \epsilon \langle \bold{r}_2,\bold{r}_1 | U^{(2)} | \bold{r}_1,\bold{r}_2 \rangle,
\end{split}
\end{equation}
where the cluster functions $U^{(2)}$ contain only the effect of the interaction,
because they are defined with unsymmetrized wave functions.
The second term in the integrand of (\ref{eq:XiPri-2nd}) describes the particle-exchange effect.
According to Eq.~(\ref{eq:U1_Uni}),
the relevant scale of the term
$\epsilon\langle \bold{r}_2 | U^{(1)}  | \bold{r}_1 \rangle
      \langle \bold{r}_1 | U^{(1)}  | \bold{r}_2 \rangle $
is the thermal length $\lambda$.
We can thus divide the effect of the two-particle correlation into
the quantum-exchange effect of the Bose-Einstein or Fermi-Dirac statistics
and the effect of the interaction with Boltzmann statistics.\\

\noindent
\textit{Example 2} (Ideal Bose gas and BEC):

\begin{figure}
\begin{picture}(220,148)
\put(2,-5){\includegraphics[width=215pt]{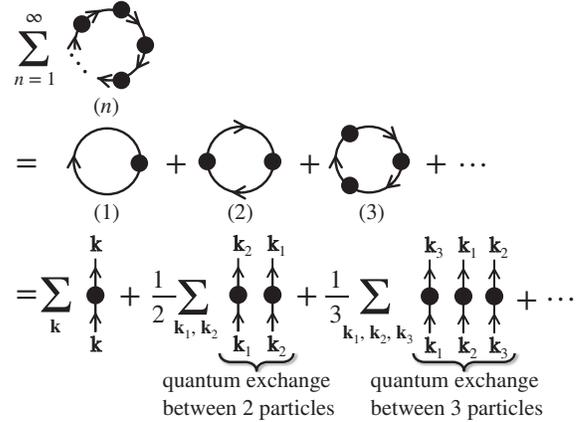}}
\end{picture}
\caption{
Sum over different primary $0$-graphs which include the $1$-vertex only
[see Eq.~(\ref{eq:XiIBose})].}
\label{fig:XiIBose}
\end{figure}
For an ideal Bose gas, 
BEC is caused by the quantum-exchange effect such as the first term in Fig.~\ref{fig:XiPri-2nd}.
For noninteracting particles, 
it is clear from Eq.~(\ref{eq:def:U})
that
$ U^{(2)}=0$,
$ U^{(3)}=0$,
etc.
Therefore, we obtain from Eq.~(\ref{eq:XiPri})
\begin{align}
& \log \Xi 
 = \sum \left[ \text{all different primary $0$-graphs} \right] \notag\\
& =  z \sum_{\bold{k}} \langle \bold{k} | U^{(1)}  | \bold{k} \rangle
      + \frac{z^2}{2} \sum_{\bold{k}_1,\bold{k}_2} \langle \bold{k}_2 | U^{(1)}  | \bold{k}_1 \rangle
      \langle \bold{k}_1 | U^{(1)}  | \bold{k}_2 \rangle \notag\\
&   \, +  \frac{z^3}{3} \sum_{\bold{k}_1,\bold{k}_2,\bold{k}_3} 
      \langle \bold{k}_2 | U^{(1)}  | \bold{k}_1 \rangle
      \langle \bold{k}_3 | U^{(1)}  | \bold{k}_2 \rangle 
      \langle \bold{k}_1 | U^{(1)}  | \bold{k}_3 \rangle   \notag\\
& \,  + \cdots ,
\label{eq:XiIBose}
\end{align}
where each term in the sum corresponds to the primary $0$-graph
in the sum in Fig.~\ref{fig:XiIBose}.
The summation extends over $\bold{k}_i \in 2\pi \mathbb{Z}^3/L$,
where we use the symbol $\mathbb{Z}=\{ 0,\pm1,\dots \}$ for a set of integers.
As mentioned earlier, 
the order (power) of fugacity $z$ indicates the number of the particles involved. 
For example,
the first term on the right-hand side of Eq.~(\ref{eq:XiIBose}) corresponds to the free motion of one particle,
the second term corresponds to the quantum-exchange effect between two bosons,
and the third term corresponds to the quantum-exchange effect among three bosons.
Substituting Eq.~(\ref{eq:U1_Mom}) in Eq.~(\ref{eq:XiIBose}), we obtain
\begin{equation}
\begin{split}
 \log \Xi
& =   \sum_{\bold{k}} \sum_{n=1}^{\infty} \frac{z^n}{n} e^{-n\beta \bold{k}^2/(2m)} \\
& =  -\sum_{\bold{k}}
 \log \left( 1 - z \, e^{-\beta \bold{k}^2/(2m)} \right),
\label{eq:XiIBose2}
\end{split}
\end{equation}
which agrees with the well-known result for ideal bosons.
At $z=e^{\beta\mu}=1$, the summation in Eq.~(\ref{eq:XiIBose2}) has a singularity at $\bold{k}=\bold{0}$, which is characteristic of a Bose-Einstein phase transition.\\

\noindent
\textit{Example 3} (Ideal Fermi gas and Fermi surface):


For ideal fermions, 
we obtain
\begin{align}
& \log \Xi
 = \sum \left[ \text{all different primary $0$-graphs} \right] \notag\\
& = z \sum_{\bold{k}} \langle \bold{k} | U^{(1)}  | \bold{k} \rangle
      - \frac{z^2}{2} \sum_{\bold{k}_1,\bold{k}_2} \langle \bold{k}_2 | U^{(1)}  | \bold{k}_1 \rangle
      \langle \bold{k}_1 | U^{(1)}  | \bold{k}_2 \rangle \notag\\
&   \, +  \frac{z^3}{3} \sum_{\bold{k}_1,\bold{k}_2,\bold{k}_3} 
      \langle \bold{k}_2 | U^{(1)}  | \bold{k}_1 \rangle
      \langle \bold{k}_3 | U^{(1)}  | \bold{k}_2 \rangle 
      \langle \bold{k}_1 | U^{(1)}  | \bold{k}_3 \rangle   \notag\\
& \,  + \cdots \label{eq:ideal_Fermi_0-graph}\\
& =  -\sum_{\bold{k}} \sum_{n=1}^{\infty} \frac{(-z)^n}{n} e^{-n\beta \bold{k}^2/(2m)} \notag\\
& =  \sum_{\bold{k}}
 \log \left( 1 + z \, e^{-\beta \bold{k}^2/(2m)} \right),\notag
\end{align}
which agrees with the known result for ideal fermions.\\

\begin{figure}
\begin{picture}(220,272)
\put(0,150){\includegraphics[width=220pt]{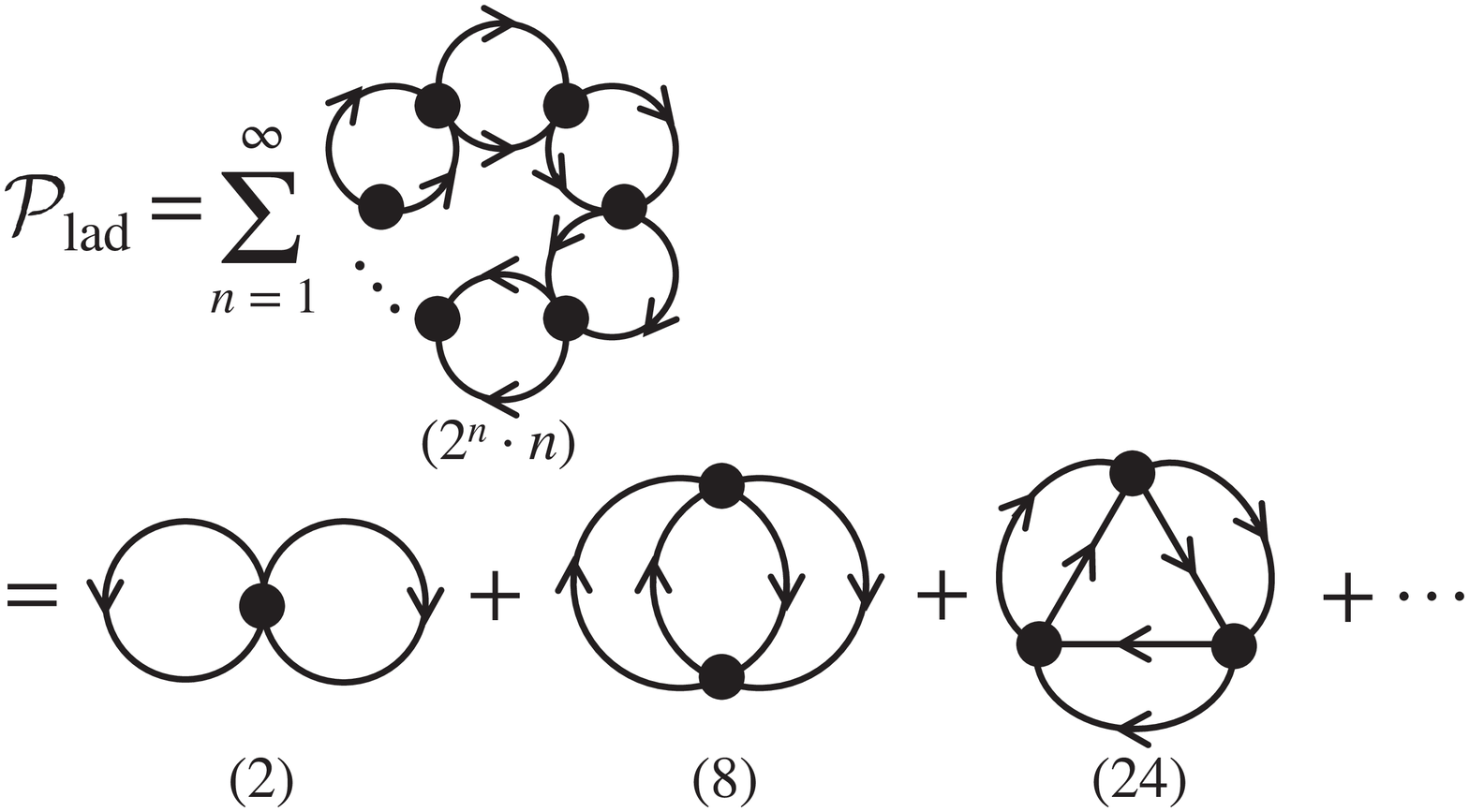}}
\put(7,-3){\includegraphics[width=183pt]{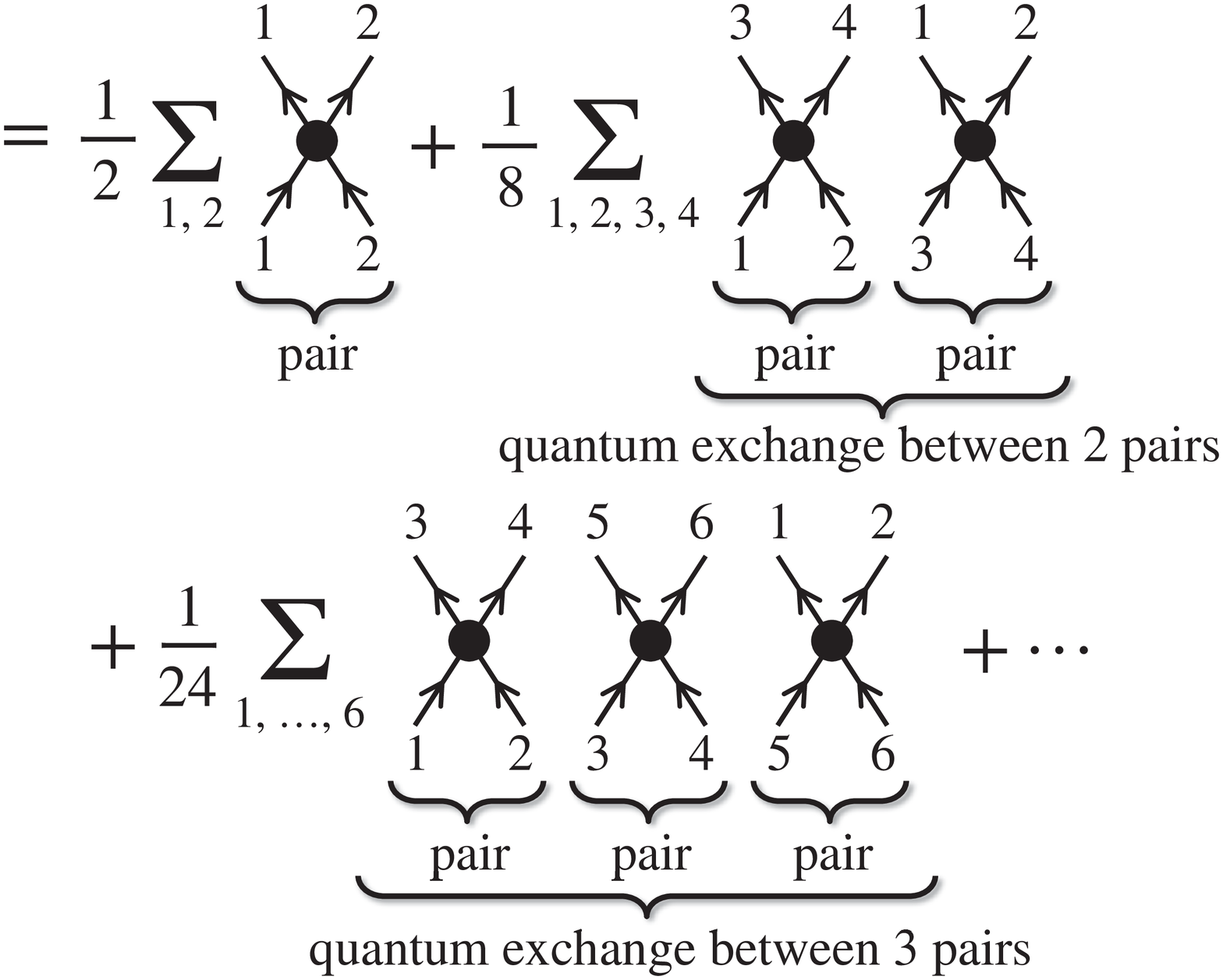}}
\end{picture}
\caption{
$\mathcal{P}_{\text{lad}}$ [Eq.~(\ref{eq:XiLad})]
as the sum over different ladder-type primary $0$-graphs.
The symmetry number is shown under each graph.}
\label{fig:XiLad}
\end{figure}

\noindent
\textit{Example 4} (Two-component Fermi gas in the tightly bound limit):


We assume that the two-body interaction $v$ is finite-ranged 
and supports one bound state corresponding to a diatomic molecular state.
Let $\psi_{\text b}(\bold{k})$ be the bound-state solution for the center-of-mass system of two particles with binding energy $E_{\text{b}}$
such that $H^{({\rm rel})} \psi_{\text{b}}(\bold{k}) = E_{\text{b}} \psi_{\text{b}}(\bold{k})$.
We assume that the binding energy is much larger than the thermal energy, i.e.,
$\left| E_{\text{b}} \right| \gg k_{\text{B}}T$.
For the Bose-Einstein distribution of diatomic molecules to develop a singularity,
we should have $\mu \leq  -\left| E_{\text{b}} \right|/2$.
Using Eq.~(\ref{eq:Urel_mom}), we obtain
\begin{equation}
\langle\bold{k}'| U^{({\rm rel})}  |\bold{k}\rangle
\simeq \psi_{\text{b}}(\bold{k}') \psi_{\text{b}}^*(\bold{k}) 
   \, e^{\beta \left| E_{\text{b}} \right|}.
\end{equation}
We can show that a general formula for a central potential $v(r)$ is given by
\begin{equation}
\begin{split}
&   \langle\bold{k}_1', \bold{k}_2'|\Upsilon_{\rm A}^{(1,1)}|\bold{k}_1, \bold{k}_2 \rangle \\
& \simeq e^{-\beta\bold{K}^2/(4m)} \delta_{\bold{K}, \bold{K}'} 
 \psi_{\text{b}}(\bold{k}') \psi_{\text{b}}^*(\bold{k}) 
   \, e^{\beta \left| E_{\text{b}} \right|}.
\end{split}
\label{eq:U2_DiBEC}
\end{equation}
The factor
$e^{-\beta\bold{K}^2 /(4m)} \delta_{\bold{K}, \bold{K}'} $
is associated with a free motion of the center of mass of molecules.
Then,
if we consider an infinite series of primary $0$-graphs that correspond to the quantum exchange of molecules, 
the sum of the infinite series leads to the BEC of diatomic molecules below the critical temperature.
We consider the sum of ladder-type primary $0$-graphs as
shown in Fig.~\ref{fig:XiLad}.
The number in the parentheses under each graph in Fig.~\ref{fig:XiLad}
is the symmetry number of the corresponding  ladder-type primary $0$-graph.
The explicit algebraic expression of the sum of the graphs is given as
\begin{equation}
\begin{split}
 \mathcal{P}_{\text{lad}} 
 = & \sum_{n=1}^{\infty} \frac{z^{2n}}{n \cdot 2^n}
 \sum_{\bold{k}_1,\dots,\bold{k}_{2n}}
   \sum_{\sigma_1,\dots,\sigma_{2n}} \\
&  \langle \bold{k}_1\sigma_1,\bold{k}_2\sigma_2 | \Upsilon_{\rm{A}}^{(2)} | \bold{k}_3\sigma_3,\bold{k}_4\sigma_4 \rangle \cdots \\
& \times 
     \langle \bold{k}_{2n-1}\sigma_{2n-1},\bold{k}_{2n}\sigma_{2n} | \Upsilon_{\rm{A}}^{(2)} | \bold{k}_1\sigma_1,\bold{k}_2\sigma_2\rangle. 
\end{split}
\label{eq:XiLad}
\end{equation}
Substituting Eq.~(\ref{eq:U2_DiBEC}) into Eq.~(\ref{eq:XiLad}), we have
\begin{align}
 \mathcal{P}_{\text{lad}} 
& = \sum_{\bold{K}} \sum_{n=1}^{\infty} \frac{1}{n} \! 
\left( z^2 e^{\beta \left| E_{\text{b}} \right|} e^{-\beta\bold{K}^2/(4m)}  \right)^n
\Bigl(  \sum_{\bold{k}} \left| \psi_{\text{b}}(\bold{k}) \right|^2  \Bigr)^n \notag\\
& =  \sum_{\bold{K}}
        \log \left( 1- z^2 e^{\beta \left| E_{\text{b}} \right|} e^{-\beta\bold{K}^2/(4m)}  
    \right).
\label{eq:XiLad2}
\end{align}
At $z=e^{- \beta \left| E_{\text{b}} \right|/2}$ (i.e., $\mu=-\left| E_{\text{b}} \right|/2$), 
the sum in Eq.~(\ref{eq:XiLad2}) develops a singularity at $\bold{K}=\bold{0}$, which is a manifestation of the Bose-Einstein phase transition of diatomic molecules.

Note that the sum of ladder-type graphs
shown in Fig.~\ref{fig:XiLad} gives rise to the quantum-exchange effect of the Bose-Einstein statistics of diatomic molecules.
In the Matsubara Green's function method,
a bound state emerges as a result of an infinite summation of ladder-type Feynmann diagrams which is interpreted as describing an effect of repeated binary collisions.
In the Lee-Yang method,
however, since the cluster function $U^{(2)}$ is evaluated from a non-perturbative bound-state solution,
the single cluster function $U^{(2)}$ includes the effect of repeated binary collisions.
In this method, we should carry out an infinite summation of ladder-type Lee-Yang graphs
to include a quantum-exchange effect of the Bose-Einstein statistics,
since the cluster function $U^{(2)}$ is defined in terms of the Boltzmann statistics.
%



\section{Criteria of ODLRO in Bose systems}
 \label{sec:Bose}
 

\subsection{Off-diagonal long-range order in Bose systems}

The defining characteristic of Bose-Einstein condensation is the occurrence of macroscopic occupation
of one-particle states \cite{PO56}.
The one-particle reduced density matrix
$\langle \bold{r}\sigma | \rho^{(1)} | \bold{r}'\sigma' \rangle
: =              \langle  \hat{\Psi}_{\sigma}^{\dagger} (\bold{r})
                \hat{\Psi}_{\sigma'} (\bold{r}')  \rangle$
may be expanded in terms of its eigenfunctions $\chi_i (\bold{r}\sigma)$
with eigenvalues $n_i$ as 
\begin{equation}
\langle \bold{r}\sigma | \rho^{(1)} | \bold{r}'\sigma' \rangle
= \sum_i n_i \,
    \chi_i^* (\bold{r}\sigma) \chi_i (\bold{r}'\sigma') .
\end{equation}
According to Penrose and Onsager \cite{PO56},
the system exhibits BEC if one or more of the eigenvalues $n_i$
are of the order of the total number of particles $N:=\langle \hat{N} \rangle$,
i.e.,
\begin{equation}
  n_i  = x_i N
  + o(N),
 \qquad\qquad (0 < x_i \leq 1).
\end{equation}
It can be shown that in an infinite system this condition is equivalent to  the emergence of an off-diagonal long-range order (ODLRO) \cite{Y62}.
A single BEC is said to emerge if one and only one eigenvalue is of the order of $N$,
all the rest being of the order of $1$.
If there exists more than one eigenvalue of the order of $N$,
the BEC is said to be fragmented \cite{Nozieres1982, Nozieres1996, Mueller2006}.

To establish the criterion of BEC in terms of the Lee-Yang cluster expansion method,
we first consider the relationship between the eigenvalues of $\rho^{(1)}$ and $\eta^{(1)}$.
It is convenient to use 
the basis set $\{ \chi_i (\bold{r}\sigma) \}$.
The matrix elements
$\rho^{(1)}_{ji}$ and
$\eta^{(1)}_{ij}$
are defined by
\begin{equation}
\begin{split}
\rho^{(1)}_{ji}
& :=
\sum_{\sigma \sigma'}
\int \!\! d^3 \bold{r}\!\! 
\int \!\! d^3 \bold{r}'
\langle \bold{r}\sigma | \rho^{(1)} | \bold{r}'\sigma' \rangle
    \chi_i (\bold{r}\sigma) \chi_j^* (\bold{r}'\sigma') \\
& = n_i \delta_{ij},
\end{split}
\end{equation}
and
\begin{equation}
\eta^{(1)}_{ij}
 :=
\sum_{\sigma \sigma'}
\int \!\! d^3 \bold{r}\!\! 
\int \!\! d^3 \bold{r}'
\langle \bold{r}'\sigma' | \eta^{(1)} | \bold{r}\sigma \rangle
    \chi_i (\bold{r}\sigma) \chi_j^* (\bold{r}'\sigma').
\end{equation}
Using Eq.~(\ref{eq:def:Eta1}),
we obtain
\begin{equation}
\eta^{(1)}_{ij}
 = \delta_{ij} + \rho^{(1)}_{ji} 
  = (n_i + 1) \delta_{ij}.
\end{equation}
 Therefore, if the system shows BEC,
 the eigenvalue of $\eta^{(1)}$ 
 is of the order of the total number of particles $N:=\langle \hat{N} \rangle$,
i.e.,
\begin{equation}
  n_i + 1  = x_i N
  + o(N),
 \qquad\qquad (0 < x_i \leq 1).
\end{equation}

\subsection{Criterion of BEC in terms of irreducible contracted graphs}

We introduce the concept of \textit{reducibility}
of a primary $1$-graph or a contracted $1$-graph.\\

\textit{Definition.}---
A primary $1$-graph or a contracted $1$-graph
 is \textit{reducible} 
if by cutting one of its \textit{internal lines}, the entire graph can be separated into
two disconnected primary $1$-graphs.\\

\textit{Definition.}---
An \textit{irreducible} primary $1$-graph or
an \textit{irreducible} contracted $1$-graph
 is 
 a primary $1$-graph or a contracted $1$-graph 
 which is not reducible.\\
 
 Note that we use the term ``irreducible" in a conventional sense.
The same term was used for a different meaning by Lee and Yang \cite{LY60-IV}.
The correspondence between the Lee-Yang's notation and ours is listed in Appendix \ref{app:notation}.

\begin{figure}
\begin{picture}(220,75)
\put(30,0){\includegraphics[width=140pt]{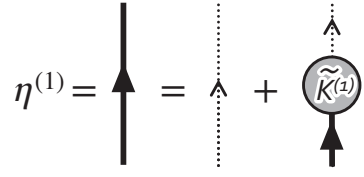}}
\end{picture}
\caption{
Expression of
$\langle \bold{r}'\sigma' | \eta^{(1)} | \bold{r}\sigma \rangle$
 in terms of an iterated equation.
}
\label{fig:Eta1Ite}
\end{figure}

In terms of irreducible contracted $1$-graphs,
$\langle \bold{r}'\sigma' | \eta^{(1)} | \bold{r}\sigma \rangle$ can be expressed as a geometric series:
\begin{align}
& \langle \bold{r}'\sigma' | \eta^{(1)} | \bold{r}\sigma \rangle
 = \langle \bold{r}'\sigma' | \eta_0^{(1)} | \bold{r}\sigma \rangle  \notag\\
& + \!\! \sum_{\sigma_1, \sigma_2} \int \!\! d^3 \bold{r}_1 d^3 \bold{r}_2 
       \langle \bold{r}'\sigma' | \eta_0^{(1)} | \bold{r}_1\sigma_1 \rangle 
       \langle \bold{r}_1\sigma_1 | \tilde{K}^{(1)} | \bold{r}_2\sigma_2 \rangle \notag\\
& \qquad \times 
       \langle \bold{r}_2\sigma_2 | \eta_0^{(1)} | \bold{r}\sigma \rangle \notag\\
& + \!\! \sum_{\sigma_1, \dots, \sigma_4} \int \!\! d^3 \bold{r}_1 \dotsi d^3 \bold{r}_4 
       \langle \bold{r}'\sigma'  | \eta_0^{(1)} | \bold{r}_1\sigma_1 \rangle 
       \langle \bold{r}_1\sigma_1 | \tilde{K}^{(1)} | \bold{r}_2\sigma_2 \rangle  \notag\\
& \qquad \times     
       \langle \bold{r}_2\sigma_2 | \eta_0^{(1)} | \bold{r}_3\sigma_3 \rangle
       \langle \bold{r}_3\sigma_3 | \tilde{K}^{(1)} | \bold{r}_4\sigma_4 \rangle 
       \langle \bold{r}_4\sigma_4 | \eta_0^{(1)} | \bold{r}\sigma \rangle 
\notag\\
& + \cdots,
\label{eq:Eta1ConGS}
\end{align}
where
\begin{align}
& \langle \bold{r}'\sigma' | \tilde{K}^{(1)} | \bold{r}\sigma \rangle \notag\\
& = \sum \bigl[ \text{all different irreducible contracted $1$-graphs} \bigr] \notag\\
& = z^2 \sum_{\bold{r}_1\sigma_1} \langle \bold{r}'\sigma', \bold{r}_1\sigma_1 | \Upsilon_\alpha^{(2)} | \bold{r}\sigma, \bold{r}_1\sigma_1 \rangle 
\langle \bold{r}_1\sigma_1 | \eta_0^{(1)} | \bold{r}_1\sigma_1 \rangle  \notag\\
& \quad + z^3 \sum_{\bold{r}_1\bold{r}_2\sigma_1\sigma_2} \langle \bold{r}'\sigma', \bold{r}_1\sigma_1, \bold{r}_2\sigma_2 | \Upsilon_\alpha^{(3)} | \bold{r}\sigma, \bold{r}_1\sigma_1, \bold{r}_2\sigma_2 \rangle \notag\\
& \qquad \qquad \times \langle \bold{r}_1\sigma_1 | \eta_0^{(1)} | \bold{r}_1\sigma_1 \rangle
\langle \bold{r}_2\sigma_2 | \eta_0^{(1)} | \bold{r}_2\sigma_2 \rangle \\
&  \quad  + \cdots . \notag
\end{align}
This geometric series is illustrated in Fig.~\ref{fig:Eta1Con}.
The matrix elements
$\eta^{(1)}_{0,ji}$ and
$\tilde{K}^{(1)}_{ij}$
are defined by
\begin{equation}
\eta^{(1)}_{0,ij}
 :=
\sum_{\sigma \sigma'}
\int \!\! d^3 \bold{r}\!\! 
\int \!\! d^3 \bold{r}'
\langle \bold{r}'\sigma' | \eta_0^{(1)} | \bold{r}\sigma \rangle
    \chi_i (\bold{r}\sigma) \chi_j^* (\bold{r}'\sigma'),
\end{equation}
and
\begin{equation}
\tilde{K}^{(1)}_{ij}
 :=
\sum_{\sigma \sigma'}
\int \!\! d^3 \bold{r}\!\! 
\int \!\! d^3 \bold{r}'
\langle \bold{r}'\sigma' | \tilde{K}^{(1)} | \bold{r}\sigma \rangle
    \chi_i (\bold{r}\sigma) \chi_j^* (\bold{r}'\sigma').
\end{equation}
We can rewrite 
Eq.~(\ref{eq:Eta1ConGS}) as
\begin{equation}
 \eta^{(1)} = \eta_0^{(1)} + \eta_0^{(1)}\tilde{K}^{(1)}  \eta^{(1)}.
\end{equation}
Then, we obtain
\begin{equation}
 \left( \eta^{(1)} \right)^{-1} =  \left( \eta_0^{(1)} \right)^{-1} - \tilde{K}^{(1)}.
\end{equation}
This equation is illustrated in Fig.~\ref{fig:Eta1Ite}.
Thus, the following theorem is established:\\

\noindent
{\it Theorem 1.} --- 

The following two criteria are equivalent:

1) The largest eigenvalue of the one-particle reduced density matrix $\rho^{(1)}$ is $O(N)$ as $N \to \infty$.

2) The smallest eigenvalue of the positive definite matrix $\left( \eta_0^{(1)} \right)^{-1} - \tilde{K}^{(1)}$ is $O(1/N)$ as $N \to \infty$.\\
 
\noindent
The former implies ODLRO and the latter can be obtained from the sum over the contracted graphs.

\subsection{Criterion of BEC in terms of irreducible primary graphs}

\begin{figure}
\begin{picture}(220,327)
\put(0,316){(a)}
\put(5,203){\includegraphics[width=205pt]{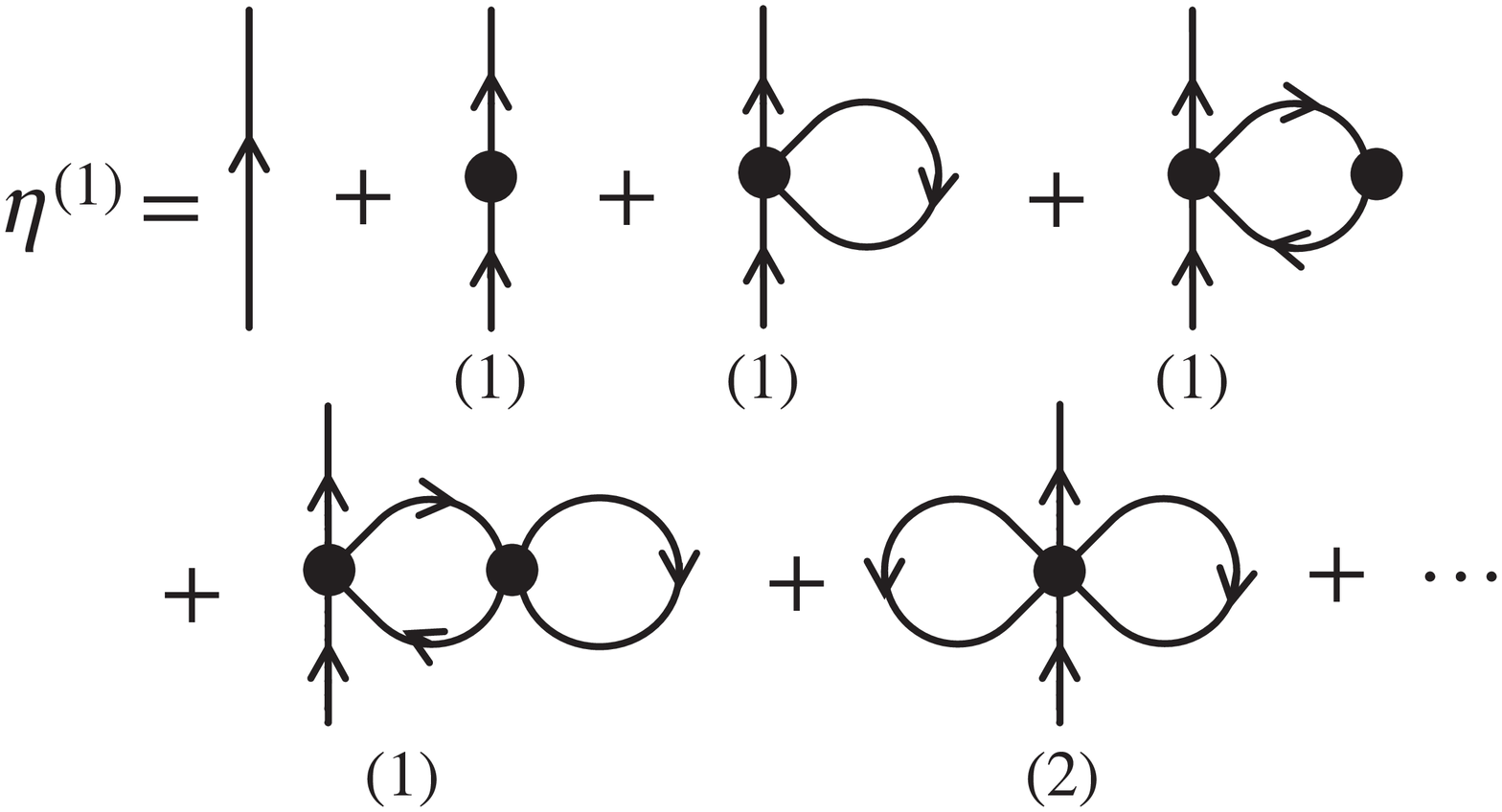}}
\put(22,138){\includegraphics[width=195pt]{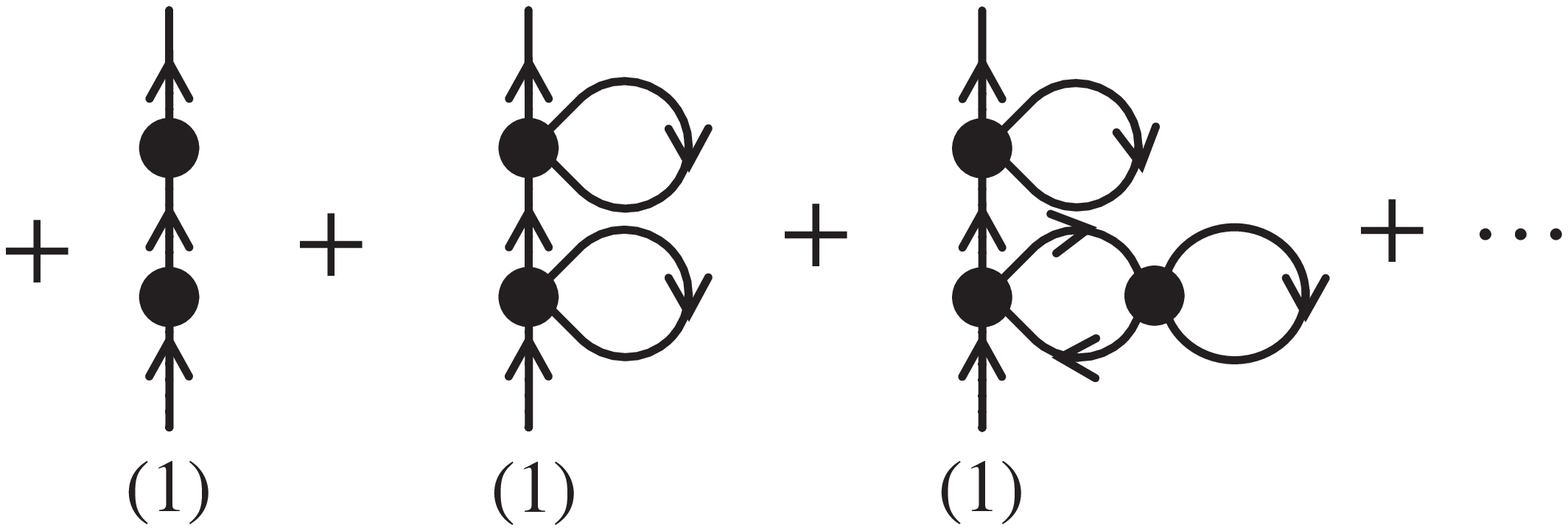}}
\put(23,68){\includegraphics[width=185pt]{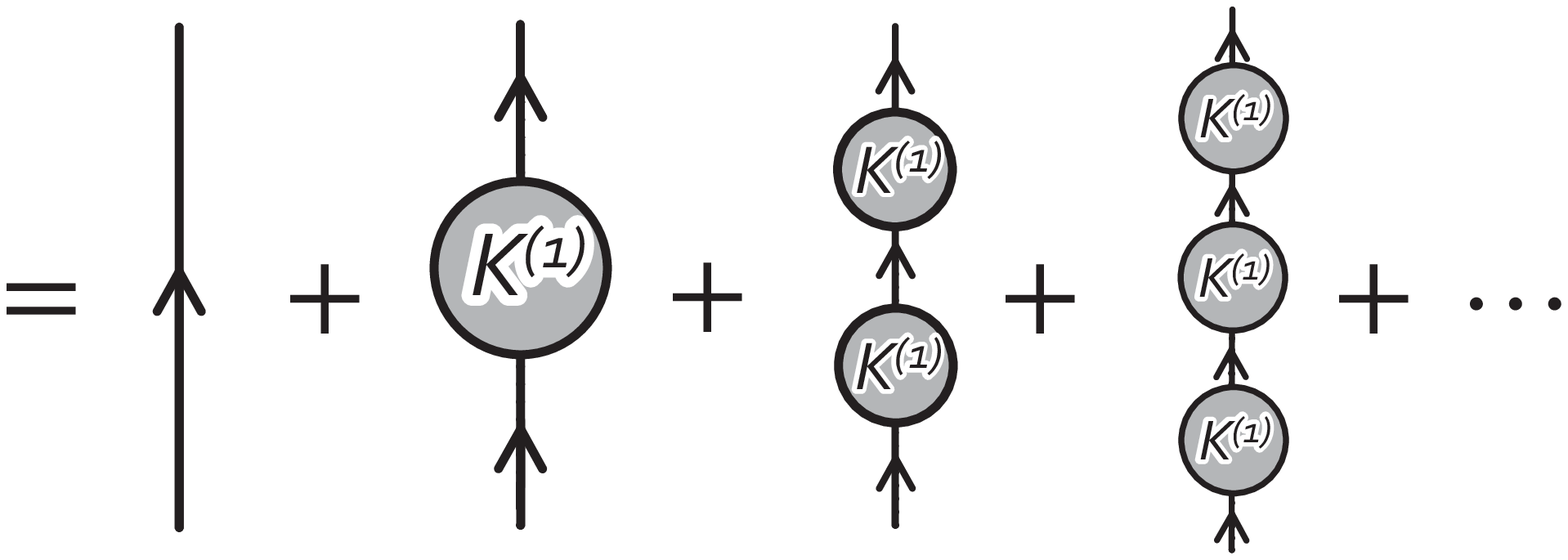}}
\put(0,60){(b)}
\put(15,0){\includegraphics[width=120pt]{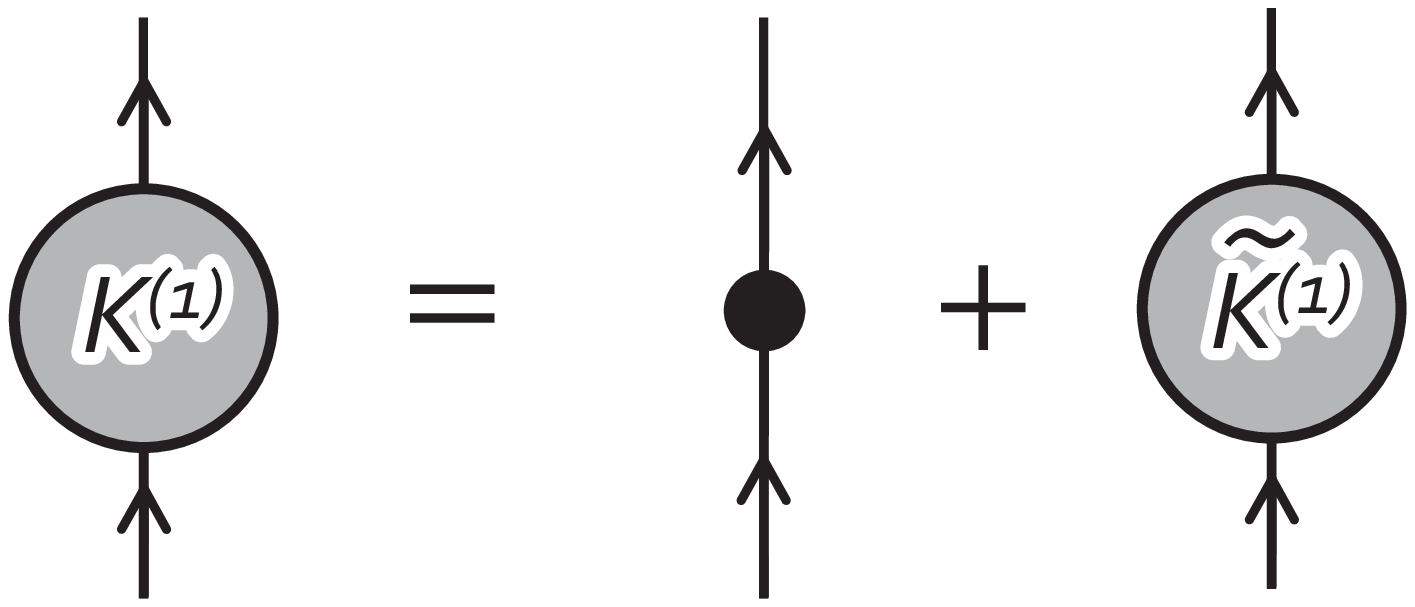}}
\end{picture}
\caption{
(a)
Expression of
$\langle \bold{r}'\sigma' | \eta^{(1)} | \bold{r}\sigma \rangle$
as the sum over all different primary $1$-graphs,
where the sum is rewritten in the form of a geometric series (last line).
The symmetry number is shown under each graph.
(b)
Expression of
$\langle \bold{r}'\sigma' | K^{(1)} | \bold{r}\sigma \rangle$ as the sum over all different irreducible primary $1$-graphs. 
}
\label{fig:Eta1Pri}
\end{figure}

We can regroup primary $1$-graphs
and establish a criterion of BEC in a more transparent manner.
In terms of irreducible primary $1$-graphs,
$\langle \bold{r}'\sigma' | \eta^{(1)} | \bold{r}\sigma \rangle$ can be expressed as a geometric series:
\begin{align}
& \langle \bold{r}'\sigma' | \eta^{(1)} | \bold{r}\sigma \rangle \notag\\
& = \delta (\bold{r}' - \bold{r}) \, \delta_{\sigma\sigma'}
  +   \langle \bold{r}'\sigma' | K^{(1)} | \bold{r}\sigma \rangle  \notag\\
&\, + \sum_{\bold{r}_1\sigma_1}
       \langle \bold{r}'\sigma' | K^{(1)} | \bold{r}_1\sigma_1 \rangle 
       \langle \bold{r}_1\sigma_1 | K^{(1)} | \bold{r}\sigma \rangle 
\notag\\
&\, + \!\!\!\! \sum_{\bold{r}_1\bold{r}_2\sigma_1\sigma_2} \!\!\!\!
       \langle \bold{r}'\sigma' | K^{(1)} | \bold{r}_1\sigma_1 \rangle 
       \langle \bold{r}_1\sigma_1 | K^{(1)} | \bold{r}_2\sigma_2 \rangle 
       \langle \bold{r}_2\sigma_2 | K^{(1)} | \bold{r}\sigma \rangle \notag\\
&\, + \cdots,
\label{eq:Eta1PriGS}
\end{align}
where
\begin{equation}
\begin{split}
& \langle \bold{r}'\sigma' | K^{(1)} | \bold{r}\sigma \rangle \\
& = \sum \bigl[ \text{all different irreducible primary $1$-graphs} \bigr] \\
& = \epsilon z \langle \bold{r}'\sigma' | U^{(1)} | \bold{r}\sigma \rangle  
    + \langle \bold{r}'\sigma' | \tilde{K}^{(1)} | \bold{r}\sigma \rangle. \\
\end{split}
\end{equation}
This geometric series is illustrated in Fig.~\ref{fig:Eta1Pri}.

Since matrices
$\langle \bold{r}'\sigma' | \eta^{(1)} | \bold{r}\sigma \rangle$
and $\langle \bold{r}'\sigma' | K^{(1)} | \bold{r}\sigma \rangle$
commute,
they can be diagonalized simultaneously.
Therefore, we can write
\begin{equation}
\langle \bold{r}\sigma | K^{(1)} | \bold{r}'\sigma' \rangle
= \sum_i \lambda_i \,
    \chi_i^* (\bold{r}\sigma) \chi_i (\bold{r}'\sigma').
\end{equation}
We can write 
Eq.~(\ref{eq:Eta1PriGS}) as
\begin{equation}
n_i +1 =   \sum_{l=0}^{\infty} \left( \lambda_i \right)^l
 =   \left( 1 - \lambda_i \right)^{-1}.
\end{equation}
Hence, we obtain
\begin{equation}
\lambda_i  = 1 - \frac{1}{n_i +1}.
\end{equation}
Thus the following theorem is established:\\

\noindent
{\it Theorem 2.} --- 

The following two criteria are equivalent:

1) The largest eigenvalue of the one-particle reduced density matrix $\rho^{(1)}$ is $O(N)$ as $N \to \infty$.

2) The largest eigenvalue of the matrix $K^{(1)}$ is $1+O(1/N)$ as $N \to \infty$.\\
 
\noindent
The former implies ODLRO and the latter can be obtained as the sum over the contracted graphs.
Thus we can use Theorems 1 and 2 to find out whether the Bose system exhibits ODLRO based on the Lee-Yang cluster expansion method.


 \subsection{Application to a uniform one-component Bose gas of hard spheres}

As an example of the formulation given in this section,
we consider  a uniform one-component dilute Bose system of hard spheres of radius $a$ 
in a gaseous phase
and evaluate the critical value $z_c$ for the BEC transition in a power series of $a$.
Now
\begin{equation}
\begin{split}
 \langle \bold{k}' | K^{(1)} | \bold{k} \rangle
& = z \langle \bold{k}' | U^{(1)} | \bold{k} \rangle  \\
&    + z^2 \sum_{\bold{q}} \langle \bold{k}', \bold{q} | \Upsilon_{\text{S}}^{(2)} | \bold{k}, \bold{q} \rangle 
    \eta_0(\bold{q})
     + O(a^2) . 
\label{eq:K1_HCBose} 
\end{split}
\end{equation}
Using an explicit form of $U^{(2)}$, which was derived in Ref. \cite{LY58-I}, we have
\begin{equation}
\begin{split}
 \langle \bold{k}', \bold{q} | \Upsilon_{\text{S}}^{(2)} | & \bold{k}, \bold{q} \rangle 
= 2  \langle \bold{k}', \bold{k}_1 | U^{(2)} | \bold{k}, \bold{k}_1 \rangle  + O(a^2) \\
& = - \delta_{\bold{k}\bold{k}'} \frac{4a\lambda^2}{V} e^{-\beta (\bold{k}_1^2 + \bold{k}^2)/(2m)} + O(a^2).
\label{eq:Up2_HCBose}
\end{split}
\end{equation}
From Eq.~(\ref{eq:Up2_HCBose}) and $\eta_0(\bold{q}) = (1-ze^{-\beta k^2/(2m)})^{-1}$,
we have
\begin{equation}
 \sum_{\bold{q}} \langle \bold{k}, \bold{q} | \Upsilon_{\text{S}}^{(2)} | \bold{k}, \bold{q} \rangle 
    \eta_0(\bold{q})
  = -  \frac{4}{z}\, \text{Li}_{\frac{3}{2}} (z)\, \frac{a}{\lambda}  + O(a^2), 
\label{eq:Up2Eta0_HCBose}
\end{equation}
where
$\text{Li}_{n} (z) := \sum_{l=1}^{\infty} z^l / l^{n}$
is the polylogarithm.
By substituting Eqs.~(\ref{eq:U1_Mom}) and (\ref{eq:Up2Eta0_HCBose}) into Eq.~(\ref{eq:K1_HCBose}),
we obtain
\begin{equation}
\begin{split}
& \langle \bold{k}' | K^{(1)} | \bold{k} \rangle \\
& = \delta_{\bold{k}\bold{k}'} e^{-\beta \bold{k}^2/(2m)} z
   \left( 1  - 4 \, \text{Li}_{\frac{3}{2}} (z)\, \frac{a}{\lambda}  \right) 
    + O(a^2) . \\
\end{split}
\end{equation}
According to Theorem 2, the critical fugacity $z_c$ satisfies the following equation: 
\begin{equation}
1 = z_c \left( 1  - 4 \, \text{Li}_{\frac{3}{2}} (z_c)\, \frac{a}{\lambda}  \right) 
    + O(a^2).
\end{equation}
Hence, we obtain
\begin{equation}
z_c = 1+ 4 \, \zeta \left( \frac{3}{2} \right) \, \frac{a}{\lambda} 
    + O(a^2),
\end{equation}
where
$ \zeta(3/2) := \text{Li}_{3/2}(1) \simeq 2.612$.
This result agrees with that obtained by Lee and Yang \cite{LY60-IV}.

\section{Criteria of ODLRO in Fermi systems}
\label{sec:Fermi}

\subsection{Off-diagonal long-range order in Fermi systems}

We consider a system of two-component Fermi particles with fixed temperature $T$ and fugacity $z$.
A general criterion of the quantum phase transition of a Fermi gas
[e.g., superconductivity] 
is characterized by the occurrence of macroscopic occupation
of two-particle states \cite{Y62}.
The two-particle reduced density matrix
\begin{equation}
\begin{split}
\rho^{(2)}_{\bold{r}_1\bold{r}_2;\bold{r}'_1\bold{r}'_2}
& := \langle \bold{r}_1\uparrow\bold{r}_2\downarrow | \mathcal{\rho}^{(2)} | \bold{r}'_1\uparrow\bold{r}'_2\downarrow \rangle  \\
& : =              \langle  \hat{\Psi}_\uparrow^{\dagger} (\bold{r}_1)
                \hat{\Psi}_\downarrow^{\dagger} (\bold{r}_2)
                \hat{\Psi}_\downarrow (\bold{r}_2')  
                \hat{\Psi}_\uparrow (\bold{r}_1')  \rangle
\end{split}
\end{equation}
may be expanded in terms of its eigenfunctions $\chi_i (\bold{r}_1\bold{r}_2)$
with eigenvalues $n^{(2)}_i$ as 
\begin{equation}
\rho^{(2)}_{\bold{r}_1\bold{r}_2;\bold{r}'_1\bold{r}'_2}
= \sum_i n^{(2)}_i 
    \chi_i^* (\bold{r}_1,\bold{r}_2) \chi_i (\bold{r}'_1,\bold{r}'_2).
\end{equation}
The system shows ODLRO if one or more of the eigenvalues $n^{(2)}_i$
are of the order of the total number of particles $N :=\langle \hat{N}_\uparrow \rangle + \langle \hat{N}_\downarrow \rangle$,
i.e.,
\begin{equation}
  n^{(2)}_i
  = x_i N + o(N),
 \qquad\qquad (0 < x_i \leq 1).
\end{equation}

To establish the criterion of ODLRO in terms of the Lee-Yang cluster expansion,
we consider the relationship between the eigenvalues of $\rho^{(2)}$ and $\eta^{(2)}$.
We define matrices $\eta^{(2)}$, $\rho^{(1\otimes 1)}$
and $\eta^{(1\otimes 1)}$ 
 as follows:
\begin{equation}
\eta^{(2)}_{\bold{r}'_1\bold{r}'_2;\bold{r}_1\bold{r}_2}
:= \langle \bold{r}_1'\uparrow\bold{r}_2'\downarrow | \eta^{(2)} | \bold{r}_1\uparrow\bold{r}_2\downarrow \rangle,
\end{equation}
\begin{equation}
\rho^{(1\otimes 1)}_{\bold{r}'_1\bold{r}'_2;\bold{r}_1\bold{r}_2}
:= \langle \bold{r}_1'\uparrow | \rho^{(1)} | \bold{r}_1\uparrow \rangle
\langle \bold{r}_2'\downarrow | \rho^{(1)} | \bold{r}_2\downarrow \rangle,
\label{eq:def:DM11}
\end{equation}
and
\begin{equation}
\eta^{(1\otimes 1)}_{\bold{r}'_1\bold{r}'_2;\bold{r}_1\bold{r}_2}
:= \langle \bold{r}_1'\uparrow | \eta^{(1)} | \bold{r}_1\uparrow \rangle
\langle \bold{r}_2'\downarrow | \eta^{(1)} | \bold{r}_2\downarrow \rangle.
\label{eq:def:Eta11}
\end{equation}
The identity matrix $I$ is
\begin{equation}
I_{\bold{r}'_1\bold{r}'_2;\bold{r}_1\bold{r}_2}
:= \delta (\bold{r}'_1 - \bold{r}_1) \, \delta (\bold{r}'_2 - \bold{r}_2 ).
\end{equation}

We first establish the following lemma.\\

\noindent
{\it Lemma} --- 
Let $n_i^{(2)}$ and $\bar{n}^{(2)}_i$
be the eigenvalues of $\rho^{(2)}$ and $\eta^{(2)}$, respectively,
and let us assume that
$n_0^{(2)} \geq n_1^{(2)} \geq n_2^{(2)} \geq \cdots$ and
$\bar{n}^{(2)}_0 \geq \bar{n}^{(2)}_1 \geq \bar{n}^{(2)}_2 \geq \cdots$.
Then, for all $i=0,1,\dots$, the following inequality holds:
\begin{equation}
\left| n_i^{(2)} - \bar{n}^{(2)}_i \right|   \leq 1.
\end{equation}

\noindent
 {\it Proof.} 
We notice that by using Eqs.~(\ref{eq:def:DM11}) and (\ref{eq:def:Eta11}),
\begin{equation}
I + \eta^{(1\otimes 1)} - \rho^{(1\otimes 1)} \geq 0 ,
\end{equation}
and
\begin{equation}
I + \rho^{(1\otimes 1)} -  \eta^{(1\otimes 1)} \geq 0.
\end{equation}
It follows from the relation
$ \eta^{(2)}-\eta^{(1\otimes 1)}
=  \rho^{(2)}-\rho^{(1\otimes 1)}$
that
$\eta^{(2)} \leq \rho^{(2)} + I$
and
$\rho^{(2)} \leq \eta^{(2)} + I$.
 
Let $A$ and $B$ be arbitrary Hermitian matrices 
and let us assume that their eigenvalues are ordered in such a manner that
$a_0 \geq a_1 \geq \cdots$
and
$b_0 \geq b_1 \geq \cdots$,
respectively.
If $A \geq B$,
then $a_i \geq b_i$ for all $i$.
Then, we have
$\bar{n}^{(2)}_i \leq n^{(2)}_i +  1$
and
$n^{(2)}_i \leq \bar{n}^{(2)}_i +  1$.
Thus, the Lemma is proved.
(Q.E.D.)\\

Then, in the normal phase, $ \bar{n}^{(2)}_i = o(N)$,
whereas in the BEC phase 
\begin{equation}
  \bar{n}^{(2)}_i
  = x_i N + o(N),
 \qquad\qquad (0 < x_i \leq 1).
\end{equation}

\subsection{Criterion of ODLRO for a two-component Fermi system in terms of irreducible graphs}

The criterion of ODLRO based on the Lee-Yang cluster expansion can be obtained for two-component Fermi systems,
following procedures similar to Bose systems. \\

\textit{Definition.}---
A contracted $2$-graph
 is called \textit{reducible} 
 if it satisfies any one of the following conditions:
 (i) By cutting one of its \textit{internal lines}, the entire graph can be separated into
two disconnected parts.
(ii) By cutting two of its \textit{internal lines} and associating $(1,2)$ with the incoming lines and 
$(1',2')$ with the outgoing lines,
the entire graph can be separated into
two disconnected contracted $2$-graphs;
one including two incoming external lines $(1,2)$ and two outgoing external lines,
and
the other including two incoming external lines and two outgoing external lines $(1',2')$.\\

\textit{Definition.}---
An \textit{irreducible} contracted $2$-graph is a contracted $2$-graph that is not reducible.\\

\begin{figure}
\begin{picture}(220,202)
\put(20,193){(a)}
\put(25,83){\includegraphics[width=170pt]{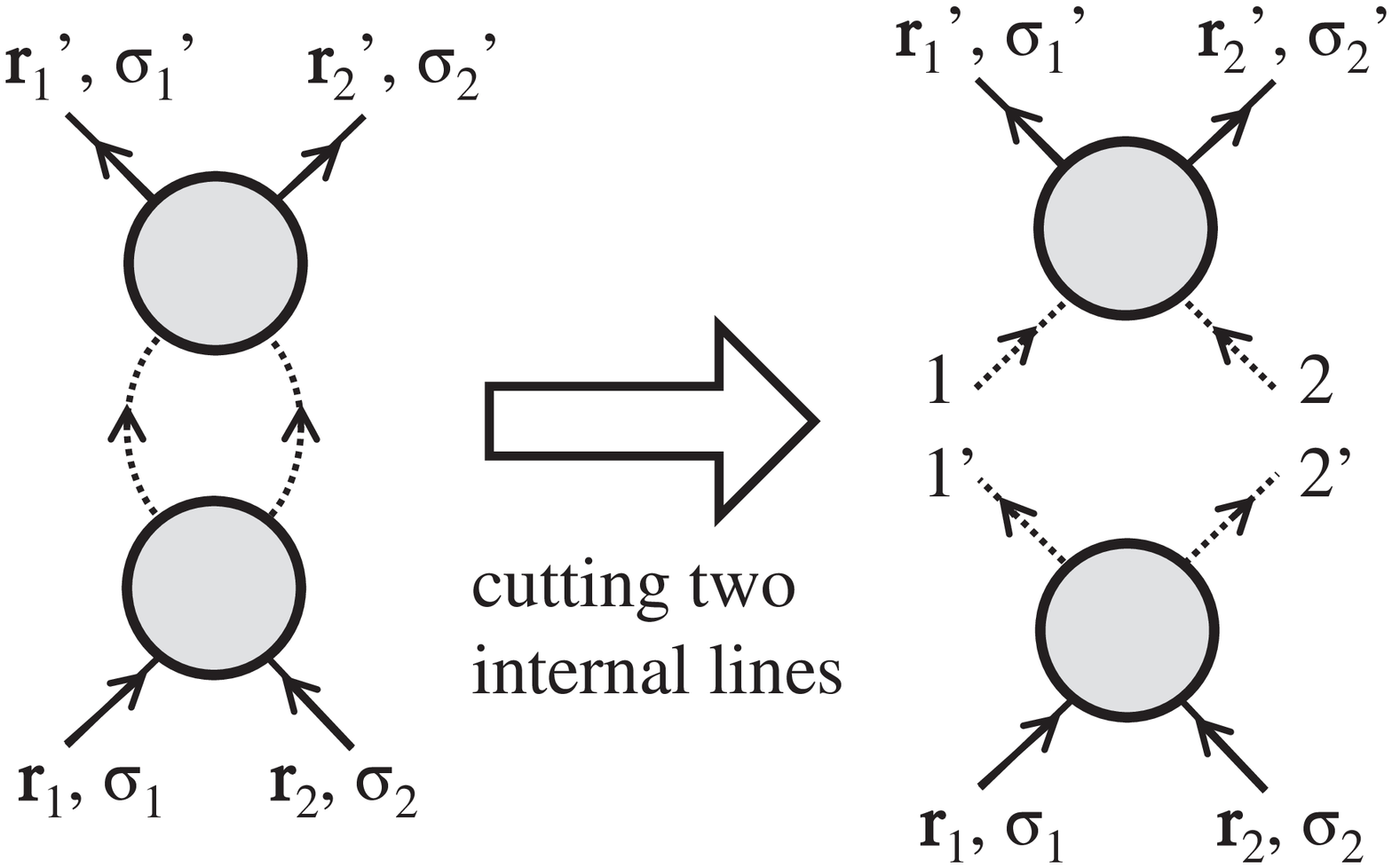}}
\put(20,73){(b)}
\put(110,73){(c)}
\put(35,-3){\includegraphics[width=135pt]{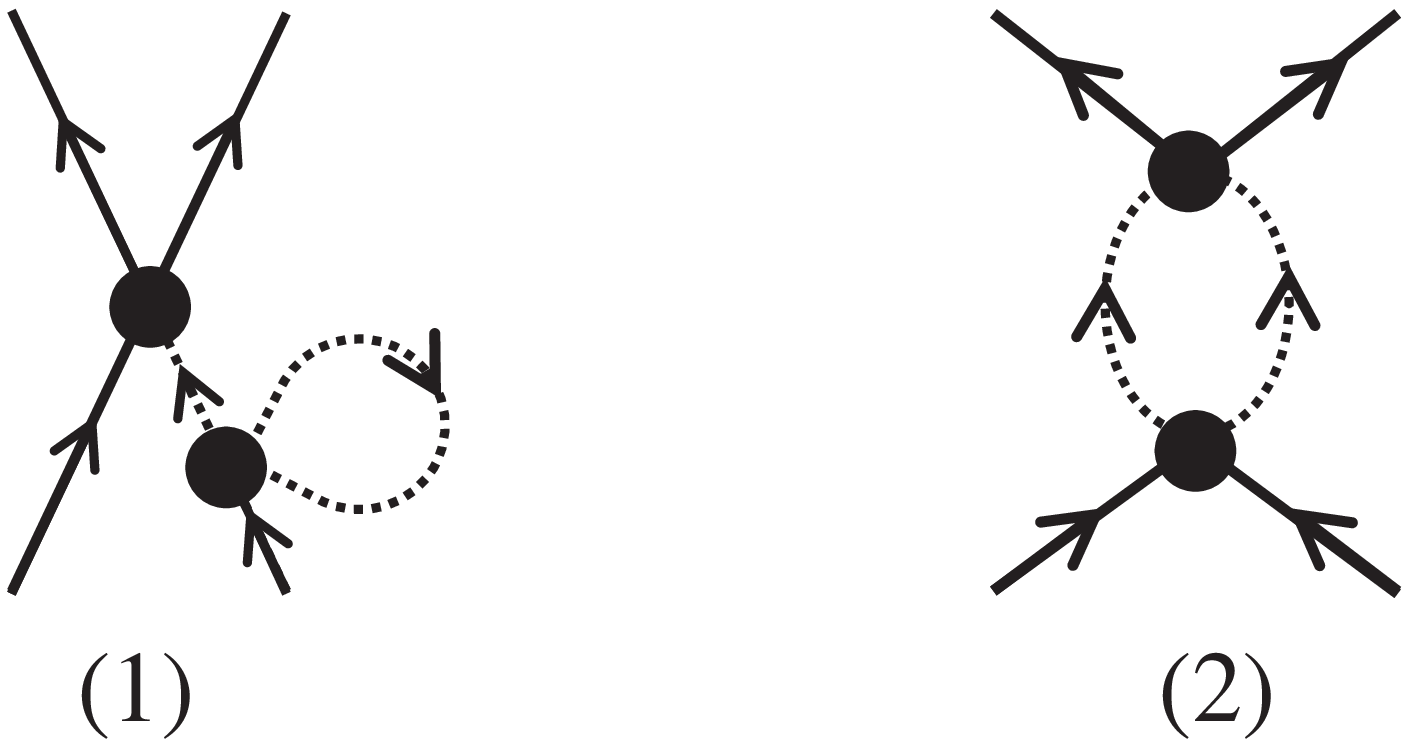}}
\end{picture}
\caption{
(a) 
One of the conditions appearing in the definition of a reducible contracted $2$-graph.
By cutting two of its internal lines and associating $(1,2)$ with the incoming lines and 
$(1',2')$ with the outgoing lines,
the entire graph can be separated into
two disconnected contracted $2$-graphs;
one including two incoming external lines $(1,2)$ and two outgoing external lines,
and
the other including two incoming external lines and two outgoing external lines $(1',2')$.
(b) An example of reducible contracted $2$-graphs. 
If we cut one of its internal lines, the entire graph can be separated into two disconnected parts.
(c) An example of reducible contracted $2$-graphs. 
If we cut two of its internal lines, the entire graph can be separated into two disconnected parts which satisfy the condition given by in (a).
}
\label{fig:Red2}
\end{figure}

The above condition (ii) and its examples are shown in Fig.~\ref{fig:Red2}.
An irreducible contracted $2$-graph is called a simple $2$-diagram in Ref. \cite{dD62}.

\begin{figure}
\begin{picture}(220,285)
\put(0,280){(a)}
\put(0,140){\includegraphics[width=190pt]{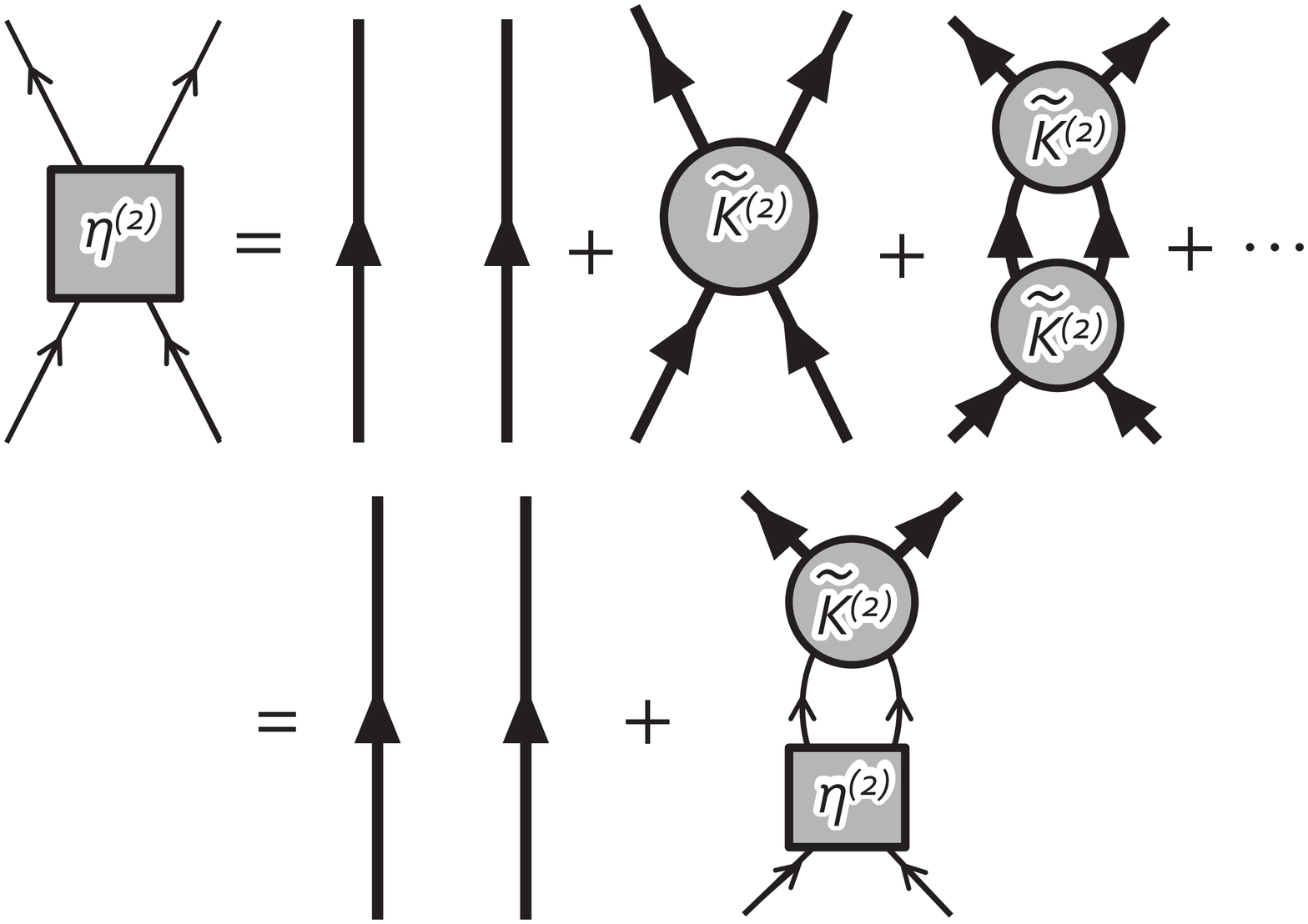}}
\put(0,130){(b)}
\put(0,0){\includegraphics[width=200pt]{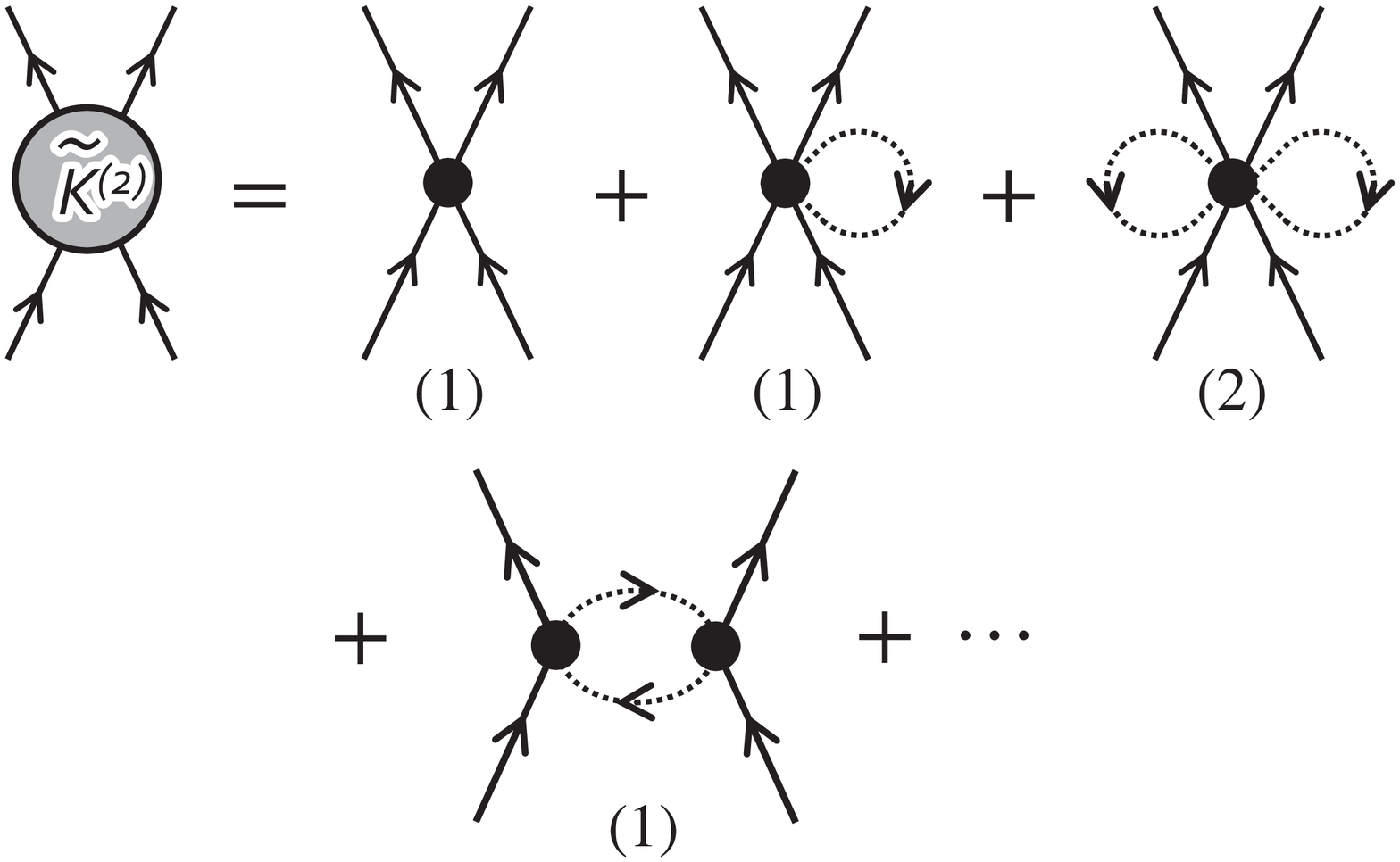}}
\end{picture}
\caption{
(a)
Expression of
$\langle 1',2' | \eta^{(2)} | 1,2 \rangle$
in terms of irreducible contracted $2$-graphs in the  form of a geometric series.
(b)
Expression of
$\langle 1',2' | \tilde{K}^{(2)} | 1,2 \rangle$ as the sum over all different irreducible contracted $2$-graphs. }
\label{fig:Eta2Irr}
\end{figure}

In terms of irreducible contracted $2$-graphs,
$\eta_{\bold{r}'_1\bold{r}'_2;\bold{r}_1\bold{r}_2}^{(2)}$
can be expressed as a geometric series:
\begin{equation}
\begin{split}
 \eta^{(2)}
 = & \eta^{(1\otimes 1)} 
+ \eta^{(1\otimes 1)} \tilde{K}^{(2)} \eta^{(1\otimes 1)}  \\
& + \eta^{(1\otimes 1)} \tilde{K}^{(2)} \eta^{(1\otimes 1)} 
\tilde{K}^{(2)} \eta^{(1\otimes 1)} 
 + \cdots,
\label{eq:Eta2IrrGS}
\end{split}
\end{equation}
%
where
\begin{align}
 \tilde{K}_{\bold{r}'_1\bold{r}'_2;\bold{r}_1\bold{r}_2}^{(2)} 
 = \sum \bigl[ 
&     \text{all different irreducible} \notag\\
&     \text{contracted $2$-graphs} \bigr]. 
\label{eq:K2Irr}
\end{align}
Equation (\ref{eq:Eta2IrrGS}) can be rewritten as 
\begin{equation}
\begin{split}
 \eta^{(2)}
 = \eta^{(1\otimes 1)} 
+ \eta^{(1\otimes 1)} \tilde{K}^{(2)} \eta^{(2)}.
\label{eq:Eta2Irr}
\end{split}
\end{equation}
Equations~(\ref{eq:Eta2IrrGS}), (\ref{eq:K2Irr}) and (\ref{eq:Eta2Irr})
are illustrated in Fig.~\ref{fig:Eta2Irr}.
%
%
%
%
From Eq.~(\ref{eq:Eta2Irr}), we obtain
\begin{equation}
\left( \eta^{(2)} \right)^{\!-1} 
 = 
\left( \eta^{(1\otimes 1)} \right)^{\!-1} -  \tilde{K}^{(2)} .
\end{equation}
Thus, the following theorem is established:\\

\noindent
{\it Theorem 3.} --- 

The following two criteria are equivalent:

1) The largest eigenvalue of the two-particle reduced density matrix $\rho^{(2)}$ is $O(N)$ as $N \to \infty$.

2) The smallest eigenvalue of $\left( \eta^{(1\otimes 1)} \right)^{\!-1} - \tilde{K}^{(2)}
$
 is $O(1/N)$ as $N \to \infty$.\\
 
\noindent
The former implies ODLRO at the two-particle level
and the latter can be obtained from the sum over the contracted graphs.

\subsection{Alternative form of the criterion of ODLRO of a two-component Fermi system}

\begin{figure}
\begin{picture}(220,280)
\put(0,272){(a)}
\put(0,135){\includegraphics[width=190pt]{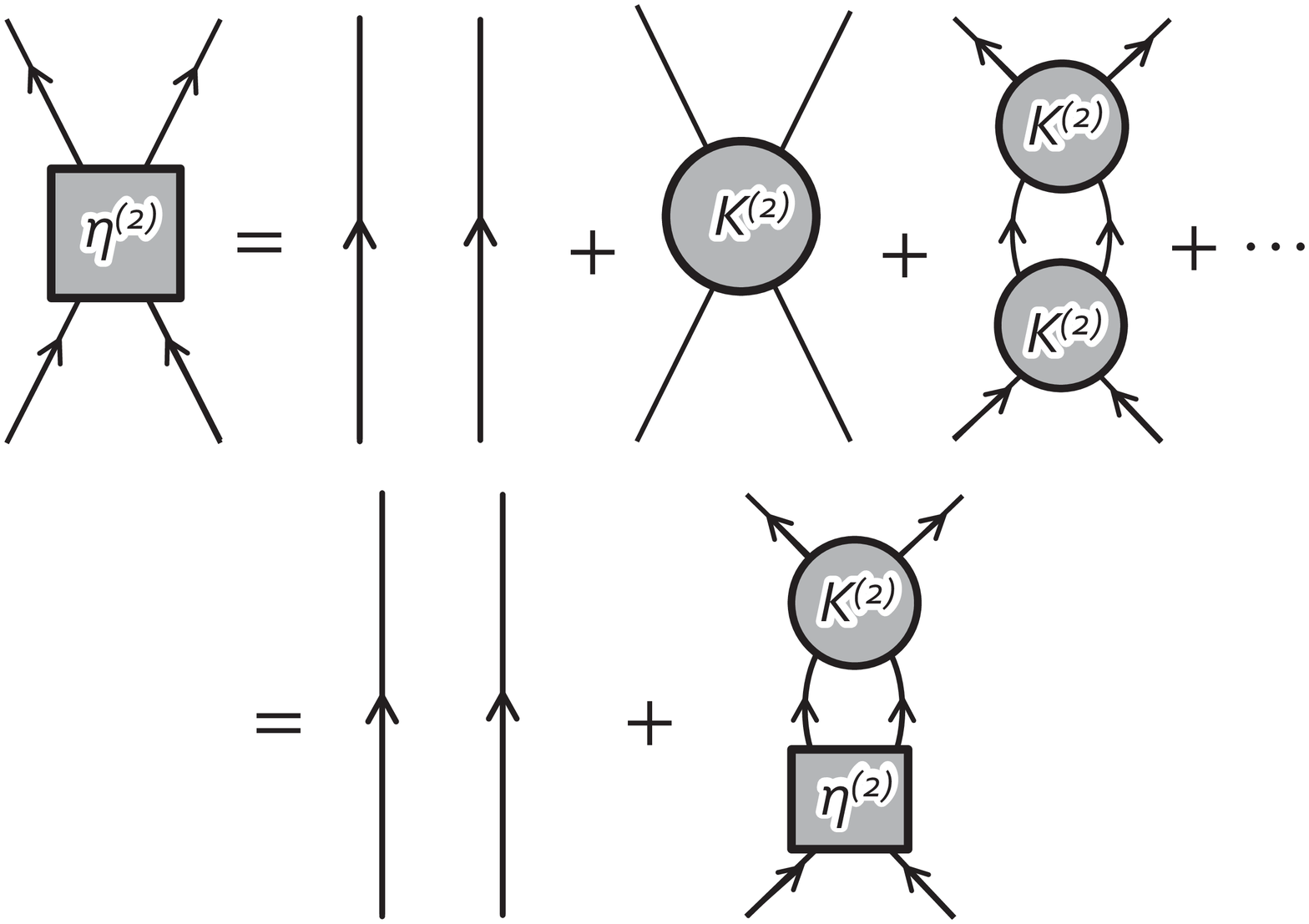}}
\put(0,122){(b)}
\put(0,0){\includegraphics[width=210pt]{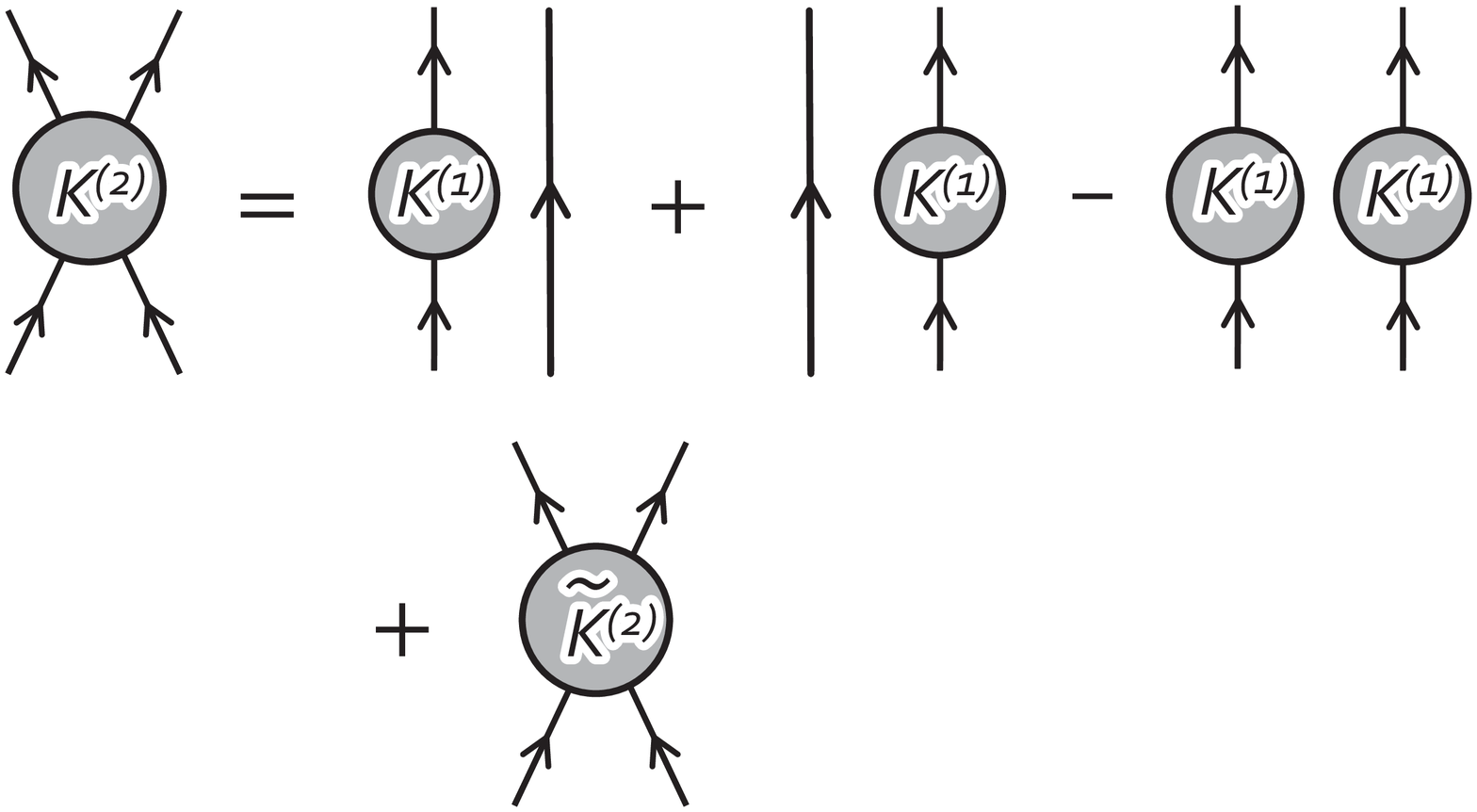}}
\end{picture}
\caption{
(a)
Expression of
$\langle 1',2' | \eta^{(2)} | 1,2 \rangle$
as the sum over all different primary $2$-graphs in the  form of a geometric series.
(b)
Expression of $\langle 1',2' | K^{(2)} | 1,2 \rangle$. 
}
\label{fig:Eta2Pri}
\end{figure}

Following a procedure similar to Sec. \ref{sec:Bose}-C,
we can rewrite 
$\eta_{\bold{r}'_1\bold{r}'_2;\bold{r}_1\bold{r}_2}^{(2)}$
in a geometric series which is different from Eq.~(\ref{eq:Eta2IrrGS}):
\begin{equation}
\begin{split}
 \eta^{(2)}
 = & I + K^{(2)} + \left( K^{(2)} \right)^2 + \left( K^{(2)} \right)^3 + \cdots,
\label{eq:Eta2AltGS}
\end{split}
\end{equation}
where
\begin{equation}
\begin{split}
 K^{(2)}_{\bold{r}'_1\bold{r}'_2;\bold{r}_1\bold{r}_2} 
 := & \eta^{(0\otimes 1)}_{\bold{r}'_1\bold{r}'_2;\bold{r}_1\bold{r}_2}  
    + \eta^{(1\otimes 0)}_{\bold{r}'_1\bold{r}'_2;\bold{r}_1\bold{r}_2}
    - \eta^{(1\otimes 1)}_{\bold{r}'_1\bold{r}'_2;\bold{r}_1\bold{r}_2} \\
&    + \tilde{K}^{(2)}_{\bold{r}'_1\bold{r}'_2;\bold{r}_1\bold{r}_2},
\end{split}
\end{equation}
\begin{equation}
\eta^{(0\otimes 1)}_{\bold{r}'_1\bold{r}'_2;\bold{r}_1\bold{r}_2}
:= \delta (\bold{r}'_1 - \bold{r}_1) \,
\langle \bold{r}_2'\downarrow | \eta^{(1)} | \bold{r}_2\downarrow \rangle,
\end{equation}
and
\begin{equation}
\eta^{(0\otimes 1)}_{\bold{r}'_1\bold{r}'_2;\bold{r}_1\bold{r}_2}
:= \langle \bold{r}_1'\uparrow | \eta^{(1)} | \bold{r}_1\uparrow \rangle
\, \delta (\bold{r}'_2 - \bold{r}_2 ).
\end{equation}
This geometric series is illustrated in Fig.~\ref{fig:Eta2Pri}.

The matrix $\eta^{(2)}_{\bold{r}'_1\bold{r}'_2;\bold{r}_1\bold{r}_2}$
may be expanded in terms of its eigenfunctions $\chi_i (\bold{r}_1,\bold{r}_2)$
with eigenvalues $\bar{n}^{(2)}_i$ as 
\begin{equation}
\eta^{(2)}_{\bold{r}'_1\bold{r}'_2;\bold{r}_1\bold{r}_2}
= \sum_i \bar{n}^{(2)}_i \,
    \bar{\chi}_i^* (\bold{r}_1,\bold{r}_2) \bar{\chi}_i (\bold{r}'_1,\bold{r}'_2).
\end{equation}
[Note that $\bar{\chi}_i(\bold{r}_1,\bold{r}_2) \not= \chi_i (\bold{r}_1,\bold{r}_2)$.]
Since matrices
$\eta^{(2)}$ and $K^{(2)}$ commute,
they can be diagonalized simultaneously.
Thus, we can write
\begin{equation}
K^{(2)}_{\bold{r}'_1\bold{r}'_2;\bold{r}_1\bold{r}_2}
= \sum_i \lambda^{(2)}_i \,
     \bar{\chi}_i^* (\bold{r}_1,\bold{r}_2) \bar{\chi}_i (\bold{r}'_1,\bold{r}'_2).
\end{equation}
We may write 
Eq.~(\ref{eq:Eta1ConGS}) as
\begin{equation}
\bar{n}^{(2)}_i =   \sum_{l=0}^{\infty} \left( \lambda^{(2)}_i \right)^l
 =   \left( 1 - \lambda^{(2)}_i \right)^{-1}.
\end{equation}
Then, we obtain
\begin{equation}
\lambda^{(2)}_i  = 1 - \frac{1}{\bar{n}^{(2)}_i}.
\end{equation}
Thus, the criterion of ODLRO at the two-particle level is stated as the largest eigenvalue of $K^{(2)}$
becoming unity as $N \to \infty$.

\noindent
{\it Theorem 4.} --- 

The following two criteria are equivalent:

1) The largest eigenvalue of the one-particle reduced density matrix $\rho^{(2)}$ is $O(N)$ as $N \to \infty$.

2) The largest eigenvalue of the matrix $K^{(2)}$ is $1+O(1/N)$ as $N \to \infty$.\\

\noindent
The former implies ODLRO at the two-particle level 
and the latter can be obtained from the sum over the contracted $2$-graphs.
Thus we can use Theorems 3 and 4 to find out whether the Fermi system exhibits ODLRO at the two-particle level based on the Lee-Yang cluster expansion method.

 \subsection{Application to a uniform two-component Fermi gas in the tightly bound limit}

As an example of Theorem 4,
we consider a uniform two-component Fermi gas in the tightly bound limit,
which was discussed in {\it Example 4} of Sec. V, and evaluate the critical value $z_c$ for the Bose-Einstein phase transition of diatomic molecules.
Using Eq. (\ref{eq:U2_DiBEC}), we have
\begin{align}
& \langle\bold{k}_1', \bold{k}_2'|K^{(2)}|\bold{k}_1, \bold{k}_2 \rangle \notag\\
& = z^2 \langle\bold{k}_1', \bold{k}_2'|\Upsilon_{\rm A}^{(1,1)}|\bold{k}_1, \bold{k}_2 \rangle 
     + z^2 \, e^{\beta \left| E_{\text{b}} \right|} \! \times \left( O \!\left(\frac{a}{\lambda}\right) + O(z) \right)  \notag\\
&   = z^2 e^{-\beta\bold{K}^2/(4m)} \delta_{\bold{K}, \bold{K}'} 
 \psi_{\text{b}}(\bold{k}') \psi_{\text{b}}^*(\bold{k}) 
   \, e^{\beta \left| E_{\text{b}} \right|}  \notag\\
&      \qquad  + z^2 \, e^{\beta \left| E_{\text{b}} \right|} \! \times \left( O \! \left(\frac{a}{\lambda}\right) + O(z) \right),
\end{align}
where $a$ is an $s$-wave scattering length between up-spin and down-spin atoms.
Here in the tightly bound limit, the following inequalities hold:
$0< z \ll 1$ and $0< a/\lambda \ll 1$.
According to Theorem 4, the critical fugacity $z_c$ satisfies the following equation: 
\begin{equation}
1 = z_c^2 \, e^{\beta \left| E_{\text{b}} \right|} \! \times \left( 1+ O \! \left(\frac{a}{\lambda}\right) + O(z) \right).
\end{equation}
Hence, we obtain
\begin{equation}
z_c = \, e^{- \beta \left| E_{\text{b}} \right|/2} \! \times \left( 1+O \! \left(\frac{a}{\lambda}\right) \right).
\end{equation}
This result agrees with that obtained in {\it Example 4} of Sec.~\ref{sec:Phys} or in Ref~\cite{OU06}.
The formulation given in this section includes the method in Ref~\cite{OU06}
and enables us to systematically compute the criteria of ODLRO at the two-particle level.

\section{summary}
\label{sec:Sum}

We have discussed the Lee-Yang cluster expansion method and its application to the criteria of ODLRO in Bose and Fermi systems.
Lee and Yang \cite{LY60-IV} and de Dominicis \cite{dD62} discussed graphical expansions of the grand partition function and the one- and two-particle reduced density matrices for the case of a one-component uniform system.
We have generalized these results for the case of multi-component uniform and trapped systems.
We have explained a physical meaning of primary $0$-graphs.
In particular, we have elucidated a physical meaning of an infinite series of the ladder-type Lee-Yang primary graphs, the sum of which leads to the BEC of the diatomic molecules below a critical temperature.
We have given the criterion of ODLRO of Bose systems at the one-particle level based on the Lee-Yang cluster expansion method.
It is expressed as a converging infinite series of irreducible contracted $1$-graphs or irreducible primary $1$-graphs.
Applications to a Bose gas of hard spheres have also been made.
Furthermore, we have given the criterion of ODLRO of Fermi systems at the two-particle level based on the Lee-Yang cluster expansion method.
This is expressed as a converging infinite series of irreducible contracted $2$-graphs or irreducible primary $2$-graphs.
Applications to a two-component Fermi gas in the tightly bound limit have also been made.
Finally, we note that the formulation given in this paper includes the method 
in Ref.~\cite{LY60-IV} for Bose systems and in Ref.~\cite{OU06} for Fermi systems, enabling us to systematically compute the criteria of ODLRO.

\begin{acknowledgements}
We thank S.~Endo, T.~N.~Ikeda, K.~Inokuchi, T.~Ohkuma H.~Tasaki and M.~Tezuka for useful discussions.
This work was supported by KAKENHI (Nos. 22340114, 22103005, 21540359 and 20102008) and by a Grant-in-Aid for the Global COE Program
``the Next Generation of Physics, Spun from Universality and
Emergence'' from MEXT of Japan. N. K. is partly supported by JSPS
through the ``Funding Program for World-Leading Innovative R\&D on
Science and Technology (FIRST Program)''.
\end{acknowledgements}

\appendix

\section{Derivations of quantum cluster expansion formulae}
\label{app:CE}

In this Appendix we derive the quantum cluster expansion formulae of the grand partition function (\ref{eq:Xi-b}) and the one- and two-particle reduced density matrices
(\ref{eq:DM1-UA-u}), (\ref{eq:DM2-UA-uu}) and (\ref{eq:DM2-UA-ud}).
We first introduce partial traces of $W_{\alpha}^{(n_\uparrow, n_\downarrow)}$ and $U_{\alpha}^{(n_\uparrow, n_\downarrow)}$,
which are denoted as
\begin{equation}
\begin{split}
& \langle \, \bold{y} \uparrow | \bar{X}_{\alpha}^{(n_\uparrow, n_\downarrow)} | \bold{x} \uparrow \, \rangle \\
& =  \frac{1}{(n_\uparrow -1) ! }\frac{1}{n_\downarrow ! }
 \int \! d^{3(n_\uparrow-1)}\bold{r}_{\uparrow} \int d^{3n_\downarrow}\bold{r}_{\downarrow} \\
& \quad \times \langle \bold{y}, 1_\uparrow,\dots , n_\uparrow -1; 1_\downarrow,\dots , n_\downarrow | \\
& \qquad X_{\alpha}^{(n_\uparrow, n_\downarrow)}  | 
 \bold{x}, 1_\uparrow,\dots , n_\uparrow -1; 1_\downarrow,\dots , n_\downarrow \rangle ,
\end{split}
\end{equation}
\begin{equation}
\begin{split}
& \langle \, \bold{y}_1 \uparrow , \bold{y}_2 \uparrow | \bar{X}_{\alpha}^{(n_\uparrow, n_\downarrow)} | \bold{x}_1 \uparrow , \bold{x}_2 \uparrow  \, \rangle \\
& :=  \frac{1}{(n_\uparrow -2) ! }\frac{1}{n_\downarrow ! }
 \int \! d^{3(n_\uparrow-2)}\bold{r}_{\uparrow} \int d^{3n_\downarrow}\bold{r}_{\downarrow} \\
& \,\, \times \langle \bold{y}_1,\bold{y}_2,1_\uparrow,\dots , n_\uparrow -2; 1_\downarrow,\dots , n_\downarrow| \\
& \quad \, X_{\alpha}^{(n_\uparrow, n_\downarrow)}  | 
  \bold{x}_1,\bold{x}_2, 1_\uparrow,\dots , n_\uparrow -2; 1_\downarrow,\dots , n_\downarrow \rangle,
\end{split}
\end{equation}
\begin{equation}
\begin{split}
& \langle \, \bold{y}_1 \uparrow , \bold{y}_2 \downarrow | \bar{X}_{\alpha}^{(n_\uparrow, n_\downarrow)} | \bold{x}_1 \uparrow , \bold{x}_2 \downarrow  \, \rangle \\
& :=  \frac{1}{(n_\uparrow -1) ! }\frac{1}{(n_\downarrow -1)! }
 \int \! d^{3(n_\uparrow-1)}\bold{r}_{\uparrow} \int d^{3(n_\downarrow -1)}\bold{r}_{\downarrow} \\
& \,\, \times \langle \bold{y}_1,1_\uparrow,\dots , n_\uparrow -1; \bold{y}_2,1_\downarrow,\dots , n_\downarrow -1| \\
& \quad \, X_{\alpha}^{(n_\uparrow, n_\downarrow)}  | 
  \bold{x}_1,1_\uparrow,\dots , n_\uparrow -1; \bold{x}_2, 1_\downarrow,\dots , n_\downarrow -1 \rangle,
\end{split}
\end{equation}
where $X_{\alpha}^{(n_\uparrow, n_\downarrow)} = W_{\alpha}^{(n_\uparrow, n_\downarrow)}$ or $U_{\alpha}^{(n_\uparrow, n_\downarrow)}$,
and
$\bar{X}_{\alpha}^{(n_\uparrow, n_\downarrow)} = \bar{W}_{\alpha}^{(n_\uparrow, n_\downarrow)}$ or $\bar{U}_{\alpha}^{(n_\uparrow, n_\downarrow)}$.

\subsection{Grand partition function: Proof of Eq.~(\ref{eq:Xi-b})}

To prove Eq.~(\ref{eq:Xi-b}), we first note that a general term in Eq.~(\ref{eq:def:UA}) for 
$W_{\alpha}^{(N_\uparrow, N_\downarrow)}$
is expressed as the sum over products of $m_{l_\uparrow, l_\downarrow}( = 0, 1, 2, \dots)$
 cluster functions $U_{\alpha}^{(l_\uparrow, l_\downarrow)}$:
\begin{align}
&  \langle 1_\uparrow,\dots , N_\uparrow; 1_\downarrow,\dots , N_\downarrow|  W_{\alpha}^{(N_\uparrow, N_\downarrow)}  | 
  1_\uparrow,\dots , N_\uparrow; 1_\downarrow,\dots , N_\downarrow \rangle
  \notag\\
& = \sum_{\{ m_{l_\uparrow, l_\downarrow} \}} \sum_{\rm{per.}}
 \underbrace{
\left(
U_{\alpha}^{(1,0)} \cdots U_{\alpha}^{(1,0)} 
\right)
}_{m_{1,0}}
\underbrace{
\left(
U_{\alpha}^{(2,0)} \cdots U_{\alpha}^{(2,0)} 
\right)
}_{m_{2,0}}
\dots \notag\\
& \quad \times
\underbrace{
\left(
U_{\alpha}^{(l_\uparrow, l_\downarrow)} \cdots U_{\alpha}^{(l_\uparrow, l_\downarrow)} 
\right)
}_{m_{l_\uparrow, l_\downarrow}}
\cdots,
\label{eq:WA}
\end{align}
where the set of integers $\{ m_{l_\uparrow, l_\downarrow} \}$ satisfies the conditions
\begin{align}
   \sum_{l_\uparrow = 0}^{N_\uparrow}
   \sum_{l_\downarrow = 0}^{N_\downarrow}
   l_\uparrow m_{l_\uparrow, l_\downarrow}
  = N_\uparrow,
 \label{eq:WA_cond1}  \\
   \sum_{l_\uparrow = 0}^{N_\uparrow}
   \sum_{l_\downarrow = 0}^{N_\downarrow}
   l_\downarrow m_{l_\uparrow, l_\downarrow}
  = N_\downarrow .
 \label{eq:WA_cond2} 
\end{align}
In Eq.~(\ref{eq:WA}), $\sum_{\{ m_{l_\uparrow, l_\downarrow} \}}$ denotes the sum over all sets $\{ m_{l_\uparrow, l_\downarrow} \}$ satisfying the conditions
(\ref{eq:WA_cond1}) and (\ref{eq:WA_cond2}),
and $\sum_{\rm{per.}}$ is the sum over all different ways of assigning 
$\{ \bold{r}_{1\uparrow},\dots , \bold{r}_{N\uparrow}, \bold{r}_{1\downarrow},\dots , \bold{r}_{N\downarrow} \}$
to each $U_{\alpha}^{(l_\uparrow, l_\downarrow)}$.
In accordance with Eq.~(\ref{eq:def:Xi}),
we integrate $W_{\alpha}^{(N_\uparrow, N_\downarrow)}$ over all the coordinates.

Integrating $W_{\alpha}^{(N_\uparrow, N_\downarrow)}$ using Eq.~(\ref{eq:WA}),
we obtain the same result for every term in the sum $\sum_{\rm{per.}}$
for fixed $\{ m_{l_\uparrow, l_\downarrow} \}$.
The number of such terms in the sum $\sum_{\rm{per.}}$ is
\begin{equation}
(N_\uparrow!) (N_\downarrow!) \Biggl[ \,
 \prod_{l_\uparrow = 0}^{N_\uparrow} 
 \prod_{l_\downarrow = 0}^{N_\downarrow }
 m_{l_\uparrow, l_\downarrow} !
  \left\{ (l_\uparrow!) (l_\downarrow!) \right\}^{m_{l_\uparrow, l_\downarrow}}
\Biggr]^{-1}.
\end{equation}
By the definition of $B^{(l_\uparrow, l_\downarrow)}$ in Eq.~(\ref{eq:b-UA}),
we have
\begin{equation}
\begin{split}
& \int \! d^{3l_\uparrow}\bold{r}_{\uparrow} \int d^{3l_\downarrow}\bold{r}_{\downarrow} 
\langle 1_\uparrow,\dots , l_\uparrow; 1_\downarrow,\dots , l_\downarrow | \\
& \qquad\qquad\qquad\quad U_{\alpha}^{(l_\uparrow, l_\downarrow)} |  1_\uparrow,\dots , l_\uparrow; 1_\downarrow,\dots , l_\downarrow \rangle \\
&= (l_\uparrow !) (l_\downarrow !)  B^{(l_\uparrow, l_\downarrow)}.
\end{split}
\end{equation}
We thus obtain
\begin{align}
& \int \! d^{3l_\uparrow}\bold{r}_{\uparrow} \int d^{3l_\downarrow}\bold{r}_{\downarrow}
   \langle 1_\uparrow,\dots , N_\uparrow; 1_\downarrow,\dots , N_\downarrow| \notag\\
& \qquad\qquad \quad W_{\alpha}^{(N_\uparrow, N_\downarrow)}  | 
  1_\uparrow,\dots , N_\uparrow; 1_\downarrow,\dots , N_\downarrow \rangle
  \label{eq:WA_tr}\\
& = (N_\uparrow!) (N_\downarrow!)  \!\!\!
\sum_{\{ m_{l_\uparrow, l_\downarrow} \} }
 \prod_{l_\uparrow = 0}^{N_\uparrow} 
 \prod_{l_\downarrow = 0}^{N_\downarrow }
 \frac{1}{m_{l_\uparrow, l_\downarrow} !}
  \left( B^{(l_\uparrow, l_\downarrow)} \right)^{m_{l_\uparrow, l_\downarrow}}.\notag
\end{align}
The grand partition function can be obtained by substituting Eq.~(\ref{eq:WA_tr}) in Eq.~(\ref{eq:def:Xi}):
\begin{align}
\Xi
&= \sum_{N_\uparrow = 0}^\infty \sum_{N_\downarrow = 0}^\infty
   \sum_{\{ m_{l_\uparrow, l_\downarrow} \} }  \notag\\
& \qquad
 \prod_{l_\uparrow = 0}^{N_\uparrow} 
 \prod_{l_\downarrow = 0}^{N_\downarrow }
 \frac{1}{m_{l_\uparrow, l_\downarrow} !}
    \left(
   z_\uparrow^{l_\uparrow}
   z_\downarrow^{l_\downarrow}
  B^{(l_\uparrow, l_\downarrow)} \right)^{m_{l_\uparrow, l_\downarrow}} \notag\\
&= 
 \prod_{l_\uparrow = 0}^{\infty} 
 \prod_{l_\downarrow = 0}^{\infty}
\sum_{m_{l_\uparrow, l_\downarrow} =0 }^{\infty}
 \frac{1}{m_{l_\uparrow, l_\downarrow} !}
  \left(
   z_\uparrow^{l_\uparrow}
   z_\downarrow^{l_\downarrow}
  B^{(l_\uparrow, l_\downarrow)} \right)^{m_{l_\uparrow, l_\downarrow}} \notag\\
&= \exp 
  \left( 
 \sum_{l_\uparrow = 0}^{\infty} 
 \sum_{l_\downarrow = 0}^{\infty}
   z_\uparrow^{l_\uparrow}
   z_\downarrow^{l_\downarrow}
  B^{(l_\uparrow, l_\downarrow)} \right).
\label{eq:Xi:App}
\end{align}

\subsection{One-particle reduced density matrix: Proofs of Eqs.~(\ref{eq:DM1-UA-u}) and (\ref{eq:DM1-UA-d})}


To prove Eq.~(\ref{eq:DM1-UA-u}),
we first rewrite the one-particle reduced density matrix using $W_{\alpha}^{(N_\uparrow, N_\downarrow)}$.
We note that
\begin{equation}
\begin{split}
& {\mathop{\mathrm{tr}}\nolimits}_{\mathcal{H}_{\alpha_\uparrow}^{(N_\uparrow)} \otimes \mathcal{H}_{\alpha_\downarrow}^{(N_\downarrow)}}
  \left( \hat{\Psi}_{\uparrow}^{\dagger} (\bold{x}) \hat{\Psi}_{\uparrow} (\bold{y})  e^{-\beta \hat{H}}  \right)  \\
& = \!\!\! \sum_{\psi_i \in \mathcal{H}_{\alpha_\uparrow}^{(N_\uparrow)} \otimes \mathcal{H}_{\alpha_\downarrow}^{(N_\downarrow)}} 
 \int \! d^{3(N_\uparrow -1)}\bold{r}_{\uparrow} \int d^{3(N_\downarrow -1)}\bold{r}_{\downarrow} \\
&\quad \times \langle \psi_i  | \hat{\Psi}_{\uparrow}^{\dagger} (\bold{x})  | 1,\dots , N-1 \rangle \\
&  \quad \times \langle 1,\dots , N-1  | \hat{\Psi}_{\uparrow} (\bold{y}) | \psi_i \rangle e^{-\beta E_i^{(N)}}
\\
& = N_{\uparrow} \cdot \!\!\!\! \sum_{\psi_i \in \mathcal{H}_{\alpha_\uparrow}^{(N_\uparrow)} \otimes \mathcal{H}_{\alpha_\downarrow}^{(N_\downarrow)}} 
\int \! d^{3(N_\uparrow -1)}\bold{r}_{\uparrow} \int d^{3N_\downarrow}\bold{r}_{\downarrow}  \\
& \quad \times \psi_i^* \left( \bold{x}, 1_\uparrow,\dots , N_\uparrow -1; 1_\downarrow,\dots , N_\downarrow  \right) \\
&  \quad \times  \psi_i \left( \bold{y},1_\uparrow,\dots , N_\uparrow -1; 1_\downarrow,\dots , N_\downarrow \right)  e^{-\beta E_i^{(N)}} \\
&   = \frac{1}{(N_\uparrow -1)! N_\downarrow !}
  \int \! d^{3(N_\uparrow -1)}\bold{r}_{\uparrow} \int d^{3N_\downarrow}\bold{r}_{\downarrow} \\
& \quad \times \langle \bold{y},1_\uparrow,\dots , N_\uparrow -1; 1_\downarrow,\dots , N_\downarrow| \\
& \qquad\,\,\, W_{\alpha}^{(N_\uparrow, N_\downarrow)}  | 
  \bold{x}, 1_\uparrow,\dots , N_\uparrow -1; 1_\downarrow,\dots , N_\downarrow \rangle.
\end{split}
\end{equation}
Then, the one-particle reduced density matrix is rewritten as
\begin{equation}
\begin{split}
& \langle \bold{x} \uparrow | \rho^{(1)} | \bold{y} \uparrow \rangle
 := \langle
 \hat{\Psi}_{\uparrow}^{\dagger} (\bold{x}) \hat{\Psi}_{\uparrow} (\bold{y}) 
   \rangle  \\
& = \frac{1}{\Xi} \sum_{N_\uparrow=1}^\infty  \sum_{N_\downarrow=0}^\infty  
  z_{\uparrow}^{N_\uparrow -1} z_{\downarrow}^{N_\downarrow}
 \langle \, \bold{y} \uparrow | \bar{W}_{\alpha}^{(N_\uparrow, N_\downarrow)} | \bold{x} \uparrow \, \rangle.
\label{eq:1DenMat_and_W}
\end{split}
\end{equation}

Using the definitions of the cluster functions (\ref{eq:def:UA}), 
$W_{\alpha}^{(N_\uparrow, N_\downarrow)}$ in Eq.~(\ref{eq:1DenMat_and_W}) 
can be expressed as the sum over products of the cluster functions:
\begin{align}
& \langle \bold{y},1_\uparrow,\dots , N_\uparrow -1; 1_\downarrow,\dots , N_\downarrow| \notag\\
& \qquad\quad W_{\alpha}^{(N_\uparrow, N_\downarrow)}  | 
  \bold{x}, 1_\uparrow,\dots , N_\uparrow -1; 1_\downarrow,\dots , N_\downarrow \rangle \notag\\
& = \sum_{n_\uparrow=1}^{N_\uparrow} \sum_{n_\downarrow=0}^{N_\downarrow} \sum_{\{ m_{l_\uparrow, l_\downarrow} \}} \sum_{\rm{per.}}\notag\\
& \quad \langle \bold{y},a_1,\dots , a_{n_\uparrow -1}; \mathfrak{a}_1,\dots , \mathfrak{a}_{n_\downarrow} | \notag\\
& \qquad \qquad U_{\alpha}^{(n_\uparrow, n_\downarrow)}  |   \bold{x}, a_1,\dots , a_{n_\uparrow -1}; \mathfrak{a}_1,\dots , \mathfrak{a}_{n_\downarrow} \rangle \notag\\
& \times \langle b_1| U_{\alpha}^{(1,0)}  | b_1 \rangle \cdots \langle b_{m_{1,0}}| U_{\alpha}^{(1,0)}  | b_{m_{1,0}} \rangle \notag\\
& \times \cdots \notag\\
& \times \langle c_1,\dots , c_{l_\uparrow};\mathfrak{c}_1,\dots , \mathfrak{c}_{l_\downarrow}
  |  U_{\alpha}^{(l_\uparrow, l_\downarrow)} | c_1,\dots , c_{l_\uparrow};\mathfrak{c}_1,\dots , \mathfrak{c}_{l_\downarrow} \rangle \notag\\ 
 & \cdots  
 \langle \, \dots ,c_{l_\uparrow m_{l_\uparrow, l_\downarrow}}; \dots ,\mathfrak{c}_{l_\downarrow m_{l_\uparrow, l_\downarrow}} | \notag\\
& \qquad\qquad U_{\alpha}^{(l_\uparrow, l_\downarrow)}   | \, \dots ,c_{l_\uparrow m_{l_\uparrow, l_\downarrow}}; \dots ,\mathfrak{c}_{l_\downarrow m_{l_\uparrow, l_\downarrow}} \rangle 
\notag\\
& \times \cdots,
\label{eq:1DenMat_WA_by_UA}
\end{align}
where $\sum_{\{ m_{l_\uparrow, l_\downarrow} \}}$ is the sum over all sets of integers $\{ m_{l_\uparrow, l_\downarrow} \}$ satisfying $N_\sigma=n_\sigma+ \sum_{l_\sigma=1}^{\infty}l_\sigma m_{l_\uparrow, l_\downarrow}$,
and $\sum_{\rm{per.}}$ is the sum over different ways of grouping as follows:
\begin{equation}
\begin{split}
&(a_1,\dots,a_{n_\uparrow -1}), \{(b_1), \dots,(b_{m_{1,0}}) \}, \dots \\
&\{ (c_1,\dots,c_{l_\uparrow}),\dots,(\dots,c_{l_\uparrow m_{l_\uparrow, l_\downarrow}})\}, \dots ,
\label{grouping:W-U1}
\end{split}
\end{equation}
where $a_1,\dots,a_{n_\uparrow -1}, b_1, \dots$ is a permutation of the coordinates
$ \{ 1_\uparrow,\dots , N_\uparrow -1\}$,
and
\begin{equation}
\begin{split}
&(\mathfrak{a}_1,\dots,\mathfrak{a}_{n_\uparrow -1}), \dots \\
&\{ (\mathfrak{c}_1,\dots,\mathfrak{c}_{l_\uparrow}),\dots,(\dots,\mathfrak{c}_{l_\uparrow m_{l_\uparrow, l_\downarrow}})\}, \dots ,
\label{grouping:W-U2}
\end{split}
\end{equation}
where $\mathfrak{a}_1,\dots,\mathfrak{a}_{n_\uparrow -1}, \dots$ is a permutation of the coordinates
$ \{ 1_\downarrow,\dots , N_\downarrow\}$.
In the groupings (\ref{grouping:W-U1}) and (\ref{grouping:W-U2}),
within each pair of round brackets the integers are arranged in ascending order and
within each curly bracket the round brackets are arranged such that their first entries follow ascending order.
%
%
%
In general, Eq.~(\ref{eq:1DenMat_WA_by_UA}) is a product of 
one $U_{\alpha}^{(n_\uparrow, n_\downarrow)}$ which includes $\bold{x}$ and $\bold{y}$ as two of its variables,
and products of cluster functions $U_{\alpha}^{(l_\uparrow, l_\downarrow)} $,
which include other particles as variables.
According to Eq.~(\ref{eq:1DenMat_and_W}),
we need to integrate $W_{\alpha}^{(N_\uparrow, N_\downarrow)}$.
Using Eq.~(\ref{eq:1DenMat_WA_by_UA}),
if we integrate $W_{\alpha}^{(N_\uparrow, N_\downarrow)}$
over coordinates $\{ 1_\uparrow,\dots , N_\uparrow -1; 1_\downarrow,\dots , N_\downarrow \}$,
we obtain the same result for every term in the sum $\sum_{\rm{per.}}$
for fixed $n_\uparrow$, $n_\downarrow$ and $\{ m_{l_\uparrow, l_\downarrow} \}$.
The number of such terms in the sum $\sum_{\rm{per.}}$ is
\begin{equation*}
\frac{(N_\uparrow -1) !N_\downarrow ! }{(n_\uparrow -1) !n_\downarrow ! }
\cdot
\biggl[ \,
 \prod_{l_\uparrow = 0}^{N_\uparrow} 
 \prod_{l_\downarrow = 0}^{N_\downarrow }
 m_{l_\uparrow, l_\downarrow} !
  \left\{ (l_\uparrow!) (l_\downarrow!) \right\}^{m_{l_\uparrow, l_\downarrow}}
\biggr]^{-1}.
\end{equation*}

We thus obtain
\begin{equation}
\begin{split}
& \langle \, \bold{y} \uparrow | \bar{W}_{\alpha}^{(N_\uparrow, N_\downarrow)} | \bold{x} \uparrow \, \rangle \\
& = 
\sum_{n_\uparrow=1}^{N_\uparrow} \sum_{n_\downarrow=0}^{N_\downarrow} \sum_{\{ m_{l_\uparrow, l_\downarrow} \}}  
\langle \, \bold{y} \uparrow | \bar{U}_{\alpha}^{(n_\uparrow, n_\downarrow)} | \bold{x} \uparrow \, \rangle \\
&\quad \times
 \prod_{l_\uparrow = 0}^{N_\uparrow} 
 \prod_{l_\downarrow = 0}^{N_\downarrow }
 \frac{1}{m_{l_\uparrow, l_\downarrow} !}
  \left( B^{(l_\uparrow, l_\downarrow)} \right)^{m_{l_\uparrow, l_\downarrow}}.
\label{eq:PTr_Wuu}
\end{split}
\end{equation}

Substituting Eq.~(\ref{eq:PTr_Wuu}) into Eq.~(\ref{eq:1DenMat_and_W}),
we obtain
\begin{equation}
\begin{split}
& \langle \bold{x} \uparrow | \rho^{(1)} | \bold{y} \uparrow \rangle \\
& = \frac{1}{\Xi} \sum_{N_\uparrow=1}^\infty  \sum_{N_\downarrow=0}^\infty 
\sum_{n_\uparrow=1}^{N_\uparrow} \sum_{n_\downarrow=0}^{N_\downarrow} \sum_{\{ m_{l_\uparrow, l_\downarrow} \}}  \\
& \qquad z_\uparrow^{n_\uparrow -1}z_\downarrow^{n_\downarrow}
  \langle \, \bold{y} \uparrow | \bar{U}_{\alpha}^{(n_\uparrow, n_\downarrow)} | \bold{x} \uparrow \, \rangle \\
& \quad \times 
 \prod_{l_\uparrow = 0}^{N_\uparrow} 
 \prod_{l_\downarrow = 0}^{N_\downarrow }
 \frac{1}{m_{l_\uparrow, l_\downarrow} !}
  \left( 
  z^{l_\uparrow} z^{l_\downarrow}
  B^{(l_\uparrow, l_\downarrow)} \right)^{m_{l_\uparrow, l_\downarrow}}.
\end{split}
\end{equation}
Using Eq.~(\ref{eq:Xi:App}), we find 
\begin{equation}
\begin{split}
&  \langle \bold{x} \uparrow | \rho^{(1)} | \bold{y} \uparrow \rangle   \\
& = 
 \sum_{n_\uparrow = 1}^\infty \sum_{n_\downarrow = 0}^\infty
  z_\uparrow^{n_\uparrow -1}z_\downarrow^{n_\downarrow}
  \langle \, \bold{y} \uparrow | \bar{U}_{\alpha}^{(n_\uparrow, n_\downarrow)} | \bold{x} \uparrow \, \rangle.
 \label{DM1-UA-u:App}
\end{split}
\end{equation}
The proof of Eq.~(\ref{eq:DM1-UA-d}) can similarly be made.


\subsection{Two-particle reduced density matrix: Proofs of Eqs.~(\ref{eq:DM2-UA-uu})-(\ref{eq:DM2-UA-ud})}


To prove Eqs.~(\ref{eq:DM2-UA-uu})-(\ref{eq:DM2-UA-ud}),
we rewrite the two-particle reduced density matrix using $W_{\alpha}^{(N_\uparrow, N_\downarrow)}$
in a manner similar to what we have done to the one-particle reduced density matrix in the preceding subsection.
The results are
\begin{equation}
\begin{split}
& \langle \bold{x}_1 \uparrow, \bold{x}_2 \uparrow | \rho^{(2)} | \bold{y}_1 \uparrow , \bold{y}_2 \uparrow \rangle \\
& := \langle
 \hat{\Psi}_{\uparrow}^{\dagger} (\bold{x}_1) \hat{\Psi}_{\uparrow}^{\dagger} (\bold{x}_2) \hat{\Psi}_{\uparrow} (\bold{y}_2) \hat{\Psi}_{\uparrow} (\bold{y}_1) 
   \rangle  \\
& = \frac{1}{\Xi} \sum_{N_\uparrow=2}^\infty  \sum_{N_\downarrow=0}^\infty    
    z_\uparrow^{N_\uparrow -2} z_\downarrow^{N_\downarrow} \\
& \quad \times \langle \, \bold{y}_1 \uparrow , \bold{y}_2 \uparrow | \bar{W}_{\alpha}^{(N_\uparrow, N_\downarrow)} | \bold{x}_1 \uparrow , \bold{x}_2 \uparrow  \, \rangle ,
\label{eq:DM2-W-uu}
\end{split}
\end{equation}
and
\begin{equation}
\begin{split}
& \langle \bold{x}_1 \uparrow, \bold{x}_2 \downarrow | \rho^{(2)} | \bold{y}_1 \uparrow , \bold{y}_2 \downarrow \rangle \\
& := \langle
 \hat{\Psi}_{\uparrow}^{\dagger} (\bold{x}_1) \hat{\Psi}_{\downarrow}^{\dagger} (\bold{x}_2) \hat{\Psi}_{\downarrow} (\bold{y}_2) \hat{\Psi}_{\uparrow} (\bold{y}_1) 
   \rangle  \\
& = \frac{1}{\Xi} \sum_{N_\uparrow=1}^\infty  \sum_{N_\downarrow=1}^\infty  
   z_\uparrow^{N_\uparrow -1} z_\downarrow^{N_\downarrow -1} \\
& \quad \times  \langle \, \bold{y}_1 \uparrow , \bold{y}_2 \downarrow | \bar{W}_{\alpha}^{(N_\uparrow, N_\downarrow)} | \bold{x}_1 \uparrow , \bold{x}_2 \downarrow  \, \rangle .
\label{eq:DM2-W-ud}
\end{split}
\end{equation}


First, we prove Eq.~(\ref{eq:DM2-UA-uu}).
We now use the definitions of the cluster functions (\ref{eq:def:UA})  and note that each $W_{\alpha}^{(N_\uparrow, N_\downarrow)}$ in Eq.~(\ref{eq:DM2-W-uu}) 
can be expressed as the sum over products of the two forms as follows:
\begin{align}
& \langle \bold{y}_1,\bold{y}_2,a_1,\dots , a_{n_\uparrow -2}; b_1,\dots , b_{n_\downarrow} | \notag\\
& \qquad U_{\alpha}^{(n_\uparrow, n_\downarrow)}  |   \bold{x}_1,\bold{x}_2,a_1,\dots , a_{n_\uparrow -2}; b_1,\dots , b_{n_\downarrow} \rangle \notag\\
& \times \langle c_1| U_{\alpha}^{(1,0)}  | c_1 \rangle \cdots \langle c_{m_{1,0}}| U_{\alpha}^{(1,0)}  | c_{m_{1,0}} \rangle \notag\\
& \times \cdots \notag\\
& \times \langle d_1,\dots , d_{l_\uparrow};e_1,\dots , e_{l_\downarrow}
  |  U_{\alpha}^{(l_\uparrow, l_\downarrow)} | d_1,\dots , d_{l_\uparrow};e_1,\dots , e_{l_\downarrow} \rangle \notag\\ 
 & \qquad \cdots  
 \langle \, \dots ,d_{l_\uparrow m_{l_\uparrow, l_\downarrow}}; \dots ,e_{l_\downarrow m_{l_\uparrow, l_\downarrow}} | \notag\\
& \qquad\qquad\qquad U_{\alpha}^{(l_\uparrow, l_\downarrow)}   | \, \dots ,d_{l_\uparrow m_{l_\uparrow, l_\downarrow}}; \dots ,e_{l_\downarrow m_{l_\uparrow, l_\downarrow}} \rangle 
\notag\\
& \times \cdots ,
\label{eq:2DenMat_WA_by_UA1}
\end{align}
and
\begin{align}
& \langle \bold{y}_1,a_1,\dots , a_{n_{1\uparrow} -1}; b_1,\dots , b_{n_{1\downarrow}} | \notag\\
& \qquad U_{\alpha}^{(n_{1\uparrow}, n_{1\downarrow})}  |   \bold{x}_1,a_1,\dots , a_{n_{1\uparrow} -1}; b_1,\dots , b_{n_{1\downarrow}} \rangle \notag\\
& \times \langle \bold{y}_2,c_1,\dots , c_{n_{2\uparrow} -1}; d_1,\dots , d_{n_{2\downarrow}} | \notag\\
& \qquad U_{\alpha}^{(n_{2\uparrow}, n_{2\downarrow})}  |   \bold{x}_2,c_1,\dots , c_{n_{2\uparrow} -1}; d_1,\dots , d_{n_{2\downarrow}} \rangle \notag\\
& \times \langle e_1| U_{\alpha}^{(1,0)}  | e_1 \rangle \cdots \langle e_{m_{1,0}}| U_{\alpha}^{(1,0)}  | e_{m_{1,0}} \rangle \notag\\
& \times \cdots \notag\\
& \times \langle f_1,\dots , f_{l_\uparrow};g_1,\dots , g_{l_\downarrow}
  |  U_{\alpha}^{(l_\uparrow, l_\downarrow)} | f_1,\dots , f_{l_\uparrow};g_1,\dots , g_{l_\downarrow} \rangle \notag\\ 
 & \qquad \cdots  
 \langle \, \dots ,f_{l_\uparrow m_{l_\uparrow, l_\downarrow}}; \dots ,g_{l_\downarrow m_{l_\uparrow, l_\downarrow}} | \notag\\
& \qquad\qquad\qquad U_{\alpha}^{(l_\uparrow, l_\downarrow)}   | \, \dots ,f_{l_\uparrow m_{l_\uparrow, l_\downarrow}}; \dots ,g_{l_\downarrow m_{l_\uparrow, l_\downarrow}} \rangle 
\notag\\
& \times \cdots .
\label{eq:2DenMat_WA_by_UA2}
\end{align}
In Eq.~(\ref{eq:2DenMat_WA_by_UA1}),
the set of integers $\{ m_{l_\uparrow, l_\downarrow} \}$ satisfies
$N_\sigma=n_\sigma+ \sum_{l_\sigma=1}^{\infty}l_\sigma m_{l_\uparrow, l_\downarrow}$,
and in Eq.~(\ref{eq:2DenMat_WA_by_UA2}) 
the set of integers $\{ m_{l_\uparrow, l_\downarrow} \}$ satisfies the conditions
$N_\sigma=n_{1\sigma}+n_{2\sigma}+  \sum_{l_\sigma=1}^{\infty}l_\sigma m_{l_\uparrow, l_\downarrow}$.
In general,  (i) the term (\ref{eq:2DenMat_WA_by_UA1}) is a product of 
one $U_{\alpha}^{(n_\uparrow, n_\downarrow)}$ which includes $\bold{x}_1$, $\bold{x}_2$, $\bold{y}_1$ and $\bold{y}_2$ as four of its variables,
and products of cluster functions $U_{\alpha}^{(l_\uparrow, l_\downarrow)} $,
which include other particles as variables,
and
(ii) the term (\ref{eq:2DenMat_WA_by_UA2}) is a product of 
one $U_{\alpha}^{(n_{1\uparrow}, n_{1\downarrow})}$ which includes $\bold{x}_1$ and $\bold{y}_1$ as two of its variables,
one $U_{\alpha}^{(n_{2\uparrow}, n_{2\downarrow})}$ which includes $\bold{x}_2$ and $\bold{y}_2$ as two of its variables
and products of cluster functions $U_{\alpha}^{(l_\uparrow, l_\downarrow)} $,
which include other particles as variables.
In accordance with Eq.~(\ref{eq:DM2-W-uu}),
we integrate $W_{\alpha}^{(N_\uparrow, N_\downarrow)}$.
Using Eqs.~(\ref{eq:2DenMat_WA_by_UA1}) and (\ref{eq:2DenMat_WA_by_UA2}),
we integrate $W_{\alpha}^{(N_\uparrow, N_\downarrow)}$
over coordinates $\{ 1_\uparrow,\dots , N_\uparrow -2; 1_\downarrow,\dots , N_\downarrow \}$.
We thus obtain
\begin{align}
& \langle \, \bold{y}_1 \uparrow , \bold{y}_2 \uparrow | \bar{W}_{\alpha}^{(N_\uparrow, N_\downarrow)} | \bold{x}_1 \uparrow , \bold{x}_2 \uparrow  \, \rangle \notag\\
& = \sum_{n_\uparrow = 2}^{N_\uparrow} \sum_{n_\downarrow = 0}^{N_\downarrow}
 \sum_{\{ m_{l_\uparrow, l_\downarrow} \}} \!\!\!
 \langle \, \bold{y}_1 \uparrow , \bold{y}_2 \uparrow | \bar{U}_{\alpha}^{(n_\uparrow, n_\downarrow)} | \bold{x}_1 \uparrow , \bold{x}_2 \uparrow  \, \rangle \notag\\
 & \quad \times
 \prod_{l_\uparrow = 0}^{N_\uparrow} 
 \prod_{l_\downarrow = 0}^{N_\downarrow }
 \frac{1}{m_{l_\uparrow, l_\downarrow} !}
  \left( B^{(l_\uparrow, l_\downarrow)} \right)^{m_{l_\uparrow, l_\downarrow}}\notag\\
& + 
\sum_{n_{1\uparrow} = 1}^{N_\uparrow} \sum_{n_{1\downarrow} = 0}^{N_\downarrow}
\sum_{n_{2\uparrow} = 1}^{N_\uparrow - n_{1\uparrow}} \sum_{n_{2\downarrow} = 0}^{N_\downarrow - n_{1\downarrow}}
 \sum_{\{ m_{l_\uparrow, l_\downarrow} \}} \notag\\
& \quad\,\, 
  \langle \, \bold{y}_1 \uparrow |  \bar{U}_{\alpha}^{(n_{1\uparrow}, n_{1\downarrow})}  | \bold{x}_1 \uparrow  \, \rangle 
  \langle \, \bold{y}_2 \uparrow |  \bar{U}_{\alpha}^{(n_{2\uparrow}, n_{2\downarrow})}  | \bold{x}_2 \uparrow  \, \rangle \notag\\
& \quad \times
 \prod_{l_\uparrow = 0}^{N_\uparrow} 
 \prod_{l_\downarrow = 0}^{N_\downarrow }
 \frac{1}{m_{l_\uparrow, l_\downarrow} !}
  \left( B^{(l_\uparrow, l_\downarrow)} \right)^{m_{l_\uparrow, l_\downarrow}}.
\label{eq:WA2-UA-uu:App}
\end{align}
Substituting Eqs.~(\ref{eq:Xi:App}) and (\ref{eq:WA2-UA-uu:App}) into Eq.~(\ref{eq:DM2-W-uu}),
we have
\begin{equation}
\begin{split}
&  \langle \bold{x}_1 \uparrow , \bold{x}_2 \uparrow | \rho^{(2)} | \bold{y}_1 \uparrow , \bold{y}_2 \uparrow \rangle  \\
& = 
 \sum_{n_\uparrow = 2}^\infty \sum_{n_\downarrow = 0}^\infty
   z_\uparrow^{n_\uparrow -2}
   z_\downarrow^{n_\downarrow}
 \langle \, \bold{y}_1 \uparrow , \bold{y}_2 \uparrow | \bar{U}_{\alpha}^{(n_\uparrow, n_\downarrow)} | \bold{x}_1 \uparrow , \bold{x}_2 \uparrow  \, \rangle \\ 
& + 
\Biggl( \,
 \sum_{n_{1\uparrow} = 1}^\infty \sum_{n_{1\downarrow} = 0}^\infty
   z_\uparrow^{n_{1\uparrow} -1}
   z_\downarrow^{n_{1\downarrow}} 
  \langle \, \bold{y}_1 \! \uparrow | 
 \bar{U}_{\alpha}^{(n_{1\uparrow}, n_{1\downarrow})} 
 | \bold{x}_1\! \uparrow  \, \rangle
 \Biggr)  \\
&  \,\,\, \times
\Biggl( \,
 \sum_{n_{2\uparrow} = 1}^\infty \sum_{n_{2\downarrow} = 0}^\infty
   z_\uparrow^{n_{2\uparrow} -1}
   z_\downarrow^{n_{2\downarrow}} 
  \langle \, \bold{y}_2 \! \uparrow | 
 \bar{U}_{\alpha}^{(n_{2\uparrow}, n_{2\downarrow})} 
 | \bold{x}_2\! \uparrow  \, \rangle
 \Biggr) .
\end{split}
\end{equation}
Then, using Eq.~(\ref{DM1-UA-u:App}) we obtain
\begin{align}
&  \langle \bold{x}_1 \uparrow , \bold{x}_2 \uparrow | \rho^{(2)} | \bold{y}_1 \uparrow , \bold{y}_2 \uparrow \rangle   \notag\\
&  = \langle \bold{x}_1 \uparrow  | \rho^{(1)} | \bold{y}_1 \uparrow \rangle
      \langle \bold{x}_2 \uparrow  | \rho^{(1)} | \bold{y}_2 \uparrow \rangle   \\
& \,\, +
 \sum_{n_\uparrow = 2}^\infty \sum_{n_\downarrow = 0}^\infty
   z_\uparrow^{n_\uparrow -2}
   z_\downarrow^{n_\downarrow}
 \langle \, \bold{y}_1 \uparrow , \bold{y}_2 \uparrow | \bar{U}_{\alpha}^{(n_\uparrow, n_\downarrow)} | \bold{x}_1 \uparrow , \bold{x}_2 \uparrow  \, \rangle . \notag
\end{align}
The proof of Eq.~(\ref{eq:DM2-UA-dd}) can simiarly be made.


Next, we prove Eq.~(\ref{eq:DM2-UA-ud}).
Following a procedure similar to the above derivation of Eq.~(\ref{eq:DM2-UA-uu}),
we obtain
\begin{align}
& \langle \, \bold{y}_1 \uparrow , \bold{y}_2 \downarrow | \bar{W}_{\alpha}^{(N_\uparrow, N_\downarrow)} | \bold{x}_1 \uparrow , \bold{x}_2 \downarrow  \, \rangle \notag\\
& = \sum_{n_\uparrow = 1}^{N_\uparrow} \sum_{n_\downarrow = 1}^{N_\downarrow}
 \sum_{\{ m_{l_\uparrow, l_\downarrow} \}} \!\!\!
 \langle \, \bold{y}_1 \uparrow , \bold{y}_2 \downarrow | \bar{U}_{\alpha}^{(n_\uparrow, n_\downarrow)} | \bold{x}_1 \uparrow , \bold{x}_2 \downarrow  \, \rangle \notag\\
 & \quad \times
 \prod_{l_\uparrow = 0}^{N_\uparrow} 
 \prod_{l_\downarrow = 0}^{N_\downarrow }
 \frac{1}{m_{l_\uparrow, l_\downarrow} !}
  \left( B^{(l_\uparrow, l_\downarrow)} \right)^{m_{l_\uparrow, l_\downarrow}}\notag\\
& + 
\sum_{n_{1\uparrow} = 1}^{N_\uparrow} \sum_{n_{1\downarrow} = 0}^{N_\downarrow}
\sum_{n_{2\uparrow} = 0}^{N_\uparrow - n_{1\uparrow}} \sum_{n_{2\downarrow} = 1}^{N_\downarrow - n_{1\downarrow}}
 \sum_{\{ m_{l_\uparrow, l_\downarrow} \}} \notag\\
& \quad\,\, 
  \langle \, \bold{y}_1 \uparrow |  \bar{U}_{\alpha}^{(n_{1\uparrow}, n_{1\downarrow})}  | \bold{x}_1 \uparrow  \, \rangle 
  \langle \, \bold{y}_2 \downarrow |  \bar{U}_{\alpha}^{(n_{2\uparrow}, n_{2\downarrow})}  | \bold{x}_2 \downarrow  \, \rangle \notag\\
& \quad \times
 \prod_{l_\uparrow = 0}^{N_\uparrow} 
 \prod_{l_\downarrow = 0}^{N_\downarrow }
 \frac{1}{m_{l_\uparrow, l_\downarrow} !}
  \left( B^{(l_\uparrow, l_\downarrow)} \right)^{m_{l_\uparrow, l_\downarrow}}.
\label{eq:WA2-UA-ud:App}
\end{align}
Substituting Eqs.~(\ref{eq:Xi:App}) and (\ref{eq:WA2-UA-ud:App}) into Eq.~(\ref{eq:DM2-W-ud}),
we find 
\begin{equation}
\begin{split}
&  \langle \bold{x}_1 \uparrow , \bold{x}_2 \downarrow | \rho^{(2)} | \bold{y}_1 \uparrow , \bold{y}_2 \downarrow \rangle   \\
& = \!\!
 \sum_{n_\uparrow = 1}^\infty \sum_{n_\downarrow = 1}^\infty
   z_\uparrow^{n_\uparrow -1}
   z_\downarrow^{n_\downarrow -1}
 \langle \, \bold{y}_1 \uparrow , \bold{y}_2 \downarrow | \bar{U}_{\alpha}^{(n_\uparrow, n_\downarrow)} | \bold{x}_1 \uparrow , \bold{x}_2 \downarrow  \, \rangle \\ 
& + 
\Biggl( \,
 \sum_{n_{1\uparrow} = 1}^\infty \sum_{n_{1\downarrow} = 0}^\infty
   z_\uparrow^{n_{1\uparrow} -1}
   z_\downarrow^{n_{1\downarrow}} 
  \langle \, \bold{y}_1 \! \uparrow | 
 \bar{U}_{\alpha}^{(n_{1\uparrow}, n_{1\downarrow})} 
 | \bold{x}_1\! \uparrow  \, \rangle
 \Biggr)  \\
&  \,\,\, \times
\Biggl( \,
 \sum_{n_{2\uparrow} = 0}^\infty \sum_{n_{2\downarrow} = 1}^\infty
   z_\uparrow^{n_{2\uparrow} }
   z_\downarrow^{n_{2\downarrow}-1} 
  \langle \, \bold{y}_2 \! \downarrow | 
 \bar{U}_{\alpha}^{(n_{2\uparrow}, n_{2\downarrow})} 
 | \bold{x}_2\! \downarrow  \, \rangle
 \Biggr) .
\end{split}
\end{equation}
Then, using Eq.~(\ref{DM1-UA-u:App}) we have
\begin{align}
&  \langle \bold{x}_1 \uparrow , \bold{x}_2 \downarrow | \rho^{(2)} | \bold{y}_1 \uparrow , \bold{y}_2 \downarrow \rangle   \notag\\
& = \langle \bold{x}_1 \uparrow  | \rho^{(1)} | \bold{y}_1 \uparrow \rangle
      \langle \bold{x}_2 \downarrow  | \rho^{(1)} | \bold{y}_2 \downarrow \rangle   \\
& + \!\!
 \sum_{n_\uparrow = 1}^\infty \sum_{n_\downarrow = 1}^\infty \!
   z_\uparrow^{n_\uparrow -1}
   z_\downarrow^{n_\downarrow -1}
 \langle \, \bold{y}_1 \uparrow , \bold{y}_2 \downarrow | \bar{U}_{\alpha}^{(n_\uparrow, n_\downarrow)} | \bold{x}_1 \uparrow , \bold{x}_2 \downarrow  \, \rangle .\notag
\end{align}

\subsection{Mathematical justification of quantum cluster expansion}

The above derivations can actually be mathematically justified.
We consider a uniform system or a trapped system.
To be rigorous, let us assume that the system is confined in an infinite potential well with a volume $V$ and the interaction $v$ has a hard repulsive core.
Then $W_{\alpha}^{(N_\uparrow, N_\downarrow)}$ vanishes for sufficiently large values of $N_\uparrow$ or $N_\downarrow$
and we denote their upper bounds as $M_\uparrow$ and $M_\downarrow$, respectively.

First, we demonstrate the validity of Eq.~(\ref{eq:Xi-b}).
The grand partition function reads
\begin{equation}
\begin{split}
\Xi
:= & \sum_{N_\uparrow = 0}^{M_\uparrow} \sum_{N_\downarrow = 0}^{M_\downarrow}
   \frac{z_\uparrow^{N_\uparrow}}{N_\uparrow!}
   \frac{z_\downarrow^{N_\downarrow}}{N_\downarrow!}
   Q^{(N_\uparrow, N_\downarrow)},
\label{eq:Xi-Q}
\end{split}
\end{equation}
where
\begin{align}
Q^{(N_\uparrow, N_\downarrow)}
:= & 
\int \! d^{3N_\uparrow}\bold{r}_{\uparrow} \int d^{3N_\downarrow}\bold{r}_{\downarrow} 
 \langle 1_\uparrow,\dots , N_\uparrow; 1_\downarrow,\dots , N_\downarrow| \notag\\
& \quad W_{\alpha}^{(N_\uparrow, N_\downarrow)}  | 
  1_\uparrow,\dots , N_\uparrow; 1_\downarrow,\dots , N_\downarrow \rangle ,
\end{align}
is a partition function.
Equation (\ref{eq:Xi-Q}) is positive and is thus a polynomial of $z_\uparrow$ and $z_\downarrow$ with no zeros on both the positive real axes.
Therefore, $\log \Xi (z_\uparrow, z_\downarrow)$ is holomorphic near the origin and along the positive real axes in both the complex $z_\uparrow$ plane and $z_\downarrow$ plane.
Near the origin this logarithm can be expanded as a power series,
and this power series is nothing but Eq.~(\ref{eq:Xi-b}).
If we understand Eq.~(\ref{eq:Xi-b}) to mean the analytic continuation,
Eq.~(\ref{eq:Xi-b}) is valid for all positive values of $z_\uparrow$ and $z_\downarrow$.

Next, we show the validity of Eq.~(\ref{eq:DM1-UA-u}).
Since $W_{\alpha}^{(N_\uparrow, N_\downarrow)}$ vanishes for sufficiently large values of $N_\uparrow$ or $N_\downarrow$
and $\Xi$ is a polynomial function of $z_\uparrow$ and $z_\downarrow$ with no zeros on their positive real axes,
from Eq.~(\ref{eq:1DenMat_and_W})
the one-particle reduced density matrix
$\langle \bold{x} \uparrow | \rho^{(1)} | \bold{y} \uparrow \rangle$
is a rational function of $z_\uparrow$ and $z_\downarrow$.
Therefore, the one-particle reduced density matrix is holomorphic near the origin and along the positive real axes in both the complex $z_\uparrow$ and $z_\downarrow$ planes.
Near the origin the one-particle reduced density matrix can be expanded as a power series,
and this power series is exactly Eq.~(\ref{eq:DM1-UA-u}).
If we understand Eq.~(\ref{eq:DM1-UA-u}) to mean the analytic continuation,
Eq.~(\ref{eq:DM1-UA-u}) is valid for all positive values of $z_\uparrow$ and $z_\downarrow$.
The two- and multi-particle reduced density matrices can be mathematically justified in a similar manner.

Note that, for the above justification, interaction $v$ needs to have a hard-repulsive core,
but the $s$-wave scattering length can be either positive or negative.


\section{$W_{\alpha}^{(l_\uparrow, l_\downarrow)}$
    in terms of $W^{(l_\uparrow, l_\downarrow)}$}
\label{app:W}

We prove Eq.~(\ref{eq:W-WA}).
The eigenfunctions of $H^{(N_\uparrow , N_\downarrow)}$ 
can be classified according to the tensor product of the irreducible representations 
of the permutation group of $N_\uparrow$ objects and that of $N_\downarrow$ objects.
If $\psi_i (1_\uparrow',\dots , N_\uparrow'; 1_\downarrow',\dots , N_\downarrow')$ 
belongs to an irreducible representation $D_\uparrow \otimes D_\downarrow$,
then $\epsilon^P \epsilon^Q  \psi_i (P(1_\uparrow'),\dots , P(N_\uparrow'); Q(1_\downarrow'),\dots , Q(N_\downarrow'))$ also belongs to the same representation $D_\uparrow \otimes D_\downarrow$. 
Hence 
\begin{equation}
\sum_{P \in S_{N_\uparrow}} \! \sum_{Q \in S_{N_\downarrow}} \!\!\!
\epsilon^P \epsilon^Q  \psi_i (P(1_\uparrow'),\dots , P(N_\uparrow'); Q(1_\downarrow'),\dots , Q(N_\downarrow'))
\label{eq:representation_perm-gr}
\end{equation}
belongs to $D_\uparrow \otimes D_\downarrow$. 
However, the function (\ref{eq:representation_perm-gr}) is symmetric. 
Hence, if $D_\uparrow \otimes D_\downarrow$ is
not the symmetric representation, (\ref{eq:representation_perm-gr}) is zero. 
On the other hand, if $D_\uparrow \otimes D_\downarrow$ is the symmetric representation,
then
\begin{align}
&\sum_{P \in S_{N_\uparrow}} \!\!\!\!\!\! \sum_{\quad Q \in S_{N_\downarrow}} \!\!\!\!\! \epsilon^P \epsilon^Q  \psi_i (P(1_\uparrow'),\dots , P(N_\uparrow'); Q(1_\downarrow'),\dots , Q(N_\downarrow')) \notag\\
& = (N_\uparrow !) (N_\downarrow !) \psi_i (1_\uparrow',\dots , N_\uparrow'; 1_\downarrow',\dots , N_\downarrow').
\end{align}
Using the definitions (\ref{eq:def:WA}) and (\ref{eq:def:W}), we obtain
Eq.~(\ref{eq:W-WA}).

\section{Correspondence between Lee-Yang and our notations}
\label{app:notation}

In this Appendix we list the correspondence between the Lee-Yang's notation \cite{LY60-IV} and ours in momentum space. 

For simplicity, we consider a one-component system.
The $\eta^{(1)}_0 (\bold{k})$ and $\eta^{(1)} (\bold{k})$ functions used in this paper are defined by $\eta^{(1)}_0 (\bold{k}) := (1-\epsilon z e^{-\beta k^2})^{-1}$ 
and $\eta^{(1)} (\bold{k}) := \langle \hat{\Psi}(\bold{k}) \hat{\Psi}^\dagger(\bold{k}) \rangle$.
The corresponding $m (\bold{k})$ and $M (\bold{k})$ functions used in the original paper of Lee and Yang \cite{LY60-IV} are defined by $m (\bold{k}) := z(1-\epsilon z e^{-\beta k^2})^{-1}$ 
and $M (\bold{k}) := z \{1 - \langle \hat{\Psi}^\dagger(\bold{k}) \hat{\Psi}(\bold{k}) \rangle \}$.
Thus we obtain
$m (\bold{k}) = z \eta^{(1)}_0 (\bold{k})$
and
$M (\bold{k}) = z \eta^{(1)} (\bold{k})$.
These relationships are listed in Table \ref{table}.

\begin{table}[htb]
\begin{center}
\caption{Correspondence between Lee-Yang and our notations}
\begin{tabular}{|l|cc|}
\hline
 & \, Lee-Yang \, & \, this paper \, \\ \hline
\, solid line & $z$ \, & $1$ \\
\, dotted line & $m (\bold{k})$ & $\eta^{(1)}_0 (\bold{k})$ \\
\, thick solid line \, & $M (\bold{k})$ & $\eta^{(1)} (\bold{k})$ \\ \hline
\, $l$-vertex & \, $\langle \cdots | \Upsilon^{(l)}| \cdots \rangle$ \quad & \quad $z^l \cdot \langle \cdots | \Upsilon^{(l)}| \cdots \rangle$ \, \\ \hline
\end{tabular}
\label{table}
\end{center}
\end{table}

\bibliographystyle{apsrev}

\end{document}